\documentclass[11pt]{amsart}
\usepackage{graphicx}
\usepackage{amsmath}
\usepackage{amssymb}
\usepackage{mathrsfs}
\usepackage[tight]{subfigure}
\usepackage{epsfig,epstopdf,psfrag}
\usepackage{hyperref,bm,color,fancyhdr}
\usepackage{algorithm,algorithmic}
\usepackage{array,mathtools}
\usepackage{booktabs}
\usepackage{natbib}
\newtheorem{theorem}{Theorem}

\newcommand{\PV}{\mathop{\rm P.V.}\nolimits}
\newcommand{\re}{\mathrm{e}}
\newcommand{\ri}{\mathrm{i}}

\providecommand{\abs}[1]{\lvert#1\rvert}
\providecommand{\norm}[1]{\lVert#1\rVert}

\newcommand{\rbs}[2]{\raisebox{#1}[0pt]{#2}}

\numberwithin{equation}{section}
\numberwithin{theorem}{section}
\numberwithin{figure}{subsection}
\numberwithin{table}{subsection}

\title[Finite-Time Singularity of 3D Euler]{Potentially Singular Solutions of the 3D Incompressible Euler Equations}
\author{Guo Luo$^{\dag}$ \and Thomas Y. Hou$^{\dag}$}
\thanks{$\dag$: Applied and Computational Mathematics, California Institute of Technology.}
\begin{document}
\maketitle

\begin{center}
\today
\end{center}

\section*{Abstract}
Whether the 3D incompressible Euler equations can develop a singularity in finite time from smooth initial data is one of the most challenging problems in mathematical fluid dynamics. This work attempts to provide an affirmative answer to this
long-standing open question from a numerical point of view, by presenting a class of potentially singular solutions to the Euler equations computed in axisymmetric geometries. The solutions satisfy a periodic boundary condition along the axial direction
and no-flow boundary condition on the solid wall. The equations are discretized in space using a hybrid 6th-order Galerkin and 6th-order finite difference method, on specially designed adaptive (moving) meshes that are dynamically adjusted to the
evolving solutions. With a maximum effective resolution of over $(3 \times 10^{12})^{2}$ near the point of the singularity, we are able to advance the solution up to $\tau_{2} = 0.003505$ and predict a singularity time of $t_{s} \approx 0.0035056$, while
achieving a \emph{pointwise} relative error of $O(10^{-4})$ in the vorticity vector $\omega$ and observing a $(3 \times 10^{8})$-fold increase in the maximum vorticity $\norm{\omega}_{\infty}$. The numerical data are checked against all major blowup
(non-blowup) criteria, including Beale-Kato-Majda, Constantin-Fefferman-Majda, and Deng-Hou-Yu, to confirm the validity of the singularity. A local analysis near the point of the singularity also suggests the existence of a self-similar blowup in the
meridian plane.

\section{Introduction}\label{sec_intro}
The celebrated 3D incompressible Euler equations in fluid dynamics describe the motion of ideal incompressible flows in the absence of external forcing. First written down by Leonhard Euler in 1757, these equations have the form
\begin{subequations}\label{eqn_eu}
\begin{align}
  u_{t} + u \cdot \nabla u & = -\nabla p, \label{eqn_eu_u} \\
  \nabla \cdot u & = 0, \label{eqn_eu_div}
\end{align}
\end{subequations}
where $u = (u_{1},u_{2},u_{3})^{T}$ is the 3D velocity vector of the fluid and $p$ is the scalar pressure. The 3D Euler equations have a rich mathematical theory, for which the interested readers may consult the excellent survey of \citet{gibbon2008} and
the references therein. This paper primarily concerns the existence or nonexistence of globally regular solutions to the 3D Euler equations, which is regarded as one of the most fundamental yet most challenging problems in mathematical fluid dynamics.

The interest in the global regularity or finite-time blowup of \eqref{eqn_eu} comes from several directions. Mathematically, the question has remained open for over 250 years and has a close connection to the Clay Millennium Prize Problem on the
Navier-Stokes equations\footnote{\texttt{http://www.claymath.org/millennium/Navier-Stokes\_Equations/navierstokes.pdf}}. Physically, the formation of a singularity in inviscid (Euler) flows may signify the onset of turbulence in viscous (Navier-Stokes)
flows, and may provide a mechanism for energy transfer to small scales. Numerically, the resolution of nearly singular flows requires special numerical techniques, which presents a great challenge to computational fluid dynamicists.

Much efforts have been devoted to the analysis of the 3D Euler equations in the past. The basic question of local well-posedness was addressed by \citet{kato1972} using vanishing-viscosity techniques. As for global regularity, the most well-known result
is due to Beale-Kato-Majda \citep{bkm1984}, which states that a smooth solution $u$ of the 3D Euler equations blows up at $t = T$ if and only if
\begin{displaymath}
  \int_{0}^{T} \norm{\omega(\cdot,t)}_{L^{\infty}}\,dt = \infty,
\end{displaymath}
where $\omega = \nabla \times u$ is the \emph{vorticity vector} of the fluid. Another important result concerning the global regularity of the 3D Euler equations is the geometric non-blowup criterion of Constantin-Fefferman-Majda \citep{cfm1996}. It
states that there can be no blowup for the 3D Euler equations if the velocity field $u$ is uniformly bounded and the vorticity direction $\xi = \omega/\abs{\omega}$ is sufficiently ``well-behaved'' near the point of the maximum vorticity. A local
Lagrangian version of the Constantin-Fefferman-Majda criterion was also proved by Deng-Hou-Yu in \citet{dhy2005}.

Besides the analytical results mentioned above, there also exists a sizable literature focusing on the numerical search for a finite-time singularity of the 3D Euler equations. Representative work in this direction include that of \citet{gs1991} and
\citet{ps1992}, who studied Euler flows with swirls in axisymmetric geometries, the work of \citet{kerr1993}, who studied Euler flows generated by a pair of perturbed antiparallel vortex tubes, and the work of \citet{bp1994}, who studied the 3D
Navier-Stokes equations using Kida's high-symmetry initial data. Another interesting piece of work is that of \citet{caflisch1993} and \citet{sc2009}, who studied axisymmetric Euler flows with \emph{complex} initial data and reported singularities in the
complex plane. The review article of \citet{gibbon2008} contains a short survey of the above results and many other interesting numerical studies.

Although finite-time singularities were frequently reported in numerical simulations of the 3D Euler equations, most such singularities turned out to be either numerical artifacts or false predictions, as a result of either insufficient resolution or
inadvertent line extrapolation procedure (more to follow on this topic in Section \ref{ssec_vfit}). Indeed, by exploiting the analogy between the 2D Boussinesq equations and the 3D axisymmetric Euler equations away from the symmetry axis, \citet{es1994}
studied the potential development of finite-time singularities in the 2D Boussinesq equations, with initial data completely analogous to those of \citet{gs1991} and \citet{ps1992}. They found no evidence for singular solutions, indicating that the
``blowups'' reported by those authors, which were located away from the axis, are almost certainly numerical artifacts. Likewise, \citet{hl2006} repeated the computation of \citet{kerr1993} with higher resolutions. Despite some ambiguity in reproducing
the initial data used by \citet{kerr1993}, they computed the solution up to $t = 19$, which is beyond the singularity time $T = 18.7$ alleged by \citet{kerr1993}. In a later work, \citet{hl2008} also repeated the computation of \citet{bp1994}. They found
that the singularity reported by those authors is likely an artifact due to under-resolution. In short, there is no conclusive numerical evidence on the existence of a finite-time singularity at the time of writing, and the question whether initially
smooth solutions to \eqref{eqn_eu} can blow up in finite time remains open.

By focusing on solutions with axial symmetry and special odd-even symmetries along the axial direction, we have carried out a careful numerical study of the 3D Euler equations in cylindrical geometries, and discovered a class of potentially singular
solutions with a ring-like singularity set on the solid boundary. The reduced computational complexity in the cylindrical geometry greatly facilitates our computations; with a specially designed adaptive mesh, we are able to achieve a maximum point
density of over $(3 \times 10^{12})^{2}$ per unit area near the point of the singularity. This allows us to achieve a $(3 \times 10^{8})$-fold increase in maximum vorticity with a \emph{pointwise} relative error of $O(10^{-4})$ in vorticity. The
numerical data are checked against all major blowup (non-blowup) criteria, including Beale-Kato-Majda, Constantin-Fefferman-Majda, and Deng-Hou-Yu, using a carefully designed line fitting procedure. A careful local analysis also suggests that the
blowing-up solution develops a self-similar structure in the meridian plane near the point of the singularity, as the singularity time is approached. Our computational method makes explicit use of the special symmetries built in the blowing-up solutions,
which eliminates symmetry-breaking perturbations and facilitates a stable computation of the singularity.

The main features of the potentially singular solutions are summarized as follows. The point of the potential singularity, which is also the point of the maximum vorticity, is always located at the intersection of the solid boundary $r = 1$ and the
symmetry plane $z = 0$. It is a \emph{stagnation point} of the flow, as a result of the special odd-even symmetries along the axial direction and the no-flow boundary condition (see \eqref{eqn_noflow}), and the vanishing velocity field at this point
could have positively contributed to the formation of the singularity given the potential regularizing effect of convection as observed by \citet{hl2009}. When viewed in the meridian plane, the point of the potential singularity is a \emph{hyperbolic
saddle} point of the flow, where the axial flow along the solid boundary marches toward the symmetry plane $z = 0$ and the radial flow marches toward the symmetry axis $r = 0$ (see Figure \ref{fig_flow}\subref{fig_flow_loc}). The axial flow brings
together the vortex lines near the solid boundary $r = 1$ and destroys the geometric regularity of the vorticity vector near the symmetry plane $z = 0$, violating the geometric non-blowup criteria of Constantin-Fefferman-Majda and Deng-Hou-Yu and leading
to the breakdown of the smooth vorticity field.

The asymptotic scalings of the various quantities involved in the potential finite-time blowup are summarized as follows. Near the predicted singularity time $t_{s}$, the scalar pressure and the velocity field remain uniformly bounded while the maximum
vorticity blows up like an inverse power-law $O(t_{s}-t)^{-\gamma}$, where $\gamma$ roughly equals $\frac{5}{2}$. Near the point of the potential singularity, namely the point of the maximum vorticity, the radial and axial components of the vorticity
vector grow roughly like $O(t_{s}-t)^{-5/2}$ while the angular vorticity grows like $O(t_{s}-t)^{-1}$. The nearly singular solution has a locally self-similar structure in the meridian plane near the point of blowup, with a rapidly collapsing support
scaling roughly like $O(t_{s}-t)^{3}$ along both the radial and the axial directions. When viewed in $\mathbb{R}^{3}$, this corresponds to a thin tube on the symmetry plane $z = 0$ evolved around the ring $r = 1$, where the radius of the tube shrinks to
zero as the singularity forms.

The rest of this paper is devoted to the study of the potential finite-time singularity and is organized as follows. Section \ref{sec_eqn} contains a brief review of the 3D Euler equations in axisymmetric form and defines the problem to be studied.
Section \ref{sec_method} gives a brief description of the numerical method that is used to track and resolve the nearly singular solutions. Section \ref{sec_result} examines the numerical data in great detail and presents evidence supporting the
existence of a finite-time singularity. Finally conclusions and discussions are given in Section \ref{sec_conclu}.

\section{Description of the Problem}\label{sec_eqn}
Recall the 3D Euler equations (see \eqref{eqn_eu})
\begin{align*}
  u_{t} + u \cdot \nabla u & = -\nabla p, \\
  \nabla \cdot u & = 0,
\end{align*}
where $u = (u_{1},u_{2},u_{3})^{T}$ is the 3D velocity vector, $p$ is the scalar pressure, and $\nabla = (\partial_{1},\partial_{2},\partial_{3})^{T}$ is the gradient operator in $\mathbb{R}^{3}$. By taking the curl on both sides, the equations can be
recast in the equivalent stream-vorticity form
\begin{displaymath}
  \omega_{t} + u \cdot \nabla \omega = \omega \cdot \nabla u,
\end{displaymath}
where $\omega = \nabla \times u$ is the 3D vorticity vector. The velocity $u$ is related to the vorticity $\omega$ via the vector-valued stream function $\psi$:
\begin{displaymath}
  -\Delta \psi = \omega,\qquad u = \nabla \times \psi.
\end{displaymath}
For flows that are symmetric about a fixed axis in space (say the $z$-axis), it is convenient to rewrite equations \eqref{eqn_eu} in cylindrical coordinates. Introducing the change of variables:
\begin{displaymath}
  x = r \cos\theta,\qquad y = r \sin\theta,\qquad z = z,
\end{displaymath}
and the decomposition
\begin{gather*}
  v(r,z) = v^{r}(r,z)\, e_{r} + v^{\theta}(r,z)\, e_{\theta} + v^{z}(r,z)\, e_{z}, \\
  e_{r} = \frac{1}{r}\, (x,y,0)^{T},\qquad e_{\theta} = \frac{1}{r}\, (-y,x,0)^{T},\qquad e_{z} = (0,0,1)^{T},
\end{gather*}
for radially symmetric vector functions $v(r,z)$, the 3D Euler equations \eqref{eqn_eu} can be written in the axisymmetric form (for details of derivation, see \citet{mb2002}):
\begin{subequations}\label{eqn_eu_axi}
\begin{align}
  u^{\theta}_{t} + u^{r} u^{\theta}_{r} + u^{z} u^{\theta}_{z} & = -\frac{1}{r}\, u^{r} u^{\theta}, \label{eqn_eu_axi_u} \\
  \omega^{\theta}_{t} + u^{r} \omega^{\theta}_{r} + u^{z} \omega^{\theta}_{z} & = \frac{2}{r}\, u^{\theta} u_{z}^{\theta} + \frac{1}{r}\, u^{r} \omega^{\theta}, \label{eqn_eu_axi_w} \\
  -\bigl[ \Delta - (1/r^{2}) \bigr] \psi^{\theta} & = \omega^{\theta}. \label{eqn_eu_axi_psi}
\end{align}
Here $u^{\theta},\ \omega^{\theta}$, and $\psi^{\theta}$ are the angular components of the velocity, vorticity, and stream function vectors, respectively. The radial ($r$) and axial ($z$) components of the velocity vector can be recovered from the
angular stream function $\psi^{\theta}$, via the relations:
\begin{equation}
  u^{r} = -\psi^{\theta}_{z},\qquad u^{z} = \frac{1}{r} (r \psi^{\theta})_{r},
  \label{eqn_eu_axi_urz}
\end{equation}
\end{subequations}
for which the incompressibility condition
\begin{displaymath}
  \frac{1}{r}\, (ru^{r})_{r} + u_{z}^{z} = 0
\end{displaymath}
is automatically satisfied. Equations \eqref{eqn_eu_axi}, together with appropriate initial and boundary conditions, completely determine the evolution of 3D axisymmetric Euler flows.

The axisymmetric Euler equations \eqref{eqn_eu_axi} have a formal singularity at $r = 0$, which sometimes is inconvenient to work with. To remove this singularity, \citet{hl2008} introduced the variables\footnote{These variables should not be confused
with the components of the velocity, vorticity, and stream function vectors.}:
\begin{displaymath}
  u_{1} = u^{\theta}/r,\qquad \omega_{1} = \omega^{\theta}/r,\qquad \psi_{1} = \psi^{\theta}/r,
\end{displaymath}
and transformed equations \eqref{eqn_eu_axi} into the form:
\begin{subequations}\label{eqn_eat}
\begin{align}
  u_{1,t} + u^{r} u_{1,r} + u^{z} u_{1,z} & = 2 u_{1} \psi_{1,z}, \label{eqn_eat_u} \\
  \omega_{1,t} + u^{r} \omega_{1,r} + u^{z} \omega_{1,z} & = (u_{1}^{2})_{z}, \label{eqn_eat_w} \\
  -\bigl[ \partial_{r}^{2} + (3/r) \partial_{r} + \partial_{z}^{2} \bigr] \psi_{1} & = \omega_{1}. \label{eqn_eat_psi}
\end{align}
In terms of the new variables, the radial and axial components of the velocity vector are given by
\begin{equation}
  u^{r} = -r \psi_{1,z},\qquad u^{z} = 2\psi_{1} + r \psi_{1,r}.
  \label{eqn_eat_urz}
\end{equation}
\end{subequations}
As shown by \citet{lw2006}, $u^{\theta},\ \omega^{\theta}$, and $\psi^{\theta}$ must all vanish at $r = 0$ if $u$ is a smooth velocity field. Thus $u_{1},\ \omega_{1}$, and $\psi_{1}$ are well defined as long as the corresponding solution to
\eqref{eqn_eu_axi} remains smooth.

We shall numerically solve the transformed equations \eqref{eqn_eat} on the cylinder
\begin{displaymath}
  D(1,L) = \Bigl\{ (r,z)\colon 0 \leq r \leq 1,\ 0 \leq z \leq L \Bigr\},
\end{displaymath}
with the initial data
\begin{subequations}\label{eqn_eat_ibc}
\begin{equation}
  u_{1}^{0}(r,z) = 100\, \re^{-30(1-r^{2})^{4}} \sin \Bigl( \frac{2\pi}{L} z \Bigr),\qquad \omega_{1}^{0}(r,z) = \psi_{1}^{0}(r,z) = 0.
  \label{eqn_eat_ic}
\end{equation}
The solution is subject to a periodic boundary condition in $z$:
\begin{equation}
  u_{1}(r,0,t) = u_{1}(r,L,t),\qquad \omega_{1}(r,0,t) = \omega_{1}(r,L,t),\qquad \psi_{1}(r,0,t) = \psi_{1}(r,L,t),
  \label{eqn_eat_bc_z}
\end{equation}
and a no-flow boundary condition on the solid boundary $r = 1$:
\begin{equation}
  \psi_{1}(1,z,t) = 0.
  \label{eqn_eat_bc_r}
\end{equation}
The pole condition
\begin{equation}
  u_{1,r}(0,z,t) = \omega_{1,r}(0,z,t) = \psi_{1,r}(0,z,t) = 0
  \label{eqn_eat_pc}
\end{equation}
\end{subequations}
is also enforced at the symmetry axis $r = 0$ to ensure the smoothness of the solution.

It is not difficult to see that the initial data \eqref{eqn_eat_ic} has the properties that $u_{1}^{0}$ is even at $z = \frac{1}{4} L,\ \frac{3}{4} L$, odd at $z = 0,\ \frac{1}{2} L$, and $\omega_{1}^{0},\ \psi_{1}^{0}$ are both odd at $z = 0,\
\frac{1}{4} L,\ \frac{1}{2} L,\ \frac{3}{4} L$. These symmetry properties are preserved by the equations \eqref{eqn_eat}, so instead of solving the problem \eqref{eqn_eat}--\eqref{eqn_eat_ibc} on the entire cylinder $D(1,L)$, it suffices to consider the
problem on the quarter cylinder $D(1,\frac{1}{4} L)$, with the periodic boundary condition \eqref{eqn_eat_bc_z} replaced by appropriate symmetry boundary conditions. It is also interesting to notice that the boundaries of $D(1,\frac{1}{4} L)$ behave like
``impermeable walls'', which is a consequence of the no-flow boundary condition \eqref{eqn_eat_bc_r} and the odd symmetry of $\psi_{1}$:
\begin{equation}
  u^{r} = -r \psi_{1,z} = 0\quad \text{on}\quad r = 1,\qquad u^{z} = 2\psi_{1} + r \psi_{1,r} = 0\quad \text{on}\quad z = 0,\, \tfrac{1}{4} L.
  \label{eqn_noflow}
\end{equation}

\section{Outline of the Numerical Method}\label{sec_method}
The potential formation of a finite-time singularity from the initial-boundary value problem \eqref{eqn_eat}--\eqref{eqn_eat_ibc} makes the numerical solution of the problem a challenging and difficult task. In this section, we describe a special mesh
adaptation strategy (Section \ref{ssec_mm}) and a B-spline based Galerkin Poisson solver (Section \ref{ssec_psolv}), which are essential to the accurate computation of the nearly singular solutions generated from \eqref{eqn_eat}--\eqref{eqn_eat_ibc}. The
overall algorithm for solving \eqref{eqn_eat}--\eqref{eqn_eat_ibc} is outlined in Section \ref{ssec_alg}.

\subsection{The Adaptive (Moving) Mesh Algorithm}\label{ssec_mm}
Singularities (blowups) are abundant in mathematical models for physical processes. Examples include the semilinear parabolic equations describing the blowup of the temperature of a reacting medium, such as a burning gas \citep{fujita1966}; the nonlinear
Schr\"{o}dinger equations describing the self-focusing of electromagnetic beams in a nonlinear medium \citep{mpss1986}; and the aggregation equations describing the concentration of interacting particles \citep{hb2010}. Often, singularities occur on
increasingly small length and time scales, which necessarily requires some form of mesh adaptation. Further, finite-time singularities usually evolve in a ``self-similar'' manner when singularity time is approached. An adaptive mesh designed for
singularity detection must also reproduce this behavior in the numerical solution.

Several methods have been proposed to compute (self-similar) singularities. \citet{mpss1986} used a dynamic rescaling algorithm to solve the cubic Schr\"{o}dinger equation. The main advantage of the method is that the rescaled equation is nonsingular and
the rescaled variable is uniformly bounded in appropriate norms. The disadvantage is that the fixed-sized mesh is spread apart by rescaling, so accuracy is inevitably lost far from the singularity.

\citet{bk1988} proposed a rescaling algorithm for the numerical solution of the semilinear heat equation, based on the idea of adaptive mesh refinement. The method repeatedly refines the mesh in the ``inner'' region of the singularity and rescales the
inner solution so that it remains uniformly bounded. The main advantage of the method is that it achieves uniform accuracy across the entire computational domain, and is applicable to more general problems. The disadvantage is that it requires \emph{a
priori} knowledge of the singularity, and is not easily adaptable to elliptic equations (especially in multiple space dimensions) due to the use of irregular mesh.

The moving mesh method of \citet{hrr1994} provides a very general framework for mesh adaptation and has been applied in various contexts, for example the semilinear heat equation \citep{bhr1996} and the nonlinear Schr\"{o}dinger equation \citep{bcr1999}.
The main idea of the method is to construct the mesh based on certain equidistribution principle, for example the equipartition of the arc length function. In one-dimension this completely determines the mesh, while in higher dimensions additional
constraints are needed to specify mesh shape and orientation. The meshes are automatically evolved with the underlying solution, typically by solving a moving mesh partial differential equation (MMPDE).

While being very general, the ``conventional'' moving mesh method has the following issues when applied to singularity detection. First, it requires explicit knowledge of the singularity, for example its scaling exponent, in order to correctly capture
the singularity \citep{hmr2008}. Second, it tends to place too many mesh points near the singularity while leaving too few elsewhere, which can cause instability. Third, mesh smoothing, an operation necessary for maintaining stability, can significantly
limit the maximum resolution power of the mesh. Finally, the moving mesh method computes only a \emph{discrete approximation} of the mesh mapping function, which can result in catastrophic loss of accuracy in the computation of a singularity (see Section
\ref{ssec_alg}).

For the particular blowup candidate considered in this paper, preliminary uniform-mesh computations suggest that the vorticity function tends to concentrate at a single point. In addition, the solution appears to remain slowly-varying and smooth outside
a small neighborhood of the singularity. These observations motivate the following special mesh adaptation strategy.

The adaptive mesh covering the computational domain $D(1,\frac{1}{4} L)$ is constructed from a pair of analytic mesh mapping functions:
\begin{displaymath}
  r = r(\rho),\qquad z = z(\eta),
\end{displaymath}
where each mesh mapping function is defined on $[0,1]$, is infinitely differentiable, and has a density that is even at both 0 and 1. The even symmetries of the mesh density ensure that the resulting mesh can be \emph{smoothly} extended to the full
cylinder $D(1,L)$. The mesh mapping functions contain a small number of parameters, which are dynamically adjusted so that along each dimension a certain fraction (e.g. 50\%) of the mesh points is placed in a small neighborhood of the singularity. Once
the mesh mapping functions are constructed, the computational domain $D(1,\frac{1}{4} L)$ is covered with a tensor-product mesh:
\begin{displaymath}
  \mathcal{G}_{0} = \Bigl\{ (r_{j},z_{i})\colon 0 \leq i \leq M,\ 0 \leq j \leq N \Bigr\},
\end{displaymath}
where
\begin{displaymath}
  r_{j} = r(jh_{r}),\ z_{i} = z(ih_{z}),\qquad h_{r} = 1/N,\ h_{z} = 1/M.
\end{displaymath}
The precise definition and construction of the mesh mapping functions are detailed in Appendix \ref{app_mm}.

The mesh is evolved using the following procedure. Starting from a reference time $t_{0}$, the ``singularity region'' $S_{0}$ at $t_{0}$ is identified as the smallest rectangle in the $rz$-plane that encloses the set
\begin{displaymath}
  D_{\delta_{0}}(t_{0}) := \Bigl\{ (r,z) \in D(1,\tfrac{1}{4} L)\colon \abs{\omega(r,z,t_{0})} \geq \delta_{0} \norm{\omega(\cdot,t_{0})}_{\infty} \Bigr\},\qquad \delta_{0} \in (0,1).
\end{displaymath}
Once $S_{0}$ is determined, an adaptive mesh $\mathcal{G}_{0}$ is fit to $S_{0}$ and the solution is advanced in the $\rho\eta$-space by one time step to $t_{1}$. The singularity region $S_{1}$ at $t_{1}$ is then computed and compared with $S_{0}$. If
the ratios between the sides of $S_{1}$ and $S_{0}$ (in either dimension) drop below a certain threshold (e.g. 80\%), which indicates the support of the maximum vorticity has shrunk by a sufficient amount, or if the maximum vorticity at $t_{1}$ is ``too
close'' to the boundaries of $S_{0}$:
\begin{equation}
  \max_{(r,z) \in \partial S_{0}} \abs{\omega(r,z,t_{1})} \geq \delta_{1} \norm{\omega(\cdot,t_{1})}_{\infty},\qquad \delta_{1} \in (\delta_{0},1),
  \label{eqn_mm_update_2}
\end{equation}
which indicates the maximum vorticity is about to leave $S_{0}$, then a new mesh $\mathcal{G}_{1}$ is computed and adapted to $S_{1}$. In the event of a mesh update, the solution is interpolated from $\mathcal{G}_{0}$ to $\mathcal{G}_{1}$ in the
$\rho\eta$-space using an 8th-order piecewise polynomial interpolation in $\rho$ and a spectral interpolation in $\eta$. The whole procedure is then repeated with $\mathcal{G}_{0}$ replaced by $\mathcal{G}_{1}$ and $t_{0}$ replaced by $t_{1}$.

We remark that the mesh update criterion \eqref{eqn_mm_update_2} is designed to prevent the peak vorticity from escaping the singularity region, as is the case in one of our earlier computations where the singularity keeps moving toward the symmetry
axis. Since in the current computation the singularity is fixed at the corner $\tilde{q}_{0} = (1,0)^{T}$, the criterion \eqref{eqn_mm_update_2} has practically no effect.

The mesh adaptation strategy described above has several advantages compared with the conventional moving mesh method. First, it can automatically resolve a self-similar singularity regardless of its scalings, provided that the singularity has a
bell-shaped similarity profile, which is what we observe in our case (see Figure \ref{fig_mesh_vort}\subref{fig_mesh_vort_re}). This is crucial to the success of our computations, because the (axisymmetric) Euler equations allow for infinitely many
self-similar scalings (see Section \ref{ssec_selfsim}), which means that the scaling exponent of the singularity cannot be determined \emph{a priori}. Second, the method always places enough mesh points (roughly 50\% along each dimension) outside the
singularity region, ensuring a well-behaved and stable mesh (see Section \ref{ssec_mm_effec}). Third, the explicit control of the mesh mapping functions eliminates the need of mesh smoothing, which allows the mesh to achieve arbitrarily high resolutions.
Finally, the analytic representation of the mesh mapping functions ensures accurate approximations of space derivatives, hence greatly improving the quality of the computed solutions (see Section \ref{ssec_alg}).

\subsection{The B-Spline Based Galerkin Poisson Solver}\label{ssec_psolv}
One of the key observations we have made from our computations is that the overall accuracy of the computed solutions depends crucially on the accuracy of the Poisson solver. Among the methods commonly used for solving Poisson equations, namely finite
difference, finite element Galerkin, and finite element collocation, we have chosen the Galerkin method, both for its high accuracy and for its rigorous theoretic framework, which makes the error analysis much easier.

We have designed and implemented a B-spline based Galerkin method for the Poisson equation \eqref{eqn_eat_psi}. Compared with the ``conventional'' Galerkin methods based on piecewise polynomials, the B-spline based method requires no mesh generation and
hence is much easier to implement. More importantly, the method can achieve \emph{arbitrary global} smoothness and approximation order with relative ease and few degrees of freedom, in contrast to the conventional piecewise polynomial based methods. This
makes the method a natural choice for our problem.

The Poisson equation \eqref{eqn_eat_psi} is solved in the $\rho\eta$-space using the following procedure. First, the equation is recast in the $\rho\eta$-coordinates:
\begin{displaymath}
  -\frac{1}{r^{3} r_{\rho}} \biggl( r^{3}\, \frac{\psi_{\rho}}{r_{\rho}} \biggr)_{\rho} - \frac{1}{z_{\eta}} \biggl( \frac{\psi_{\eta}}{z_{\eta}} \biggr)_{\eta} = \omega,\qquad (\rho,\eta) \in [0,1]^{2},
\end{displaymath}
where for clarity we have written $\psi$ for $\psi_{1}$ and $\omega$ for $\omega_{1}$. Next, the equation is multiplied by $r^{3} r_{\rho} z_{\eta} \phi$ for a suitable test function $\phi \in V$ (to be defined below) and is integrated over the domain
$[0,1]^{2}$. After a routine integration by parts, this yields the desired weak formulation of \eqref{eqn_eat_psi}, which reads: find $\psi \in V$ such that
\begin{subequations}\label{eqn_psolv}
\begin{align}
  a(\psi,\phi) & := \int_{[0,1]^{2}} \biggl[ \frac{\psi_{\rho}}{r_{\rho}} \frac{\phi_{\rho}}{r_{\rho}} + \frac{\psi_{\eta}}{z_{\eta}} \frac{\phi_{\eta}}{z_{\eta}} \biggr] r^{3} r_{\rho} z_{\eta}\,d\rho\,d\eta \nonumber \\
  & = \int_{[0,1]^{2}} \omega \phi r^{3} r_{\rho} z_{\eta}\,d\rho\,d\eta =: f(\phi),\qquad \forall \phi \in V, \label{eqn_psolv_v}
\end{align}
where (recall the odd symmetry of $\psi$ at $z = 0,\ \frac{1}{4} L$)
\begin{align*}
  V & = \text{span} \Bigl\{ \phi \in H^{1}[0,1]^{2}\colon \phi(-\rho,\eta) = \phi(\rho,\eta), \\
  &\hspace{1.56in} \phi(1,\eta) = 0,\ \phi(\rho,\ell-\eta) = -\phi(\rho,\ell+\eta),\ \forall \ell \in \mathbb{Z} \Bigr\}.
\end{align*}

To introduce Galerkin approximation, define the finite-dimensional subspace of \emph{weighted uniform B-splines} \citep{hollig2003} of even order $k$:
\begin{displaymath}
  V_{h} := V_{w,h}^{k} = \text{span} \Bigl\{ w(\rho) b_{j,h_{r}}^{k}(\rho) b_{i,h_{z}}^{k}(\eta) \Bigr\} \cap V,
\end{displaymath}
where $w(\rho)$ is a nonnegative weight function of order 1 vanishing on $\rho = 1$:
\begin{displaymath}
  w(\rho) \sim (1-\rho),\qquad \rho \to 1^{-},
\end{displaymath}
and $b_{\ell,h}^{k}(s) = b^{k}((s/h) - (\ell-k/2))$ is the shifted and rescaled uniform B-spline of order $k$. The Galerkin formulation then reads: find $\psi_{h} \in V_{h}$ such that
\begin{equation}
  a(\psi_{h},\phi_{h}) = f(\phi_{h}),\qquad \forall \phi_{h} \in V_{h}.
  \label{eqn_psolv_vh}
\end{equation}
\end{subequations}
With suitably chosen basis functions of $V_{h}$, this gives rise to a symmetric, positive definite linear system $Ax = b$ which can be solved to yield the Galerkin solution $\psi_{h}$. The details are given in Appendix \ref{app_bsp}.

The parameters used in our computations are $k = 6$ and $w(\rho) = 1-\rho^{2}$.

Using the theory of quasi-interpolants, it can be shown that
\begin{equation}
  \int_{[0,1]^{2}} \abs{\nabla \psi - \nabla \psi_{h}}^{2} r^{3}\,dr\,dz \leq C_{0} C_{rz} h_{r}^{k-1} h_{z}^{k-1} \int_{[0,1]^{2}} \sum_{\abs{\alpha} \leq k-1} \abs{\tilde{\partial}^{\alpha} \nabla \psi}^{2} r^{3}\,dr\,dz,
  \label{eqn_psolv_gerr}
\end{equation}
where $\nabla = (\partial_{r},\partial_{z})^{T},\ \tilde{\partial}^{\alpha} = \partial_{\rho}^{\alpha_{1}} \partial_{\eta}^{\alpha_{2}}$ are differential operators in $rz$- and $\rho\eta$-planes, respectively, $C_{rz}$ is a mesh-mapping dependent
constant, and $C_{0}$ is an absolute constant. In our computations, the constant $C_{rz}$ is observed to be very close to 1 for all times, which confirms the stability of the Galerkin solver.

The detailed error analysis of the Poisson solver will be reported in a separate paper.

\subsection{The Overall Algorithm}\label{ssec_alg}
Given an adaptive mesh $\mathcal{G}_{0}$ and the data $(u_{1},\omega_{1})$ defined on it, the solution is advanced on $\mathcal{G}_{0}$ using the following procedure. First, the Poisson equation \eqref{eqn_eat_psi} is solved for $\psi_{1}$ in the
$\rho\eta$-space using a 6th-order B-spline based Galerkin method (Section \ref{ssec_psolv}). Second, the 2D velocity $\tilde{u} = (u^{r},u^{z})^{T}$ is evaluated at the grid points using \eqref{eqn_eat_urz}. Third, an adaptive time step $\delta_{t}$ is
computed on $\mathcal{G}_{0}$ so that the CFL condition is satisfied with a suitably small CFL number $\nu$ (e.g. 0.5), and the relative growth of the solution in one step remains below a certain threshold $\epsilon_{t}$ (e.g. 5\%). Finally, the solution
$(u_{1},\omega_{1})$ is advanced by $\delta_{t}$ using an explicit 4th-order Runge-Kutta method, and the mesh $\mathcal{G}_{0}$ is adapted to the new solution if necessary (Section \ref{ssec_mm}).

In the last step of the algorithm, the evolution equations for $u_{1}$ and $\omega_{1}$ are semi-discretized in the $\rho\eta$-space, where the space derivatives are expressed in the $\rho\eta$-coordinates and are approximated using 6th-order centered
difference schemes, e.g.
\begin{displaymath}
  v_{r}(r_{j},z_{i}) =: (v_{r})_{ij} = \frac{(v_{\rho})_{ij}}{(r_{\rho})_{j}} \approx \frac{1}{(r_{\rho})_{j}} (Q_{\rho,6} v_{i,\cdot})_{j},\qquad v = u_{1}\quad \text{or}\quad \omega_{1}.
\end{displaymath}
Here, as usual,
\begin{displaymath}
  Q_{\rho,6} := D_{\rho,0} \biggl\{ I - \frac{1}{6}\, h_{r}^{2} D_{\rho,+} D_{\rho,-} + \frac{1}{30}\, h_{r}^{4} D_{\rho,+}^{2} D_{\rho,-}^{2} \biggr\},
\end{displaymath}
denotes the standard 6th-order centered approximation to $\partial_{\rho}$, and
\begin{displaymath}
  (D_{\rho,\pm} v_{i,\cdot})_{j} := \pm \frac{1}{h_{r}} (v_{i,j \pm 1} - v_{i,j}),\qquad (D_{\rho,0} v_{i,\cdot})_{j} := \frac{1}{2h_{r}} (v_{i,j+1} - v_{i,j-1}),
\end{displaymath}
denote the standard forward, backward, and centered difference operators, respectively. Note that the derivative $r_{\rho}$ of the mesh mapping function is computed directly from the analytic representation of $r$ without any difference approximation.
This is crucial for the accurate evaluation of $v_{r}$, especially in ``singularity regions'' where the inverse mesh density $r_{\rho}$ is close to 0 and is nearly constant (Appendix \ref{app_mm}; in particular, see \eqref{eqn_mm_alpha}). When $r_{\rho}$
is small and nearly constant, a high-order difference approximation of $r_{\rho}$ tends to be contaminated by catastrophic cancellation, and the discretely approximated values of $r_{\rho}$ can have large relative errors or even become \emph{negative},
causing failures of the entire computation. By computing $r_{\rho}$ directly from the analytic representation of $r$, this problem is avoided and the solution is ensured to be accurately approximated even in regions where the singularity is about to form
and where $r_{\rho} \approx c \ll 1$. This also explains why the conventional moving mesh method is not suitable for singularity computations where high accuracy is demanded, because the method only computes a \emph{discrete approximation} of the mesh
mapping function, which necessarily requires a difference approximation of $r_{\rho}$ in the evaluation of a space derivative $v_{r}$. Without mesh smoothing, this can cause instability, but with mesh smoothing the mesh resolution will be inevitably
limited, which is undesired.

The centered difference formulas described above need to be supplemented by numerical boundary conditions near $\rho,\eta = 0,1$. Along the $\eta$-dimension, the symmetry condition
\begin{displaymath}
  v_{-i,j} = -v_{i,j},\quad v_{M+i,j} = \pm v_{M-i,j},\qquad 1 \leq i \leq 3,\ 0 \leq j \leq N,
\end{displaymath}
is used near $\eta = 0$ and $\eta = 1$, where the $+$ sign applies to $u_{1}$ and the $-$ sign applies to $\omega_{1}$. Along the $\rho$-dimension, the symmetry condition
\begin{align*}
  v_{i,-j} = v_{i,j},&\qquad 0 \leq i \leq M,\ 1 \leq j \leq 3, \\
  \intertext{is used near the axis $\rho = 0$ and the extrapolation condition}
  (D_{\rho,-}^{7} v_{i,\cdot})_{N+j} = 0,&\qquad 0 \leq i \leq M,\ 1 \leq j \leq 3,
\end{align*}
is applied near the solid boundary $\rho = 1$\footnote{While a 6th-order extrapolation condition $(D_{\rho,-}^{6} v_{i,\cdot})_{N+j} = 0$ suffices to maintain a formal 6th-order accuracy for the overall scheme, we choose the higher-order extrapolation
condition for better accuracy.}. The extrapolation condition is known to be GKS stable for linear hyperbolic problems \citep[Theorem 13.1.3]{gko1995}, and is expected to remain stable when applied to the Euler equations as long as the underlying solution
is sufficiently smooth.

Using the superconvergence properties of the Poisson solver at the grid points (to be proved elsewhere), it can be shown that the overall algorithm is formally 6th-order accurate in space and 4th-order accurate in time. The details of this error analysis
will be reported in a separate paper.

\section{Numerical Results}\label{sec_result}
We have numerically solved the initial-boundary value problem \eqref{eqn_eat}--\eqref{eqn_eat_ibc} on the quarter cylinder $D(1,\frac{1}{24})$ (with $L = \frac{1}{6}$). The results suggest that the solution develops a singularity in finite time and we
shall provide, in what follows, ample evidence to support this finding. We start with an overview of our computations in Section \ref{ssec_mm_effec}--\ref{ssec_1st_sign} where the effectiveness of the adaptive mesh is demonstrated and the first sign of a
finite-time singularity is given. After a careful resolution study of the computed solutions in Section \ref{ssec_res}, we proceed to Section \ref{ssec_vfit}--\ref{ssec_efit} where the asymptotic scalings of the vorticity moments are analyzed in great
detail. The results indicate the divergence of the time integral of the maximum vorticity, and hence the blowup of the computed solutions. This conclusion is further confirmed in Section \ref{ssec_geosd}, where the geometric structures of the vorticity
direction field are analyzed and the consistency between the blowing-up solutions and the various geometric non-blowup criteria is demonstrated. Once the existence of a finite-time singularity is confirmed, we move on to Section \ref{ssec_selfsim} where
the locally self-similar structure of the blowing-up solutions is examined. The discussion is concluded in Section \ref{ssec_intrp} with a physical interpretation of the finite-time singularity, where the driving mechanism behind the blowup is
investigated.

\subsection{Effectiveness of the Adaptive Mesh}\label{ssec_mm_effec}
We have numerically solved the problem \eqref{eqn_eat}--\eqref{eqn_eat_ibc} on meshes of size $256k \times 256k$ where $k = 4,5,6,7,8$. In each computation, the solution is initialized on a uniform mesh, which is then adjusted to the initial data using
the adaptive mesh algorithm described in Section \ref{ssec_mm}. Once an ``optimal'' mesh is obtained, the solution is advanced indefinitely in time using the method described in Section \ref{sec_method}, until either the time step drops below $10^{-12}$,
or the minimum mesh spacing drops below $\epsilon_{r} = 10^{-15}$ (in $r$) or $\epsilon_{z} = 10^{-15} (\frac{1}{4} L)$ (in $z$), whichever happens first.

\begin{figure}[h]
  \centering
  \subfigure[$rz$-plane]{
    \includegraphics[scale=0.415]{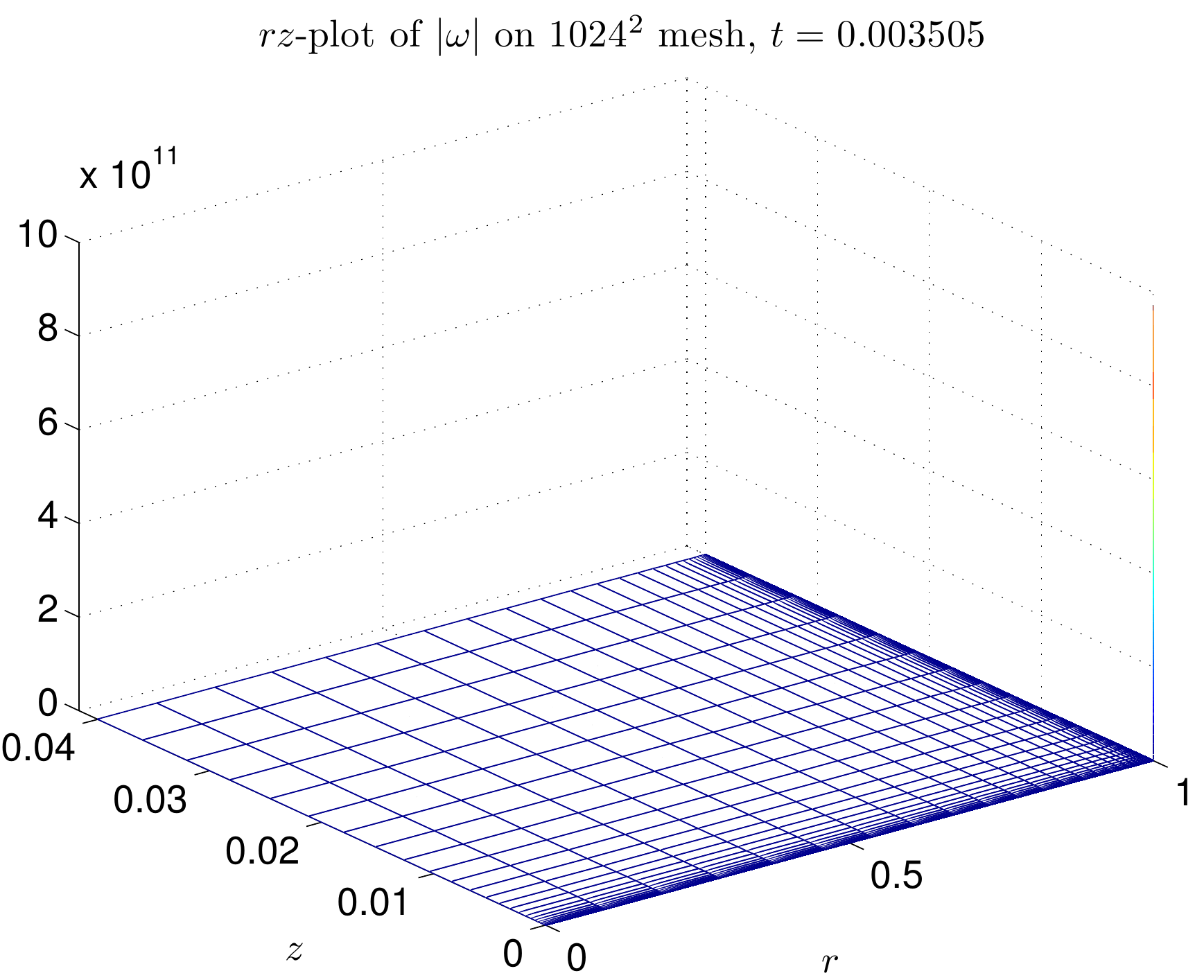}
    \label{fig_mesh_vort_rz}
  }\quad
  \subfigure[$\rho\eta$-plane]{
    \includegraphics[scale=0.415]{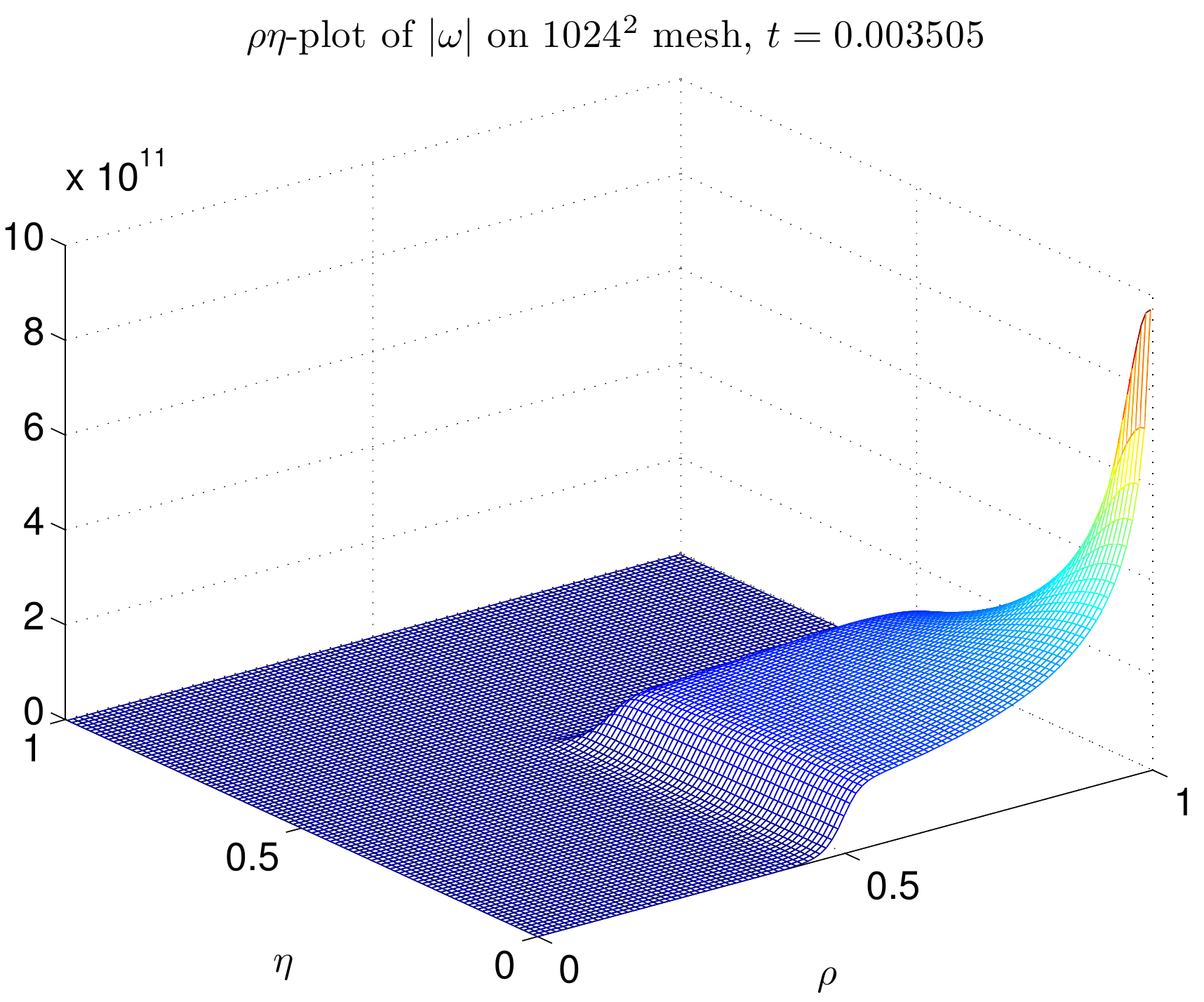}
    \label{fig_mesh_vort_re}
  }
  \caption{The vorticity function $\abs{\omega}$ on the $1024 \times 1024$ mesh at $t = 0.003505$, in (a) $rz$-coordinates and (b) $\rho\eta$-coordinates, where for clarity only one-tenth of the mesh lines are displayed along each dimension.}
  \label{fig_mesh_vort}
\end{figure}

Table \ref{tab_stop_time} shows the stopping time $t_{e}$ and the cause of termination for each computation. As indicated by the \emph{mostly decreasing} stopping time (with respect to the increasing resolution) and the vanishing minimum mesh spacings,
the solution seems to develop a very singular structure in finite time.
\begin{table}[h]
  \centering
  \caption{Stopping time $t_{e}$ and cause of termination, where $(\delta_{r},\delta_{z})$ denote the minimum mesh spacing in $r$ and $z$, respectively.}
  \label{tab_stop_time}
  \begin{tabular}{*{3}{>{$}c<{$}}}
    \toprule
    \text{Mesh size} & t_{e} & \text{Cause of termination} \\
    \midrule
    1024 \times 1024 & 0.0035055667206 & \delta_{r} < \epsilon_{r}\ \text{and}\ \delta_{z} < \epsilon_{z} \\
    1280 \times 1280 & 0.0035055581996 & \delta_{z} < \epsilon_{z} \\
    1536 \times 1536 & 0.0035055522856 & \delta_{z} < \epsilon_{z} \\
    1792 \times 1792 & 0.0035055523092 & \delta_{r} < \epsilon_{r}\ \text{and}\ \delta_{z} < \epsilon_{z} \\
    2048 \times 2048 & 0.0035055472037 & \delta_{r} < \epsilon_{r}\ \text{and}\ \delta_{z} < \epsilon_{z} \\
    \bottomrule
  \end{tabular}
\end{table}%
To determine the nature of the singular structure and to see how well the adaptive mesh resolves it, we plot in Figure \ref{fig_mesh_vort} the vorticity function $\abs{\omega}$ computed on the $1024 \times 1024$ mesh at $t = 0.003505$, in both the
$rz$-coordinates (Figure \ref{fig_mesh_vort}\subref{fig_mesh_vort_rz}) and the $\rho\eta$-coordinates (Figure \ref{fig_mesh_vort}\subref{fig_mesh_vort_re}). The $rz$-plot suggests that the singular structure could be a point-singularity at the corner
$\tilde{q}_{0} = (1,0)^{T}$. The $\rho\eta$-plot, on the other hand, shows that a good portion of the mesh points (roughly 50\% along each dimension) are consistently placed in regions where $\abs{\omega}$ is comparable with the maximum vorticity
$\norm{\omega}_{\infty}$, hence demonstrating the effectiveness of the adaptive mesh in capturing the potential singularity.
To obtain a quantitative measure of the maximum resolution power achieved by the adaptive mesh, we define the mesh compression ratios
\begin{displaymath}
  p_{\infty} := \frac{L}{4 z'(\eta_{\infty})},\qquad q_{\infty} := \frac{1}{r'(\rho_{\infty})},
\end{displaymath}
and the effective mesh resolutions
\begin{displaymath}
  M_{\infty} := p_{\infty} M = \frac{LM}{4 z'(\eta_{\infty})},\qquad N_{\infty} := q_{\infty} N = \frac{N}{r'(\rho_{\infty})},
\end{displaymath}
at the location $(\rho_{\infty},\eta_{\infty})^{T} \equiv (1,0)^{T}$ of the maximum vorticity $\norm{\omega}_{\infty}$. The values of these metrics computed at $t = 0.003505$ are summarized in Table \ref{tab_effec_res}.
\begin{table}[h]
  \centering
  \caption{Mesh compression ratios $(p_{\infty},q_{\infty})$ and effective mesh resolutions $(M_{\infty},N_{\infty})$ at the location of the maximum vorticity at $t = 0.003505$.}
  \label{tab_effec_res}
  \begin{tabular}{*{5}{>{$}c<{$}}}
    \toprule
     & \multicolumn{4}{c}{$t = 0.003505$} \\
    \cmidrule{2-5}
    \rbs{1.75ex}{Mesh size} & p_{\infty} & M_{\infty} & q_{\infty} & N_{\infty} \\
    \midrule
    1024 \times 1024 & 1.9456 \times 10^{9} & 1.9923 \times 10^{12} & 1.6316 \times 10^{9} & 1.6708 \times 10^{12} \\
    1280 \times 1280 & 1.9530 \times 10^{9} & 2.4999 \times 10^{12} & 1.6285 \times 10^{9} & 2.0844 \times 10^{12} \\
    1536 \times 1536 & 1.9444 \times 10^{9} & 2.9866 \times 10^{12} & 1.6328 \times 10^{9} & 2.5079 \times 10^{12} \\
    1792 \times 1792 & 1.9504 \times 10^{9} & 3.4951 \times 10^{12} & 1.6344 \times 10^{9} & 2.9288 \times 10^{12} \\
    2048 \times 2048 & 1.9503 \times 10^{9} & 3.9942 \times 10^{12} & 1.6330 \times 10^{9} & 3.3444 \times 10^{12} \\
    \bottomrule
  \end{tabular}
\end{table}%

\begin{figure}[h]
  \centering
  \subfigure[full view]{
    \includegraphics[scale=0.415]{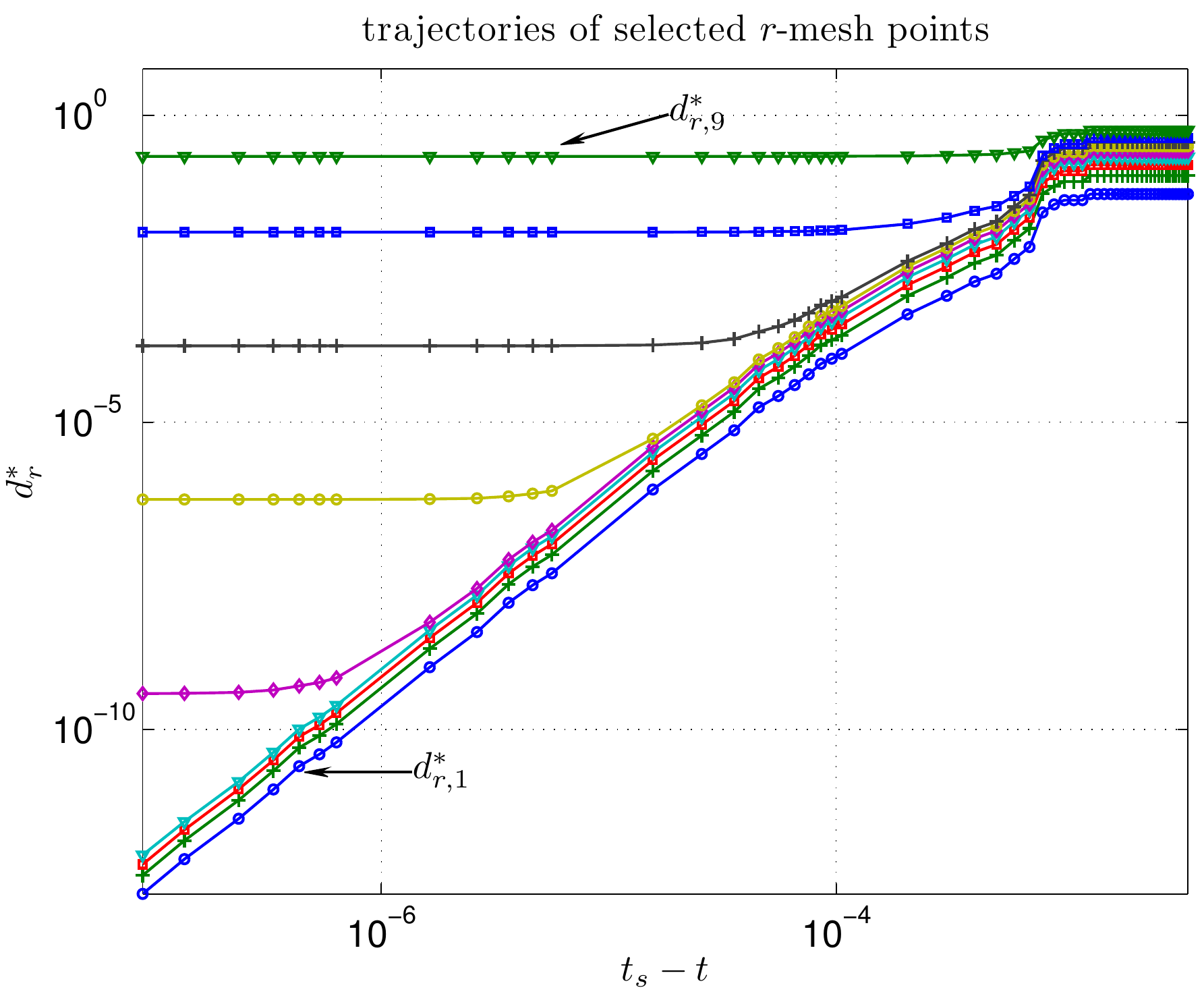}
    \label{fig_pct_msh_ra}
  }\quad
  \subfigure[detailed view]{
    \includegraphics[scale=0.415]{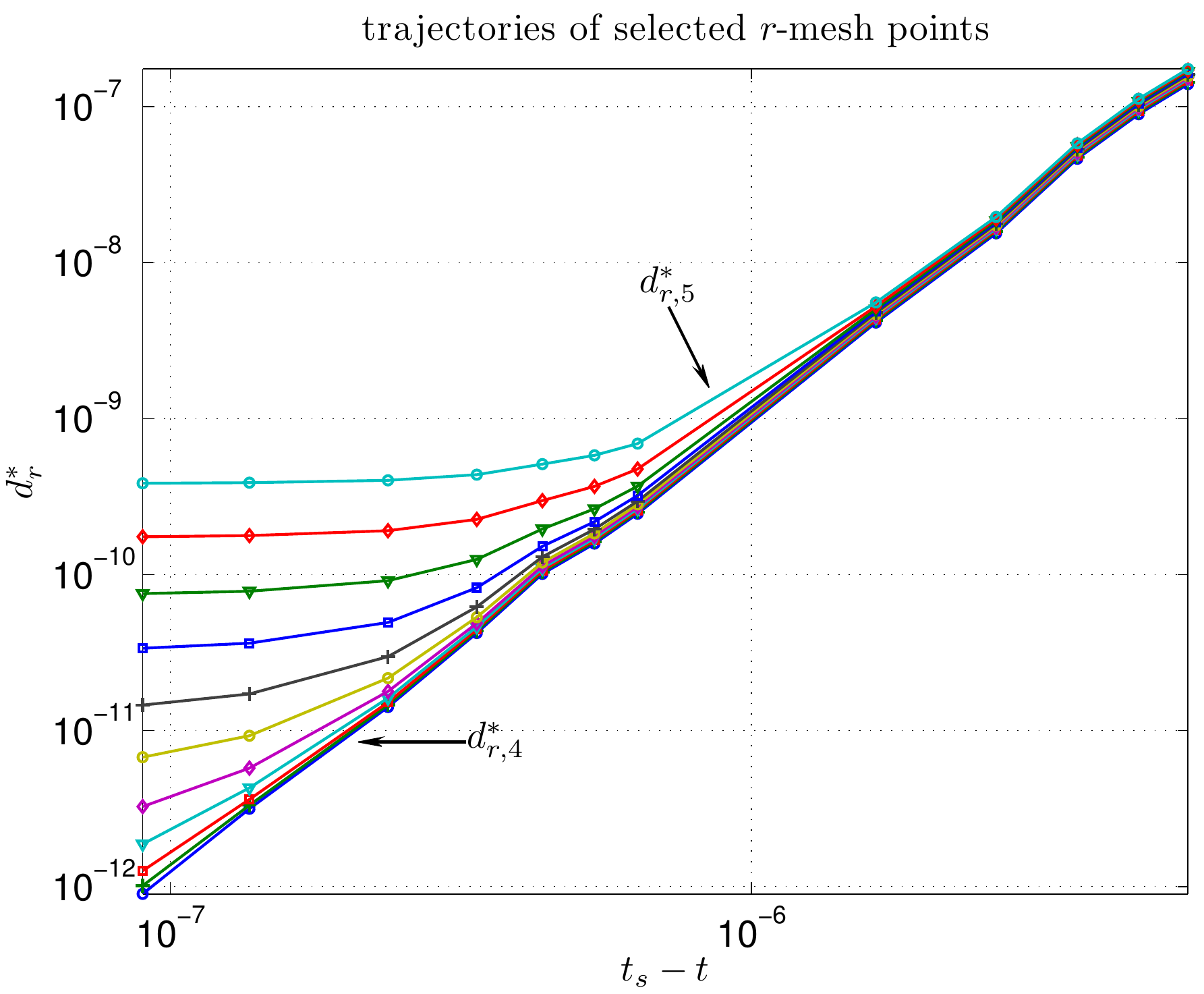}
    \label{fig_pct_msh_rs}
  }
  \caption{Trajectories of selected $r$-mesh points on the $2048 \times 2048$ mesh, in log-log scale (see text for explanation). The last time instant shown in the figure is $t_{e}$, the stopping time.}
  \label{fig_pct_msh_r}
\end{figure}

The above analysis confirms the effectiveness of the adaptive mesh in the ``inner region'' where the vorticity function $\abs{\omega}$ is most singular, but it says nothing about the quality of the mesh outside the inner region. To address this issue, we
plot in Figure \ref{fig_pct_msh_r}\subref{fig_pct_msh_ra} the trajectories of the $r$-mesh points
\begin{displaymath}
  \rho_{j}^{*} := \rho_{\infty} - \frac{j}{10} \equiv 1 - \frac{j}{10},\qquad j = 1,\dotsc,9,
\end{displaymath}
which can be viewed as ``Lagrangian markers'' equally spaced (in $\rho$) away from the location of the maximum vorticity $\rho_{\infty} \equiv 1$. The ordinate of the figure represents the distance between the selected mesh points and the location of the
maximum vorticity,
\begin{displaymath}
  d_{r,j}^{*} := \frac{1}{1} \bigl[ r(\rho_{\infty}) - r(\rho_{j}^{*}) \bigr] \equiv 1 - r(\rho_{j}^{*}),\qquad j = 1,\dotsc,9,
\end{displaymath}
expressed as a fraction of the total length of the computational domain (1 in this case). The abscissa of the figure represents $t_{s}-t$ where $t_{s}$ is the predicted singularity time (see Section \ref{ssec_vfit}). As is clear from the figure, the 40\%
mesh points that lie closest to $\rho_{\infty}$ are always placed in the inner region while the 50\% points farthest away from $\rho_{\infty}$ eventually move into the outer region. The 10\% points lying between the inner and outer regions belong to the
``transition region'' and are shown in greater detail in Figure \ref{fig_pct_msh_r}\subref{fig_pct_msh_rs}. A similar analysis applied to the adaptive mesh along the $z$-dimension shows that the $z$-mesh has a completely similar character.

\begin{figure}[h]
  \centering
  \subfigure[full view]{
    \includegraphics[scale=0.415]{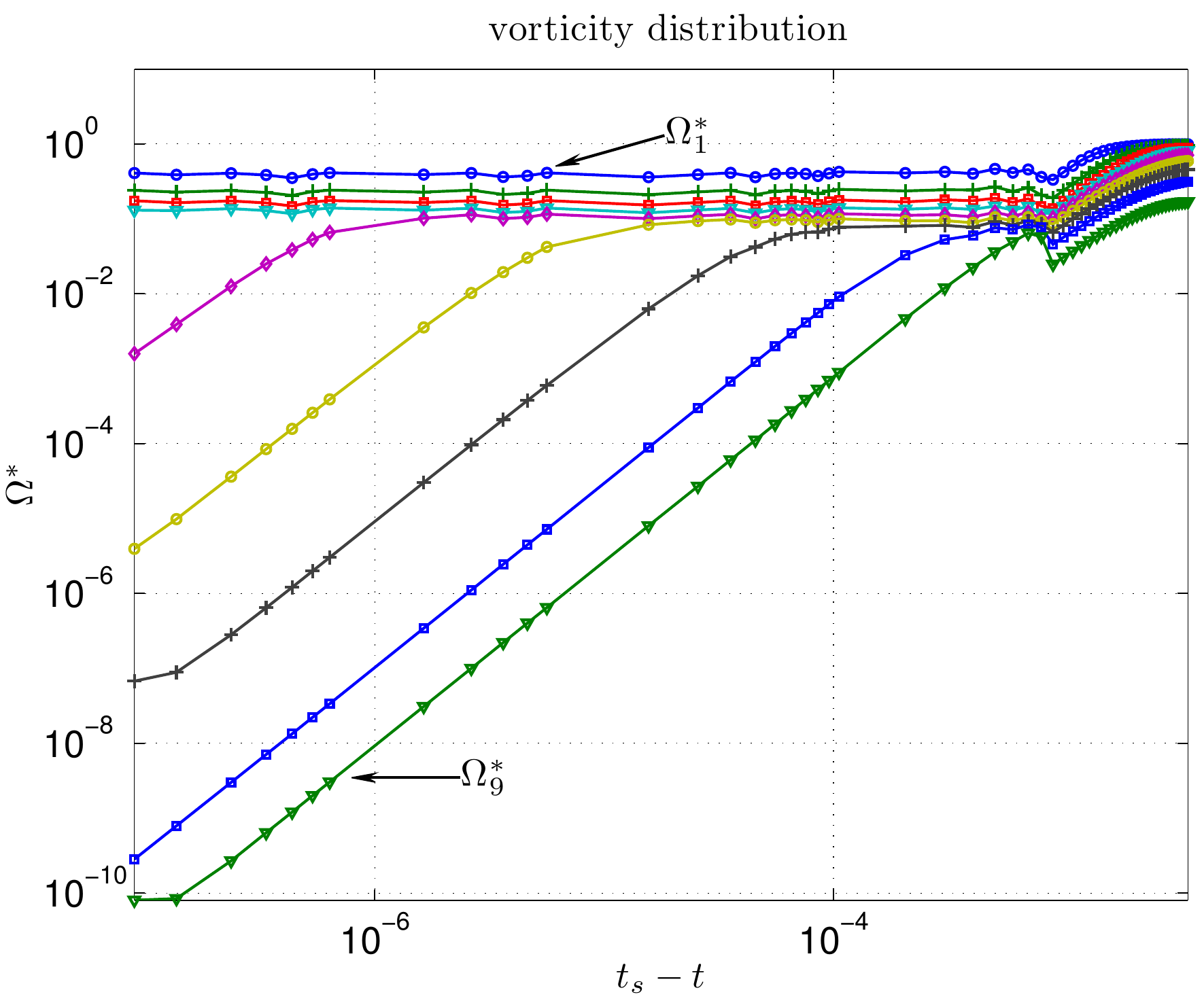}
    \label{fig_pct_vorta}
  }\quad
  \subfigure[detailed view]{
    \includegraphics[scale=0.415]{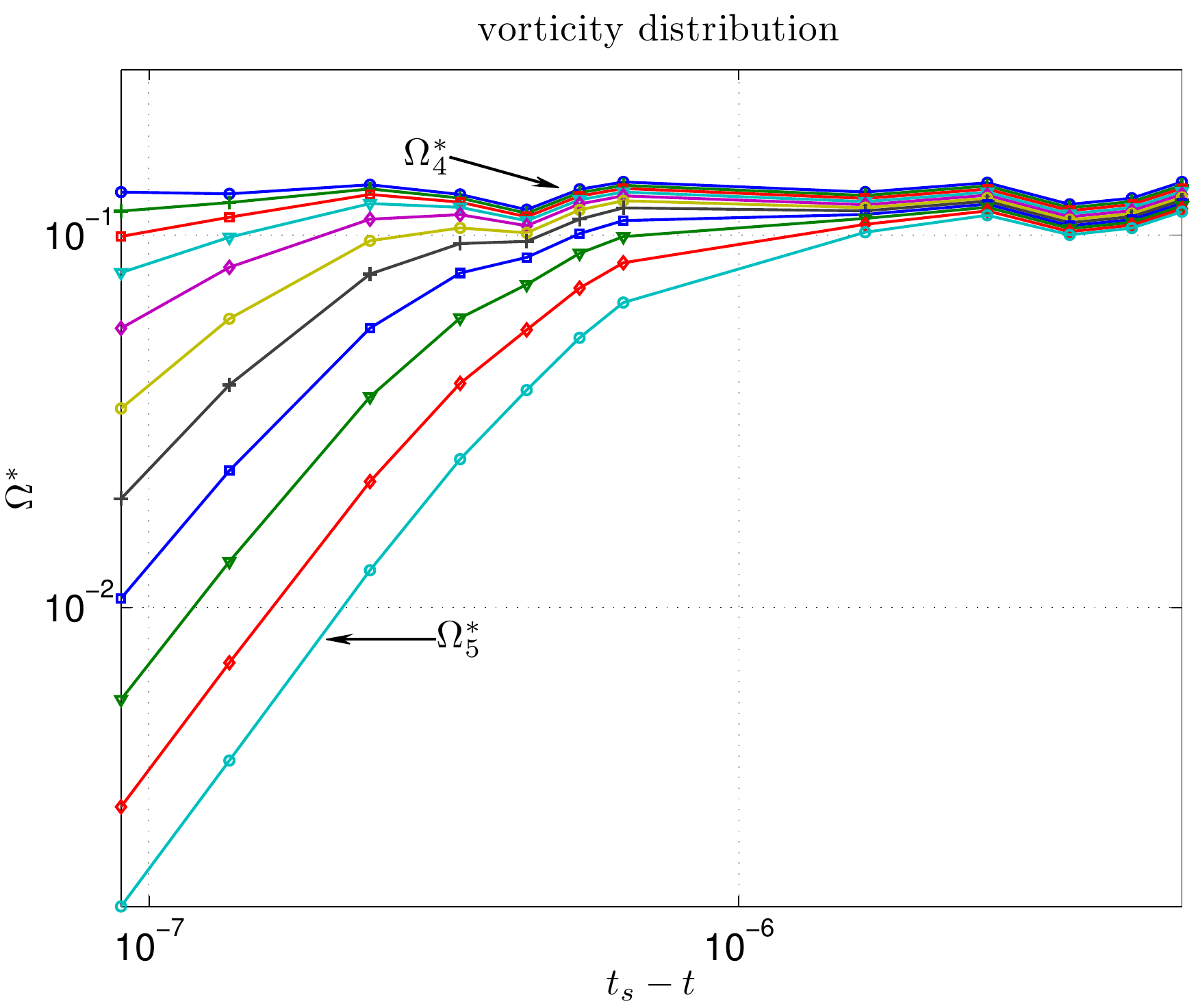}
    \label{fig_pct_vorts}
  }
  \caption{Vorticity distribution on the $2048 \times 2048$ mesh, in log-log scale (see text for explanation). The last time instant shown in the figure is $t_{e}$, the stopping time.}
  \label{fig_pct_vort}
\end{figure}

To see how well the solution is resolved in the transition region, we define
\begin{displaymath}
  \Omega_{j}^{*} := \frac{1}{\norm{\omega}_{\infty}} \sup_{(r,z) \in D_{j}^{*}} \abs{\omega(r,z)},\qquad j = 1,\dotsc,9,
\end{displaymath}
where
\begin{displaymath}
  D_{j}^{*} = D(1,\tfrac{1}{4} L) \setminus [r(\rho_{j}^{*}),1] \times [0,z(\eta_{j}^{*})],\qquad \eta_{j}^{*} := \eta_{\infty} + \frac{j}{10} \equiv \frac{j}{10},
\end{displaymath}
is the portion of the quarter cylinder $D(1,\frac{1}{4} L)$ outside the region $[r(\rho_{j}^{*}),1] \times [0,z(\eta_{j}^{*})]$. As is clear from Figure \ref{fig_pct_vort}\subref{fig_pct_vorta}, the values of $\Omega_{j}^{*}$ stay nearly constant for $j
\leq 4$ and steadily decay for $j \geq 5$, consistent with the observation that the 40\% points lying closest to $(\rho_{\infty},\eta_{\infty})$ belong to the inner region while the 50\% points farthest away from $(\rho_{\infty},\eta_{\infty})$ belong to
the outer region. Within the transition region where the rest 10\% points belong to, the vorticity function $\abs{\omega}$ varies smoothly from $10^{-3} \norm{\omega}_{\infty}$ to $10^{-1} \norm{\omega}_{\infty}$ (Figure
\ref{fig_pct_vort}\subref{fig_pct_vorts}). This suggests that the adaptive mesh produces a nearly uniform representation of the computed solution across the entire computational domain, hence confirming its efficacy.

To analyze the performance of the Poisson solver, in particular that of the linear solve $Ax = b$, we define as in \citet{add1989} the componentwise backward errors of the first and second kind:
\begin{displaymath}
  \omega_{i} = \max_{j} \frac{\abs{A^{(i)} \hat{x} - b^{(i)}}_{j}}{(\abs{A^{(i)}} \abs{\hat{x}} + f^{(i)})_{j}},\qquad i = 1,2,
\end{displaymath}
and the componentwise condition numbers of the first and second kind $\kappa_{\omega_{1}},\ \kappa_{\omega_{2}}$. Here $\hat{x}$ is the numerical approximation to the exact solution $x$ and
\begin{displaymath}
  f^{(1)} = \abs{b^{(1)}},\qquad f^{(2)} = \abs{A^{(2)}} e \norm{\hat{x}}_{\infty},\quad \text{$e$ is the vector of all ones}.
\end{displaymath}
The equations in the linear system are classified as follows: let $w = \abs{A} \abs{\hat{x}} + \abs{b}$ be the vector of denominators in the definition of $\omega_{i}$. If $w_{j} > \tau_{j}$ for a user-defined threshold $\tau_{j}$, then the $j$-th
equation is said to belong to the first category ($i = 1$); otherwise it is said to belong to the second category ($i = 2$). To leading-order approximation, the error $\delta x = \hat{x}-x$ of the linear solve satisfies \citep{add1989}
\begin{equation}
  \frac{\norm{\delta x}_{\infty}}{\norm{x}_{\infty}} \leq \omega_{1} \kappa_{\omega_{1}} + \omega_{2} \kappa_{\omega_{2}}.
  \label{eqn_psolv_berr}
\end{equation}
Compared with the standard norm-based error metrics, the error predicted by \eqref{eqn_psolv_berr} tends to give a \emph{much tighter} bound for the actual error, especially when $A$ is badly row-scaled \citep{add1989}.

Table \ref{tab_psolv_berr} shows the backward errors \eqref{eqn_psolv_berr} as well as other related error metrics computed for the linear system associated with the Poisson solve \eqref{eqn_psolv_vh}.
\begin{table}[h]
  \centering
  \caption{Backward errors of the linear solve $Ax = b$ associated with \eqref{eqn_psolv_vh} at $t = 0.003505$.}
  \label{tab_psolv_berr}
  \begin{tabular}{*{6}{>{$}c<{$}}}
    \toprule
     & \multicolumn{5}{c}{$t = 0.003505$} \\
    \cmidrule{2-6}
    \rbs{1.75ex}{Mesh size} & \omega_{1} & \kappa_{\omega_{1}} & \omega_{2} & \kappa_{\omega_{2}} & \norm{\delta x}_{\infty}/\norm{x}_{\infty} \\
    \midrule
    256 \times 256 & 4.2456 \times 10^{-12} & 974.28 & 5.0563 \times 10^{-20} & 1.6772 \times 10^{7} & 4.1372 \times 10^{-9} \\
    512 \times 512 & 5.8812 \times 10^{-15} & 1247.29 & 1.8902 \times 10^{-23} & 2.3027 \times 10^{7} & 7.3360 \times 10^{-12} \\
    768 \times 768 & 1.0843 \times 10^{-15} & 1788.84 & 2.1290 \times 10^{-23} & 5.2033 \times 10^{7} & 1.9407 \times 10^{-12} \\
    1024 \times 1024^{\dag} & 1.4721 \times 10^{-15} & 6748.83 & 6.4433 \times 10^{-23} & 9.2646 \times 10^{7} & 9.9408 \times 10^{-12} \\
    \bottomrule
    \multicolumn{6}{l}{\footnotesize{$\dag$: For technical reasons, the analysis is restricted to meshes of size no larger than $1024 \times 1024$.}}
  \end{tabular}
\end{table}%
It can be observed that both condition numbers $\kappa_{\omega_{1}},\ \kappa_{\omega_{2}}$ grow roughly like $h^{-2}$ where $h := \min\{h_{r},h_{z}\} = \min\{1/M,1/N\}$ is the (uniform) mesh spacing \emph{in the $\rho\eta$-space}. It can also be observed
that the value of $\kappa_{\omega_{2}}$ is considerably larger than that of $\kappa_{\omega_{1}}$, but the backward error $\omega_{2}$ is so small that the net contribution of $\omega_{2} \kappa_{\omega_{2}}$ is negligible compared with that of
$\omega_{1} \kappa_{\omega_{1}}$. As a result, the backward error bound $\norm{\delta x}_{\infty}/\norm{x}_{\infty}$ of the computed solution remains uniformly small for all meshes.

The backward error analysis as described above is applied only to meshes of size no larger than $1024 \times 1024$, due to a technical restriction of the linear solve package that we use. To complete the picture, we also carry out a \emph{forward} error
analysis where the error of the linear solve $Ax = b$ as well as that of the discrete problem \eqref{eqn_psolv_vh} are estimated directly using a three-step procedure. First, the approximate solution $\hat{x}$ of the linear system is taken as the exact
solution and a new right-hand side $\hat{b} = A \hat{x}$ is computed from $\hat{x}$ using 128-bit (quadruple-precision) arithmetic\footnote{Implemented using GNU's GMP library.}. Second, the linear system $Ax = \hat{b}$ with the new right-hand side
$\hat{b}$ is solved numerically, yielding an approximate solution $\tilde{x}$. Finally, the reference and the approximate stream functions $\hat{\psi}_{h},\ \tilde{\psi}_{h}$ are assembled from the solution vectors $\hat{x},\ \tilde{x}$, and the relative
errors of $\hat{x}$ as well as that of $\hat{\psi}_{h},\ \hat{\psi}_{h,r},\ \hat{\psi}_{h,z}$ are computed. The results of this error analysis are summarized in Table \ref{tab_psolv_ferr}.
\begin{table}[h]
  \centering
  \caption{Forward errors of the linear solve $Ax = b$ and of the discrete problem \eqref{eqn_psolv_vh} at $t = 0.003505$.}
  \label{tab_psolv_ferr}
  \begin{tabular}{*{5}{>{$}c<{$}}}
    \toprule
     & \multicolumn{4}{c}{Sup-norm relative error at $t = 0.003505$} \\
    \cmidrule{2-5}
    \rbs{1.75ex}{Mesh size} & \hat{x} & \hat{\psi}_{h} & \hat{\psi}_{h,r} & \hat{\psi}_{h,z} \\
    \midrule
    1024 \times 1024 & 3.9638 \times 10^{-14} & 1.6697 \times 10^{-14} & 4.2104 \times 10^{-12} & 5.3310 \times 10^{-12} \\
    1280 \times 1280 & 4.1397 \times 10^{-14} & 1.1431 \times 10^{-14} & 5.6280 \times 10^{-12} & 6.4547 \times 10^{-12} \\
    1536 \times 1536 & 7.0504 \times 10^{-14} & 4.8934 \times 10^{-14} & 1.1191 \times 10^{-11} & 9.3045 \times 10^{-12} \\
    1792 \times 1792 & 4.3910 \times 10^{-14} & 1.1045 \times 10^{-14} & 9.4986 \times 10^{-12} & 1.4097 \times 10^{-11} \\
    2048 \times 2048 & 6.9127 \times 10^{-14} & 3.3393 \times 10^{-14} & 1.2582 \times 10^{-11} & 1.4449 \times 10^{-11} \\
    \bottomrule
  \end{tabular}
\end{table}%
As is clear from the table, the Poisson solver is numerically stable despite the very high compression ratios achieved by the adaptive mesh (Table \ref{tab_effec_res}). Combined with the discretization error estimate \eqref{eqn_psolv_gerr}, this
establishes the convergence of the Poisson solver under mesh refinement.

\subsection{First Sign of Singularity}\label{ssec_1st_sign}
\begin{figure}[h]
  \centering
  \includegraphics[scale=0.42]{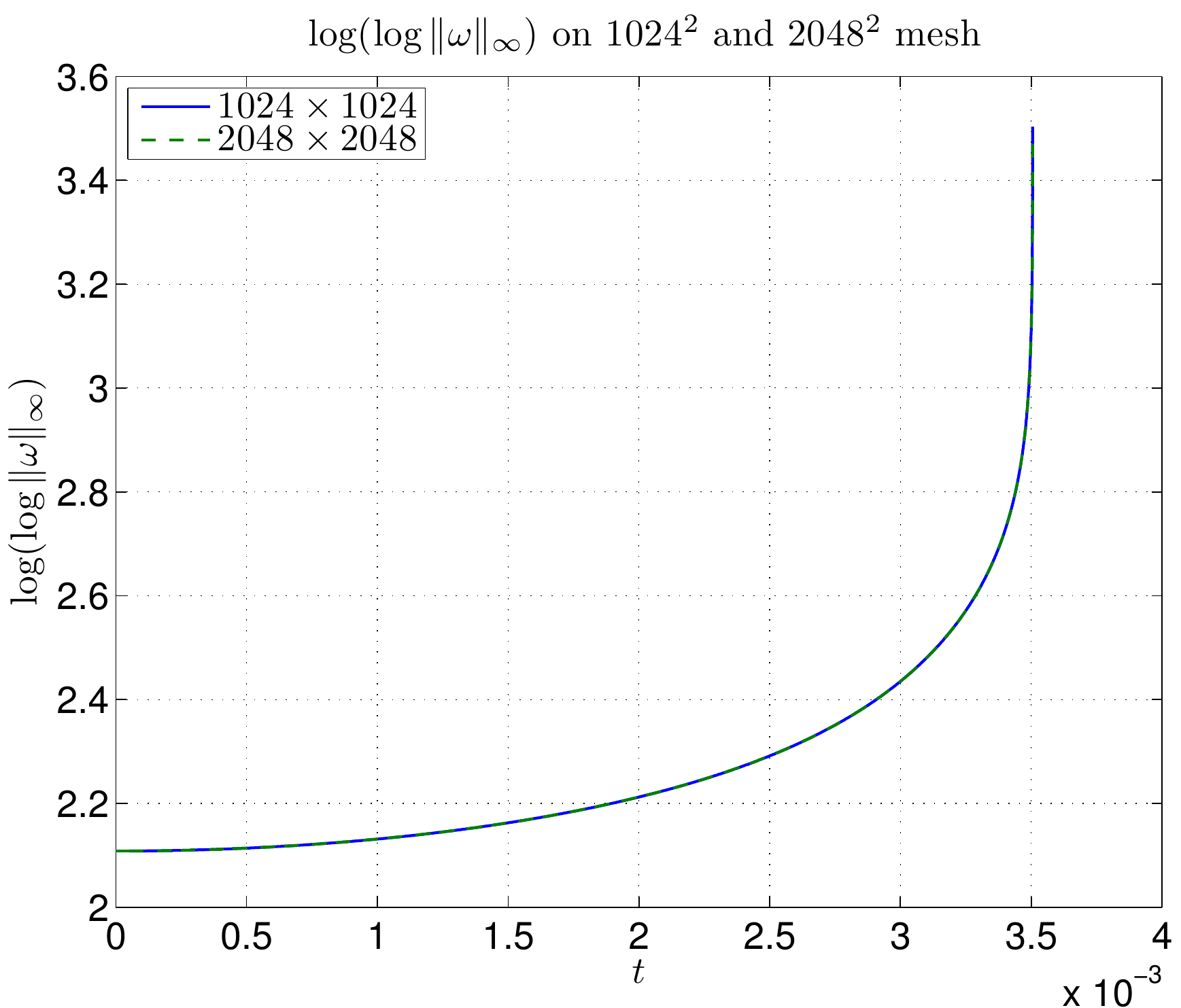}
  \caption{The double logarithm of the maximum vorticity $\log(\log \norm{\omega}_{\infty})$ computed on the $1024 \times 1024$ and the $2048 \times 2048$ mesh. The two curves overlap and are virtually indistinguishable from each other (see Section
  \ref{ssec_res} below for a detailed resolution study on the nearly singular solutions).}
  \label{fig_loglog_vort}
\end{figure}

Now we examine more closely the nature of the singular structure observed in the vorticity function $\abs{\omega}$ (see Figure \ref{fig_mesh_vort}). We first report in Table \ref{tab_time_step}--\ref{tab_max_vort} the (variable) time steps $\delta_{t}$
and the maximum vorticity $\norm{\omega}_{\infty}$ recorded at selected time instants $t$. We also plot in Figure \ref{fig_loglog_vort} the double logarithm of the maximum vorticity, $\log(\log \norm{\omega}_{\infty})$, computed on the coarsest $1024
\times 1024$ and the finest $2048 \times 2048$ mesh.
\begin{table}[h]
  \centering
  \caption{Time step $\delta_{t}$ at selected time $t$.}
  \label{tab_time_step}
  \begin{tabular}{*{6}{>{$}c<{$}}}
    \toprule
     & \multicolumn{5}{c}{$\delta_{t}$} \\
    \cmidrule{2-6}
    \rbs{1.75ex}{Mesh size} & t = 0^{\dag} & t = 0.003 & t = 0.0034 & t = 0.0035 & t = 0.003505 \\
    \midrule
    1024 \times 1024 & 1 \times 10^{-6} & 2.8754 \times 10^{-7} & 4.9502 \times 10^{-8} & 2.8831 \times 10^{-9} & 2.4240 \times 10^{-10} \\
    1280 \times 1280 & 1 \times 10^{-6} & 2.3120 \times 10^{-7} & 3.9636 \times 10^{-8} & 2.2983 \times 10^{-9} & 2.5772 \times 10^{-10} \\
    1536 \times 1536 & 1 \times 10^{-6} & 1.9165 \times 10^{-7} & 3.2907 \times 10^{-8} & 1.9165 \times 10^{-9} & 2.2223 \times 10^{-10} \\
    1792 \times 1792 & 1 \times 10^{-6} & 1.6578 \times 10^{-7} & 2.8451 \times 10^{-8} & 1.6418 \times 10^{-9} & 1.9122 \times 10^{-10} \\
    2048 \times 2048 & 1 \times 10^{-6} & 1.4509 \times 10^{-7} & 2.4046 \times 10^{-8} & 1.4367 \times 10^{-9} & 2.0272 \times 10^{-10} \\
    \bottomrule
    \multicolumn{6}{l}{\footnotesize{$\dag$: The maximum time step allowed in our computations is $10^{-6}$.}}
  \end{tabular}
\end{table}%
\begin{table}[h]
  \centering
  \caption{Maximum vorticity $\norm{\omega}_{\infty}$ at selected time $t$.}
  \label{tab_max_vort}
  \begin{tabular}{*{6}{>{$}c<{$}}}
    \toprule
     & \multicolumn{5}{c}{$\norm{\omega}_{\infty}$} \\
    \cmidrule{2-6}
    \rbs{1.75ex}{Mesh size} & t = 0 & t = 0.003 & t = 0.0034 & t = 0.0035 & t = 0.003505 \\
    \midrule
    1024 \times 1024 & 3.7699 \times 10^{3} & 9.0847 \times 10^{4} & 4.3127 \times 10^{6} & 5.8438 \times 10^{9} & 1.2416 \times 10^{12} \\
    1280 \times 1280 & 3.7699 \times 10^{3} & 9.0847 \times 10^{4} & 4.3127 \times 10^{6} & 5.8423 \times 10^{9} & 1.2407 \times 10^{12} \\
    1536 \times 1536 & 3.7699 \times 10^{3} & 9.0847 \times 10^{4} & 4.3127 \times 10^{6} & 5.8417 \times 10^{9} & 1.2403 \times 10^{12} \\
    1792 \times 1792 & 3.7699 \times 10^{3} & 9.0847 \times 10^{4} & 4.3127 \times 10^{6} & 5.8415 \times 10^{9} & 1.2401 \times 10^{12} \\
    2048 \times 2048 & 3.7699 \times 10^{3} & 9.0847 \times 10^{4} & 4.3127 \times 10^{6} & 5.8413 \times 10^{9} & 1.2401 \times 10^{12} \\
    \bottomrule
  \end{tabular}
\end{table}%
It can be observed from these results that, for each computation, there exists a short time interval right before the stopping time $t_{e}$ in which the solution grows tremendously. This can be readily inferred from the sharp decrease in the time step
$\delta_{t}$ (Table \ref{tab_time_step}) as well as the super-double-exponential growth of the maximum vorticity $\norm{\omega}_{\infty}$ (Table \ref{tab_max_vort}, Figure \ref{fig_loglog_vort}). In addition, the nearly singular solution seems to
converge under mesh refinement (Table \ref{tab_max_vort}). These behaviors are characteristic of a blowing-up solution and may be viewed as the first sign of a finite-time singularity.

\subsection{Resolution Study}\label{ssec_res}
Of course, neither a rapidly decreasing time step nor a fast growing vorticity is sufficient evidence for the existence of a finite-time singularity. To obtain more convincing evidence, a much more thorough analysis is needed, which, in the first place,
requires a careful examination of the accuracy of the computed solutions.

There are several well-established, ``standard'' methods in the literature to gauge the quality of an Euler computation:
\begin{itemize}
  \item energy conservation: it is well-known that, under suitable regularity assumptions, the solutions of the Euler equations conserve the \emph{kinetic energy}
  \begin{displaymath}
    E = \frac{1}{2} \int_{D(1,L)} \abs{u}^{2}\,dx = \frac{1}{2} \int_{0}^{1} \int_{0}^{L} \bigl( \abs{u_{1}}^{2} + \abs{\psi_{1,r}}^{2} + \abs{\psi_{1,z}}^{2} \bigr) r^{3}\,dr\,dz;
  \end{displaymath}
  thus a widely used ``quality indicator'' for Euler computations is the relative change of the energy integral $E$ over time;

  \item enstrophy and enstrophy production rate: another widely accepted ``error indicator'' for Euler computations is the \emph{enstrophy} integral
  \begin{displaymath}
    \mathcal{E} = \int_{D(1,L)} \abs{\nabla u}^{2}\,dx = \int_{D(1,L)} \abs{\omega}^{2}\,dx
  \end{displaymath}
  and the \emph{enstrophy production rate} integral
  \begin{displaymath}
    \mathcal{E}_{p} := \frac{d\mathcal{E}}{dt} = 2 \int_{D(1,L)} \omega \cdot S \omega\,dx,\qquad S = \frac{1}{2} \bigl( \nabla u + \nabla u^{T} \bigr);
  \end{displaymath}
  these quantities are not conserved over time, but their convergence under mesh refinement provides partial evidence on the convergence of the underlying numerical solutions;

  \item energy spectra: for problems defined on periodic domains, it is also a common practice to perform convergence analysis on the \emph{energy spectra} of the periodic velocity field $u$:
  \begin{displaymath}
    E_{p}(k) = \sum_{\abs{\ell} \in (k-1/2,k+1/2]} \abs{\hat{u}_{\ell}}^{2},
  \end{displaymath}
  and use the results as a measure of the quality of the underlying solutions; here, as usual, $\hat{u}_{\ell}$ denotes the vector Fourier coefficients of the velocity $u$, which on an $L_{1} \times L_{2} \times L_{3}$ periodic box $B$ is defined by
  \begin{displaymath}
    \hat{u}_{\ell} = \frac{1}{\abs{B}} \int_{B} u \re^{-\ri \ell \cdot x}\,dx = \frac{1}{L_{1} L_{2} L_{3}} \int_{0}^{L_{3}} \re^{-\ri \ell_{3} x_{3}} \int_{0}^{L_{2}} \re^{-\ri \ell_{2} x_{2}} \int_{0}^{L_{1}} u \re^{-\ri \ell_{1} x_{1}}\,dx_{1}\,dx_{2}\,dx_{3};
  \end{displaymath}

  \item maximum vorticity: perhaps one of the most important quantities in the regularity theory of the Euler equations, the \emph{maximum vorticity}
  \begin{displaymath}
    \norm{\omega}_{\infty} := \norm{\omega}_{L^{\infty}(D(1,L))} = \max_{(r,z) \in D(1,L)} \abs{\omega(r,z)}
  \end{displaymath}
  is closely monitored in most Euler computations, and its convergence under mesh refinement is also frequently used as a ``quality indicator'' for the underlying numerical simulations;

  \item conservation of circulation: in a more recent work, \citet{bk2008} proposed to use the relative change of the \emph{circulation}
  \begin{displaymath}
    \Gamma = \oint_{C} u \cdot ds
  \end{displaymath}
  around selected material curves $C$ as an ``error indicator'' for the underlying numerical solutions; the idea is that, according to Kelvin's circulation theorem, the circulation around any closed material curve $C$ is conserved by an Euler flow, hence
  the same should be expected for a numerical solution as well; while conservation of circulation is a physically important principle, its numerical confirmation is not always plausible because it is not always clear how to choose the ``representative''
  material curves $C$; in addition, it is generally not easy to follow a material curve in an Euler flow, since most such simulations are performed on Eulerian meshes while tracking a material curve requires the use of a Lagrangian mesh.
\end{itemize}

We argue that none of the above ``quality indicators'' is adequate in the context of singularity detection. Admittedly, energy, enstrophy, and circulation are all physically significant quantities, and without doubt they should all be accurately resolved
in any ``reasonable'' Euler simulations. On the other hand, it is also important to realize that these quantities are \emph{global} quantities and do not measure the accuracy of a numerical solution at any particular point or in any particular subset of
the computational domain. Since blowing-up solutions of the Euler equations must be characterized by rapidly growing vorticity \citep{bkm1984}, and in most cases such intense vorticity amplification is realized in spatial regions with rapidly
\emph{collapsing} support \citep{kerr1993,hl2006}, it is crucial that the accuracy of a numerically detected blowup candidate be measured by \emph{local} error metrics such as the pointwise (sup-norm) error. When restricted to bounded domains, the
pointwise error is stronger than any other global error metrics in the sense that the latter can be easily bounded in terms of the former, while the converse does not hold true in general. Consequently, the pointwise error provides the most stringent
measure for the quality of a blowup candidate, both near the point of blowup and over the entire (bounded) computational domain.

Arguing in a similar manner, we see that neither energy spectra nor maximum vorticity gives an adequate measure of error for a potentially blowing-up solution. On the one hand, the construction of an energy spectra removes the phase information and
reduces the dimension of the data from three to one, leaving only an incomplete picture of a solution and hence of its associated error. On the other hand, maximum vorticity, albeit significant in its own right, does not tell anything about a solution
except at the point where the vorticity magnitude attained its maximum.

In view of the above considerations, we shall gauge the quality of our Euler simulations at any fixed time instant $t$ using the sup-norm relative errors of the computed solutions $(u_{1},\omega_{1},\psi_{1})$. More specifically, we shall estimate the
error of a given solution, say $u_{1}$, by comparing it with a ``reference solution'', say $\hat{u}_{1}$, that is computed at the same time $t$ on a finer mesh. The reference solution $\hat{u}_{1}$ is first interpolated to the coarse mesh on which
$u_{1}$ is defined. Then the maximum difference between the two solutions is computed and the result is divided by the maximum of $\abs{\hat{u}_{1}}$ (measured on the finer mesh) to yield the desired relative error.

We check the accuracy of our computations in five steps.

\subsubsection{Code Validation on Test Problems}\label{ssec2_res_test}
First, we apply the numerical method described in Section \ref{sec_method} to a test problem with known exact solutions and artificially generated external forcing terms (Appendix \ref{app_test}). The exact solutions are chosen to mimic the behavior of
the blowing-up Euler solution computed from \eqref{eqn_eat}--\eqref{eqn_eat_ibc}, and numerical experiments on successively refined meshes confirm the 6th-order convergence of the overall method (Table \ref{tab_test_err}).
\begin{table}[h]
  \centering
  \caption{Sup-norm relative error and numerical order of convergence (see \eqref{eqn_num_ord}) of the transformed primitive variables $(u_{1},\omega_{1},\psi_{1})$, computed for a test problem with known exact solutions. The absolute size of each
  variable, measured on the finest $768 \times 768$ mesh, is indicated in the last row ``Sup-norm'' of the table.}
  \label{tab_test_err}
  \begin{tabular}{*{7}{>{$}c<{$}}}
    \toprule
     & \multicolumn{6}{c}{Sup-norm relative error at $t = 0.029$} \\
    \cmidrule{2-7}
    \rbs{1.75ex}{Mesh size} & u_{1} & \text{Order} & \omega_{1} & \text{Order} & \psi_{1} & \text{Order} \\
    \midrule
    128 \times 128 & 1.2252 \times 10^{-4} & - & 6.2554 \times 10^{-5} & - & 4.7084 \times 10^{-1} & - \\
    256 \times 256 & 2.9249 \times 10^{-6} & 5.39 & 4.6254 \times 10^{-7} & 7.08 & 2.5819 \times 10^{-3} & 7.51 \\
    384 \times 384 & 2.6925 \times 10^{-7} & 5.88 & 1.8224 \times 10^{-8} & 7.98 & 2.2455 \times 10^{-4} & 6.02 \\
    512 \times 512 & 4.8713 \times 10^{-8} & 5.94 & 1.9185 \times 10^{-9} & 7.83 & 3.9857 \times 10^{-5} & 6.01 \\
    640 \times 640 & 1.3293 \times 10^{-8} & 5.82 & 3.1179 \times 10^{-10} & 8.14 & 1.0217 \times 10^{-5} & 6.10 \\
    768 \times 768 & 4.4301 \times 10^{-9} & 6.03 & 8.8603 \times 10^{-11} & 6.90 & 3.6163 \times 10^{-6} & 5.70 \\
    \midrule
    \text{Sup-norm} & 1.0000 \times 10^{-6} & - & 4.8900 \times 10^{3} & - & 1.1036 \times 10^{-10} & - \\
    \bottomrule
  \end{tabular}
\end{table}%

\subsubsection{Resolution Study on Transformed Primitive Variables}\label{ssec2_res_prim}
Second, we perform a resolution study on the actual solutions of problem \eqref{eqn_eat}--\eqref{eqn_eat_ibc} at various time instants $t$, up to the time $t = 0.003505$ shortly before the simulations terminate. For each $256k \times 256k$ mesh except
for the finest one, we compare the solution $(u_{1},\omega_{1},\psi_{1})$ computed on this mesh with the reference solution $(\hat{u}_{1},\hat{\omega}_{1},\hat{\psi}_{1})$ computed at the same time $t$ on the finer $[256(k+1)] \times [256(k+1)]$ mesh,
and compute the sup-norm relative error using the procedure described above. For each $256k \times 256k$ mesh except for the coarsest one, we also compute, for each error $e_{k}$ defined on this mesh, the numerical order of convergence
\begin{equation}
  \beta_{k} = \log_{k/(k-1)} \Bigl( \frac{e_{k-1}}{e_{k}} \Bigr).
  \label{eqn_num_ord}
\end{equation}
Here, the error $e_{k}$ is understood as a function of the (uniform) mesh spacing $h_{r} = h_{z} = 1/(256k)$ in the $\rho\eta$-space, and is assumed to admit an asymptotic expansion in powers of $h_{r}$ and $h_{z}$. Under suitable regularity assumptions
on the underlying exact solutions and with suitable choices of time steps, it can be shown that $\beta_{k}$ converges to its theoretical value (6 in this case) as $k \to \infty$.

\begin{figure}[p]
  \centering
  \subfigure[sup-norm relative error of $u_{1}$]{
    \includegraphics[scale=0.415]{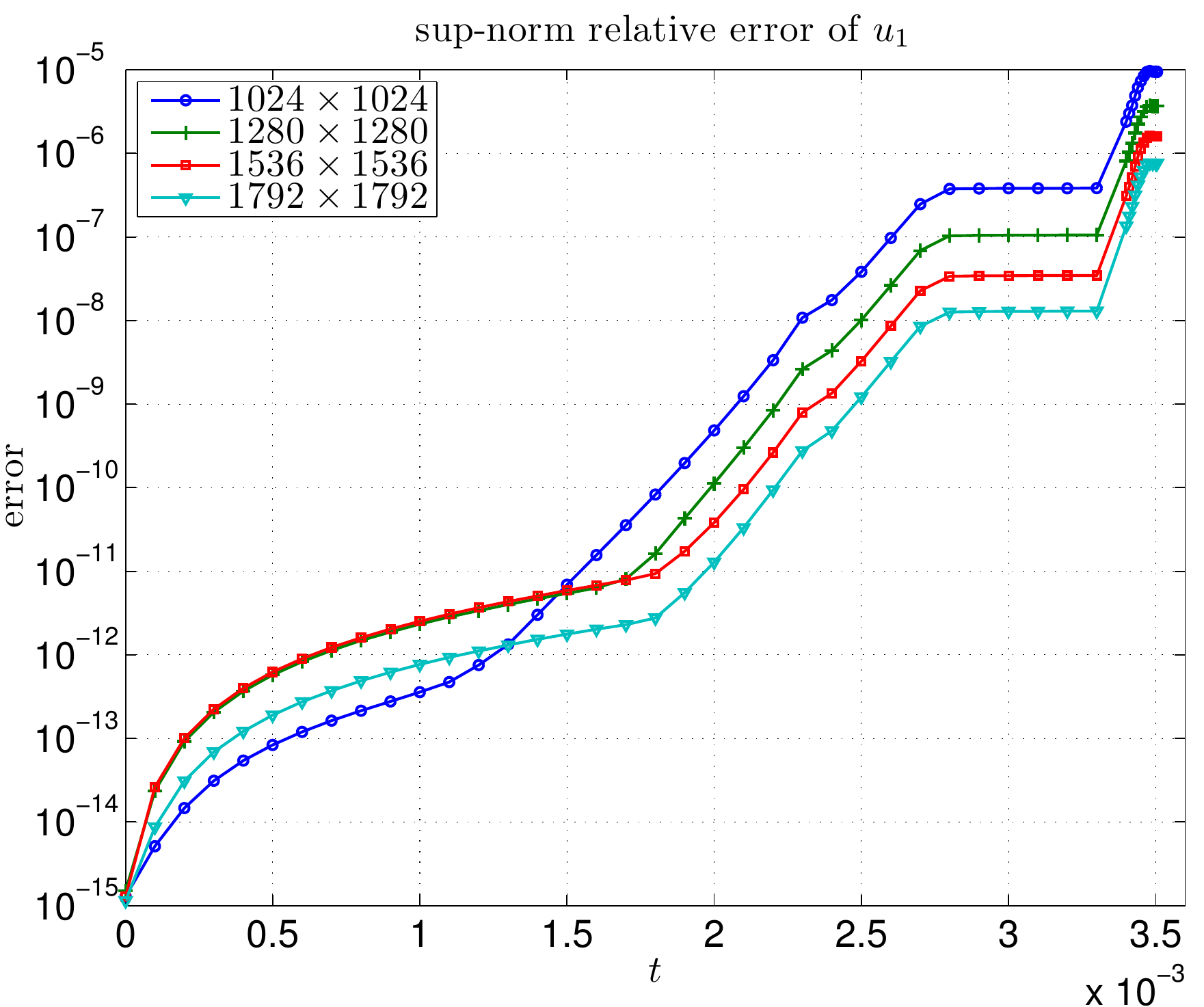}
  }\quad
  \subfigure[numerical order of $u_{1}$ in sup-norm]{
    \includegraphics[scale=0.415]{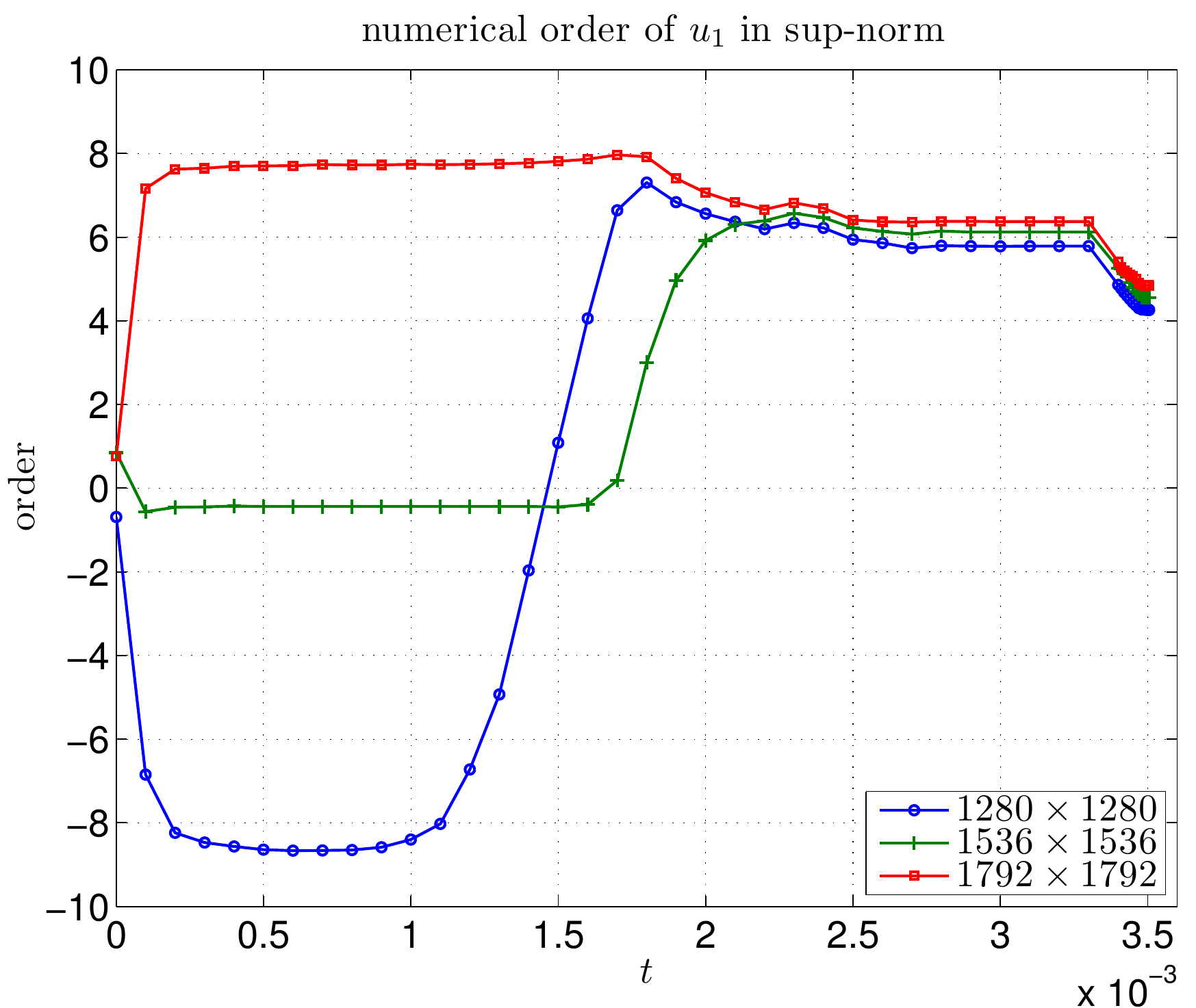}
  }
  \subfigure[sup-norm relative error of $\omega_{1}$]{
    \includegraphics[scale=0.415]{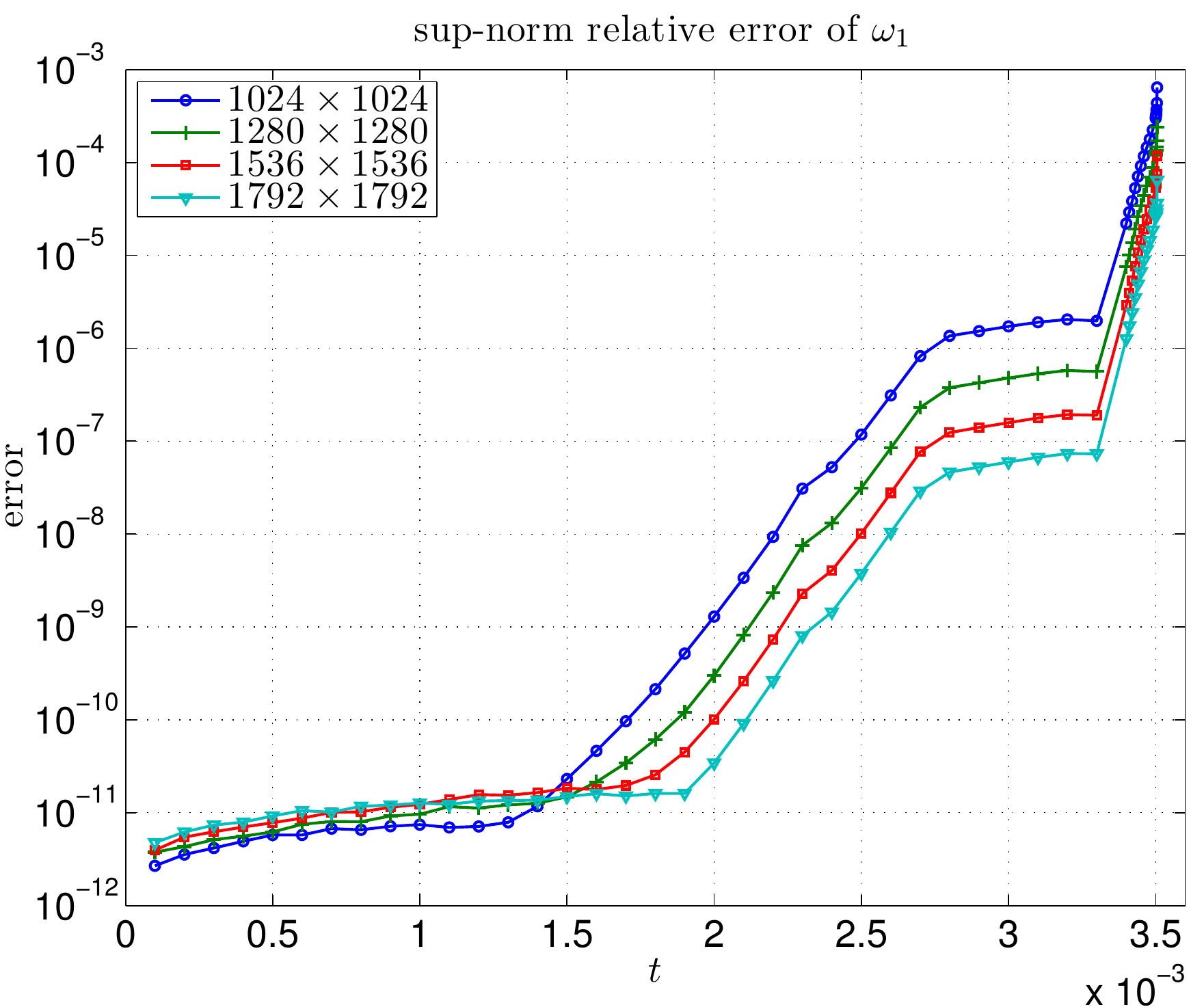}
  }\quad
  \subfigure[numerical order of $\omega_{1}$ in sup-norm]{
    \includegraphics[scale=0.415]{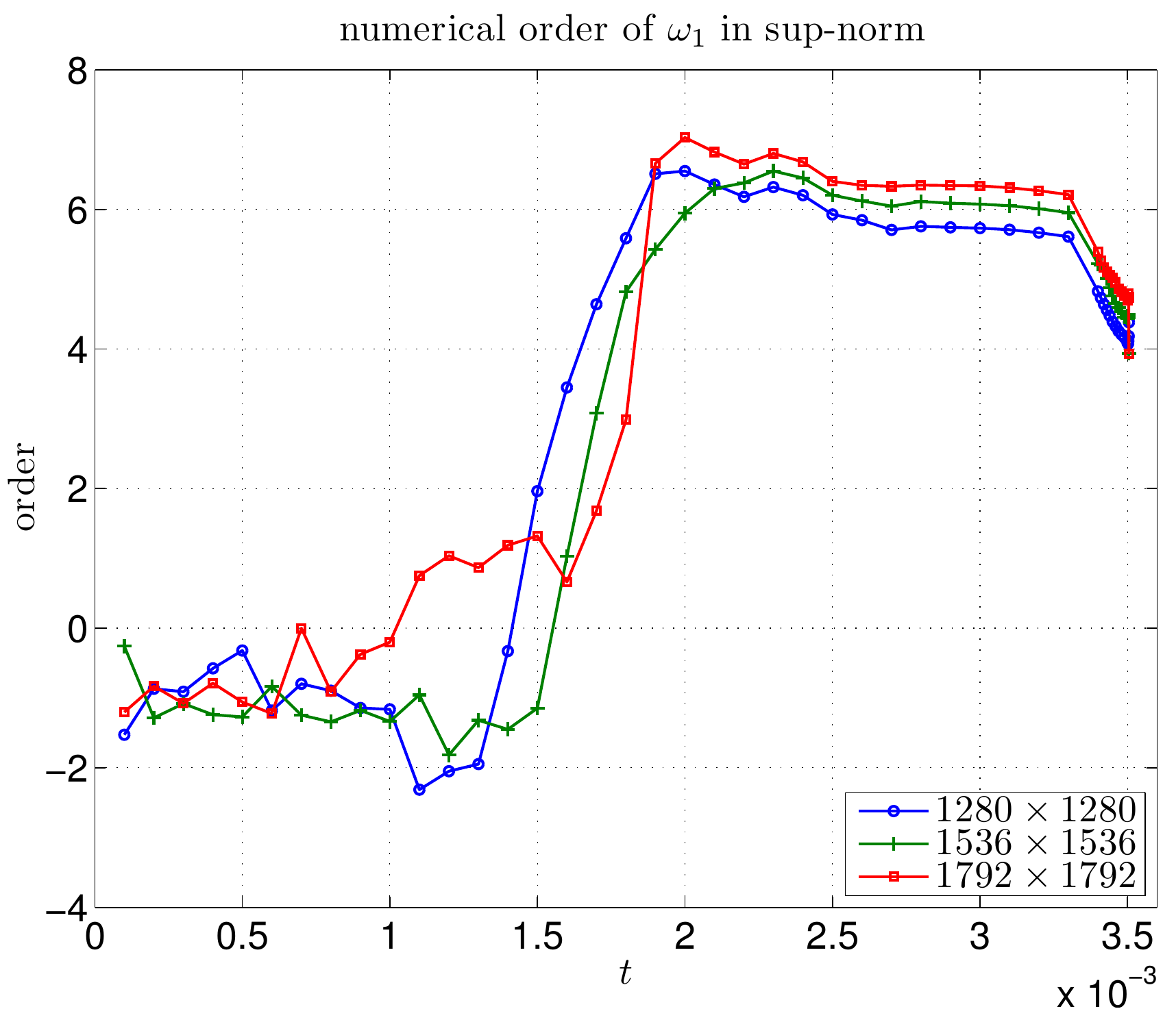}
  }
  \subfigure[sup-norm relative error of $\psi_{1}$]{
    \includegraphics[scale=0.415]{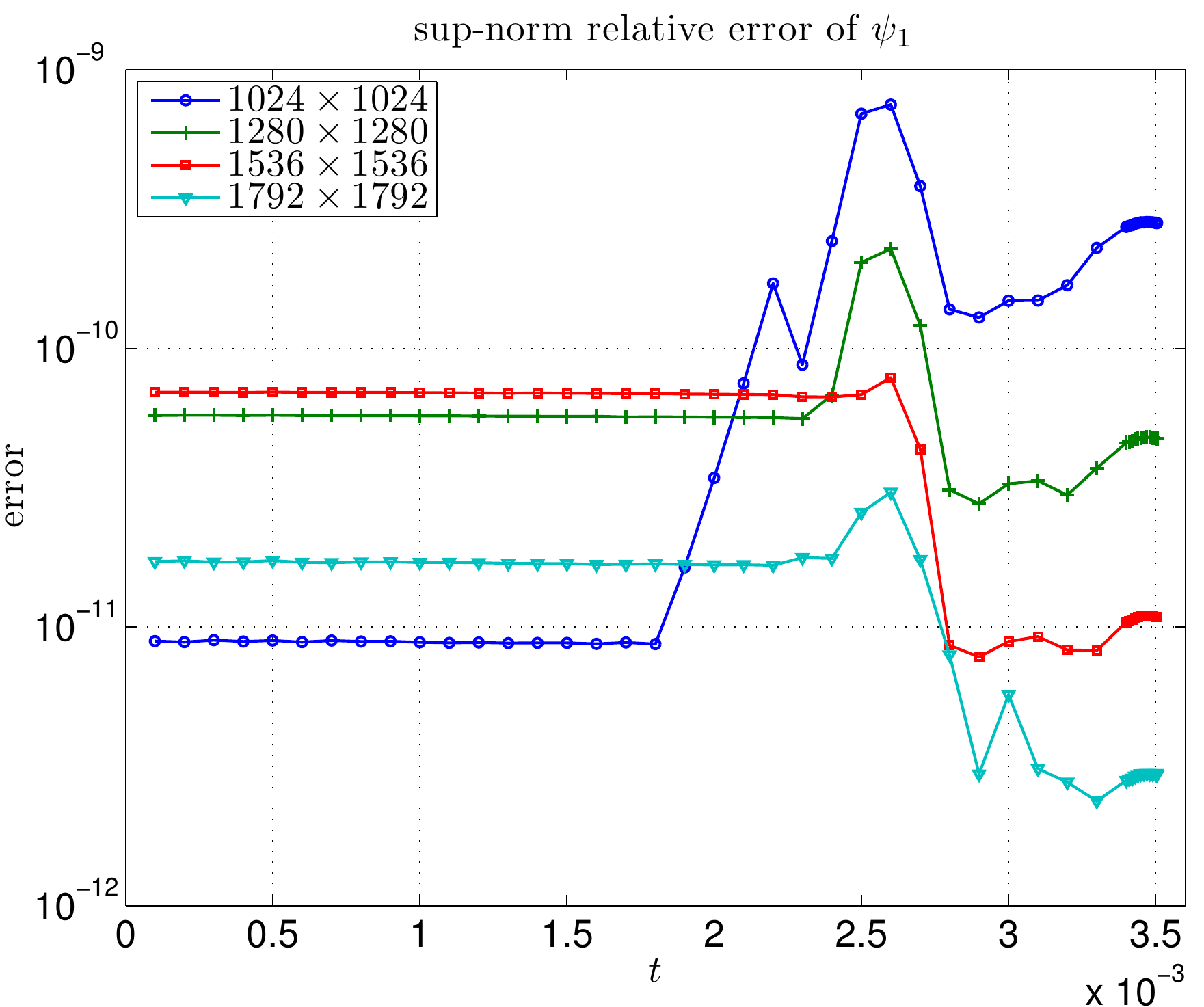}
  }\quad
  \subfigure[numerical order of $\psi_{1}$ in sup-norm]{
    \includegraphics[scale=0.415]{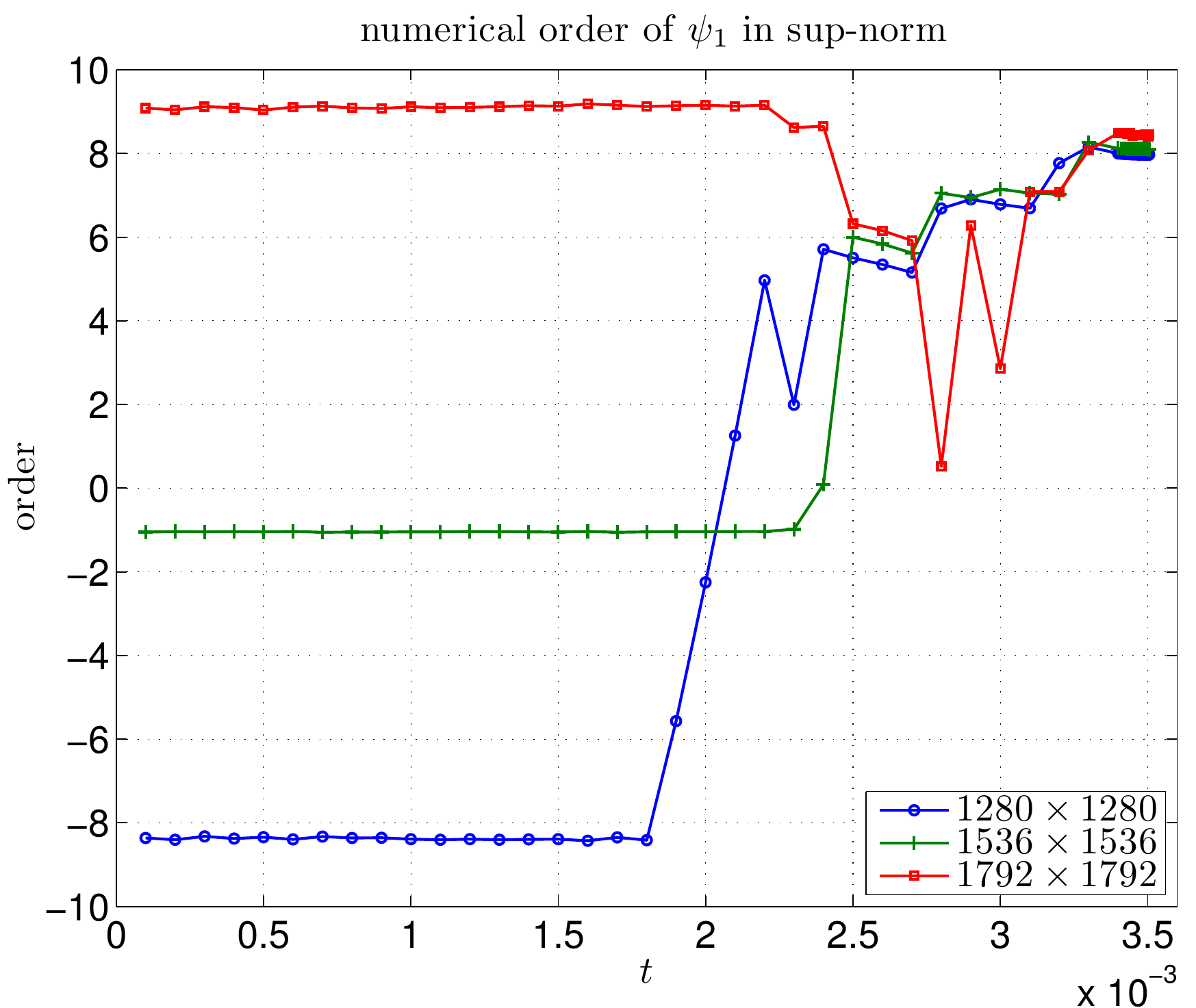}
  }
  \caption{Resolution study in space: (a)(c)(e) sup-norm relative error and (b)(d)(f) numerical order in sup-norm of the transformed primitive variables $(u_{1},\omega_{1},\psi_{1})$. The last time instant shown in the figure is $t = 0.003505$.}
  \label{fig_res_prim}
\end{figure}

The results of the resolution study on the primitive variables $(u_{1},\omega_{1},\psi_{1})$ among the five mesh resolutions are summarized in Figure \ref{fig_res_prim}. To examine more closely the errors at the times when the solutions are about to
``blow up'', we also report in Table \ref{tab_res_prim} the estimated sup-norm errors and numerical orders at $t = 0.003505$. It can be observed from these results that, for small $t$, specifically for $t \lessapprox 0.0015$, the solutions are well
resolved even on the coarsest $1024 \times 1024$ mesh, and further increase in mesh size does not lead to further improvement of the sup-norm errors. For $0.0015 \lessapprox t \lessapprox 0.0033$, the errors first grow exponentially in time and then
level off after $t \approx 0.0028$. The numerical orders estimated on this time interval roughly match their theoretical values 6, confirming the full-order convergence of the computed solutions. For $t \gtrapprox 0.0033$, the exponential growth of the
sup-norm errors resumes at an accelerated pace, in correspondence with the strong, nonlinear amplifications of the underlying solutions observed in this stage. The numerical orders estimated for $u_{1}$ and $\omega_{1}$ decline slightly from 6 to 4, as a
result of the rapidly growing discretization error in time (Figure \ref{fig_res_time}), while the ones for $\psi_{1}$ increase slightly from 6 to 8, thanks most likely to the superconvergence property of the B-spline based Poisson solver at grid points
(Section \ref{ssec_psolv}). Based on these observations, we conclude that the primitive variables computed on the finest two meshes have at least \emph{four significant digits} up to and including the time $t = 0.003505$ shortly before the singularity
forms. To the best of our knowledge, this level of accuracy has never been observed in previous numerical studies (see also Table \ref{tab_vfit_cmp}).
\begin{table}[h]
  \centering
  \caption{Sup-norm relative error and numerical order of convergence of the transformed primitive variables $(u_{1},\omega_{1},\psi_{1})$ at $t = 0.003505$. The absolute size of each variable, measured on the finest $2048 \times 2048$ mesh, is indicated
  in the last row ``Sup-norm'' of the table.}
  \label{tab_res_prim}
  \begin{tabular}{*{7}{>{$}c<{$}}}
    \toprule
     & \multicolumn{6}{c}{Sup-norm relative error at $t = 0.003505$} \\
    \cmidrule{2-7}
    \rbs{1.75ex}{Mesh size} & u_{1} & \text{Order} & \omega_{1} & \text{Order} & \psi_{1} & \text{Order} \\
    \midrule
    1024 \times 1024 & 9.4615 \times 10^{-6} & - & 6.4354 \times 10^{-4} & - & 2.8180 \times 10^{-10} & - \\
    1280 \times 1280 & 3.6556 \times 10^{-6} & 4.26 & 2.4201 \times 10^{-4} & 4.38 & 4.7546 \times 10^{-11} & 7.97 \\
    1536 \times 1536 & 1.5939 \times 10^{-6} & 4.55 & 1.1800 \times 10^{-4} & 3.94 & 1.0873 \times 10^{-11} & 8.09 \\
    1792 \times 1792 & 7.5561 \times 10^{-7} & 4.84 & 6.4388 \times 10^{-5} & 3.93 & 2.9518 \times 10^{-12} & 8.46 \\
    \midrule
    \text{Sup-norm} & 1.0000 \times 10^{2} & - & 1.0877 \times 10^{6} & - & 2.1610 \times 10^{-1} & - \\
    \bottomrule
  \end{tabular}
\end{table}%

\subsubsection{Resolution Study on Vorticity Vector}\label{ssec2_res_vort}
\begin{figure}[h]
  \centering
  \subfigure[sup-norm relative error]{
    \includegraphics[scale=0.415]{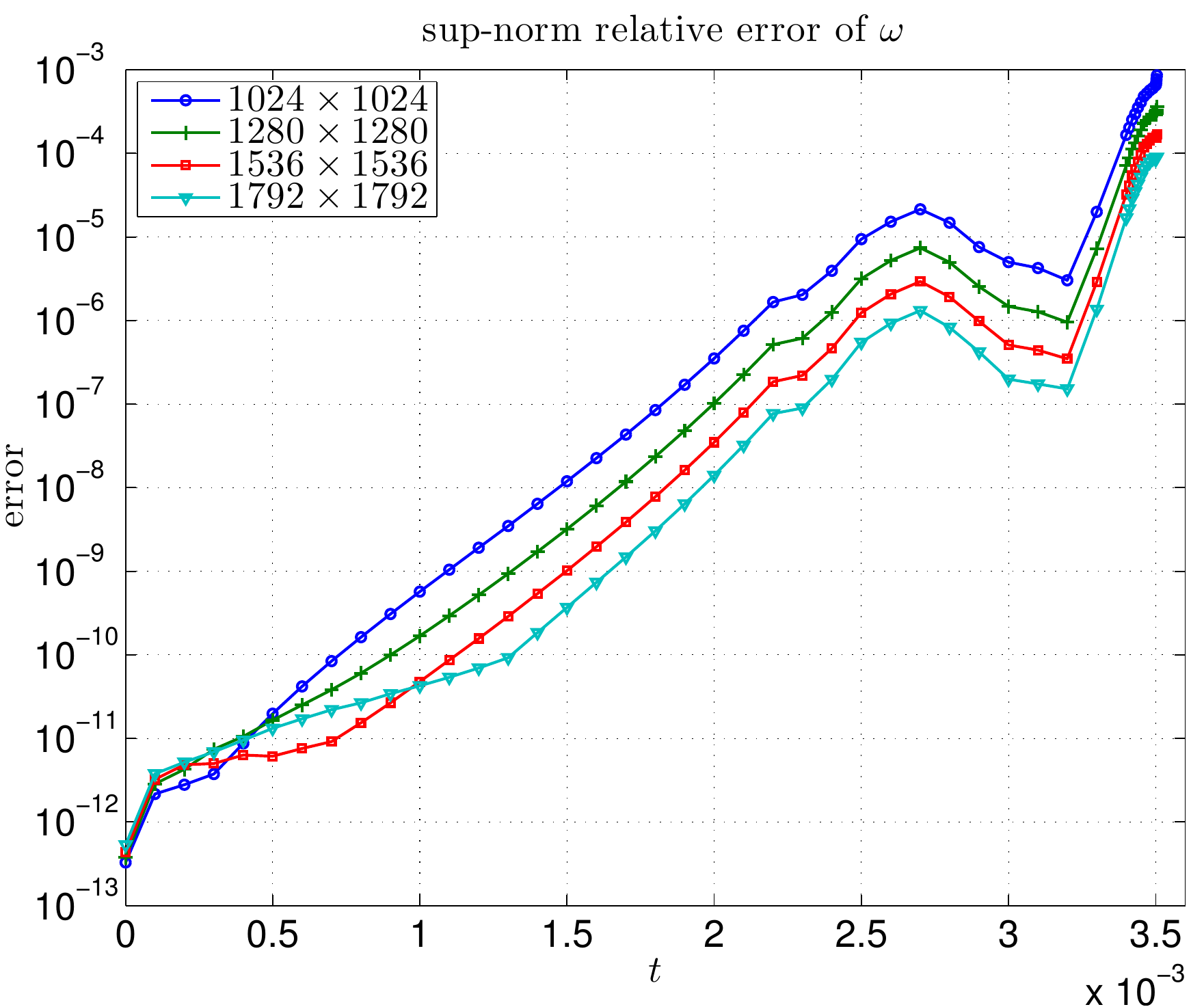}
  }\quad
  \subfigure[numerical order in sup-norm]{
    \includegraphics[scale=0.415]{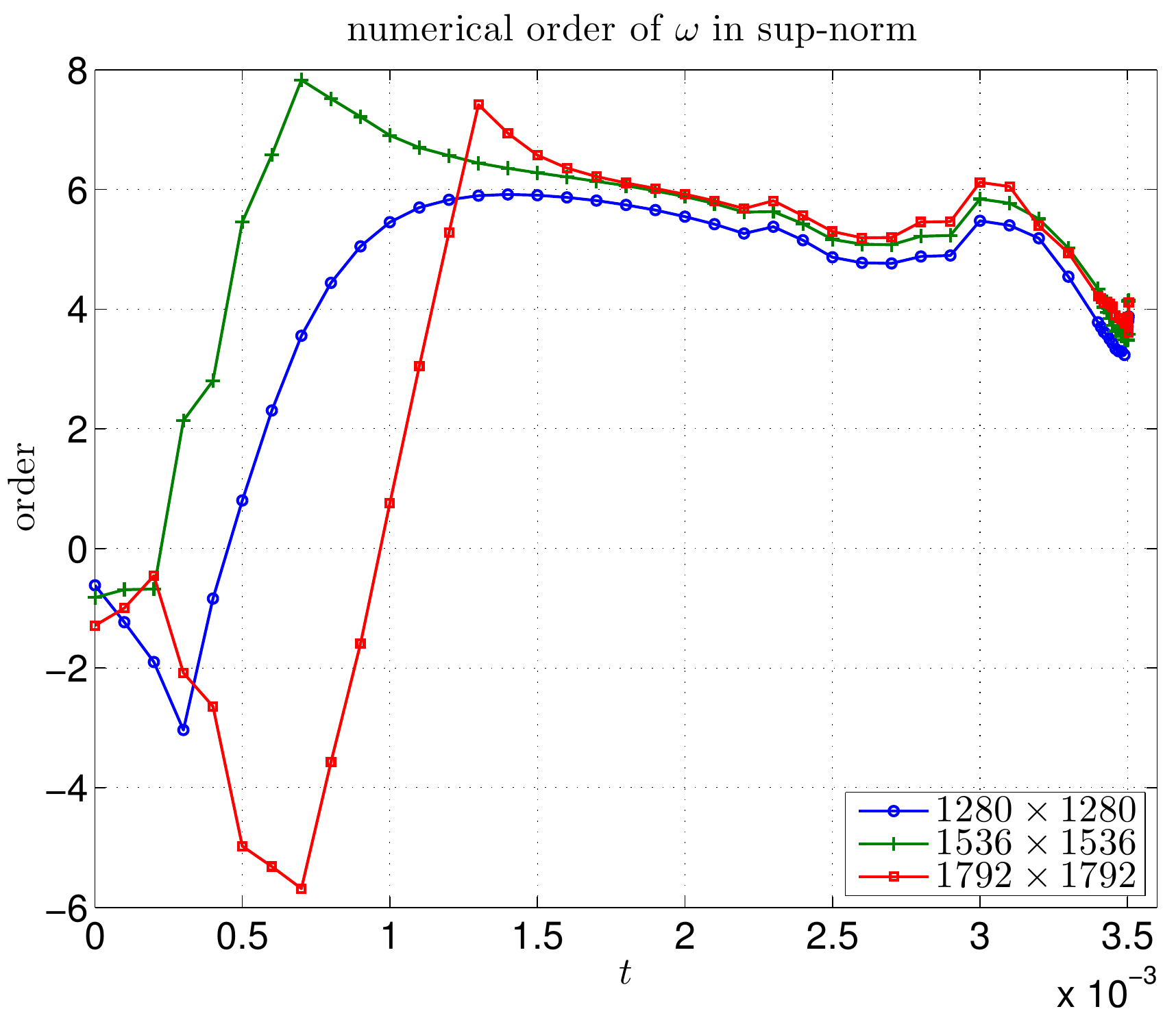}
  }
  \caption{Resolution study in space: (a) sup-norm relative error and (b) numerical order in sup-norm of the vorticity vector $\omega$. The last time instant shown in the figure is $t = 0.003504$.}
  \label{fig_res_vort}
\end{figure}

Since the Beale-Kato-Majda criterion suggests that the vorticity vector $\omega$ controls the blowup of smooth Euler solutions, we next perform a resolution study on $\omega$ to see how well it is resolved in our computations. The procedure is almost
identical to that described for the primitive variables $(u_{1},\omega_{1},\psi_{1})$, except that the difference between a vorticity vector $\omega$ and its reference value $\hat{\omega}$ needs to be measured in a suitable vector norm. By choosing the
usual Euclidean norm, we have
\begin{align*}
  \abs{\omega - \hat{\omega}} & = \Bigl[ (\omega^{r}-\hat{\omega}^{r})^{2} + (\omega^{\theta}-\hat{\omega}^{\theta})^{2} + (\omega^{z}-\hat{\omega}^{z})^{2} \Bigr]^{1/2} \\
  & = \Bigl[ (ru_{1,z} - r\hat{u}_{1,z})^{2} + (r\omega_{1,z} - r\hat{\omega}_{1,z})^{2} + (2u_{1} + ru_{1,r} - 2\hat{u}_{1} - r\hat{u}_{1,r})^{2} \Bigr]^{1/2}.
\end{align*}
The resulting sup-norm errors and numerical orders are summarized in Figure \ref{fig_res_vort} and Table \ref{tab_res_vort}. These results will be used below in Section \ref{ssec_vfit} in the computation of the asymptotic scalings of the nearly singular
solutions.
\begin{table}[h]
  \centering
  \caption{Sup-norm relative error and numerical order of convergence of the vorticity vector $\omega$ at selected time $t$. The absolute size of $\omega$, measured on the finest $2048 \times 2048$ mesh, is indicated in the last row ``Sup-norm'' of the
  table.}
  \label{tab_res_vort}
  \begin{tabular}{*{5}{>{$}c<{$}}}
    \toprule
     & \multicolumn{4}{c}{Sup-norm relative error of $\omega$} \\
    \cmidrule{2-5}
    \rbs{1.75ex}{Mesh size} & t = 0.003504 & \text{Order} & t = 0.003505 & \text{Order} \\
    \midrule
    1024 \times 1024 & 8.5671 \times 10^{-4} & - & 1.1352 \times 10^{-3} & - \\
    1280 \times 1280 & 3.6084 \times 10^{-4} & 3.87 & 4.5801 \times 10^{-4} & 4.07 \\
    1536 \times 1536 & 1.6929 \times 10^{-4} & 4.15 & 2.3050 \times 10^{-4} & 3.77 \\
    1792 \times 1792 & 8.9837 \times 10^{-5} & 4.11 & 3.3212 \times 10^{-4} & -^{\dag} \\
    \midrule
    \text{Sup-norm} & 1.2209 \times 10^{11} & - & 1.2401 \times 10^{12} & - \\
    \bottomrule
    \multicolumn{5}{l}{\footnotesize{$\dag$: Round-off error begins to dominate.}}
  \end{tabular}
\end{table}%

\subsubsection{Resolution Study on Global Quantities}\label{ssec2_res_glob}
The next step in our resolution study is to examine the ``conventional'' error indicators defined using global quantities such as energy $E$, enstrophy $\mathcal{E}$, enstrophy production rate $\mathcal{E}_{p}$\footnote{All these integrals are
discretized in the $\rho\eta$-space using the 6th-order composite Boole's rule.}, maximum vorticity $\norm{\omega}_{\infty}$\footnote{We define $\norm{\omega}_{\infty}$ simply as the maximum value of $\abs{\omega}$ on the discrete mesh points (i.e. no
interpolation is used to find the ``precise'' maximum). In view of the highly effective adaptive mesh, this does not cause any loss of accuracy. In addition, for the specific initial data \eqref{eqn_eat_ic}, $\norm{\omega}_{\infty}$ is always attained at
$\tilde{q}_{0} = (1,0)^{T}$ which is always a mesh point.}, and circulation $\Gamma$. As we already pointed out, conservation of circulation is physically important but is difficult to check in practice, because it requires selection and tracking of
representative material curves which is not always easy. On the other hand, in axisymmetric flows the total circulation along the circular contours
\begin{displaymath}
  C = \Bigl\{ (x,y,z) \in \mathbb{R}^{3}\colon x^{2}+y^{2} = r^{2} < 1,\ \text{$z$ a constant} \Bigr\}
\end{displaymath}
is easily found to be $\Gamma = 2\pi r^{2} u_{1}$. Thus as an alternative to conservation of circulation, we choose to monitor the extreme circulations
\begin{displaymath}
  \Gamma_{1} = 2\pi \min_{(r,z) \in D(1,L)} r^{2} u_{1}(r,z),\qquad \Gamma_{2} = 2\pi \max_{(r,z) \in D(1,L)} r^{2} u_{1}(r,z),
\end{displaymath}
which must be conserved over time according to Kelvin's circulation theorem.

We study the errors of the above-mentioned global quantities as follows. For conserved quantities such as kinetic energy and extreme circulations, the maximum (relative) change
\begin{displaymath}
  \norm{\delta Q}_{\infty,t} = \max_{s \in [0,t]} \abs{\delta Q(s)}
\end{displaymath}
over the interval $[0,t]$ is computed, where
\begin{displaymath}
  \delta Q(t) =
  \biggl\{\! \begin{array}{ll}
    Q(0)^{-1} \bigl[ Q(t) - Q(0) \bigr], & \text{if $Q(0) \neq 0$} \\
    Q(t) - Q(0), & \text{if $Q(0) = 0$}
  \end{array}.
\end{displaymath}
For other nonconservative quantities, the relative error
\begin{displaymath}
  \frac{1}{\hat{Q}(t)} \abs{Q(t) - \hat{Q}(t)}
\end{displaymath}
is computed where $Q$ denotes global quantities computed on a $256k \times 256k$ mesh and $\hat{Q}$ represents reference values obtained on the finer $[256(k+1)] \times [256(k+1)]$ mesh. The resulting errors and numerical orders at $t = 0.003505$ are
summarized in Table \ref{tab_res_consv}--\ref{tab_res_glob}.

As a side remark, we note that the error of the maximum vorticity $\norm{\omega}_{\infty}$ is always a \emph{lower bound} of the error of the vorticity vector $\omega$. This is a direct consequence of the triangle inequality
\begin{displaymath}
  \bigl\lvert \norm{\omega}_{\infty} - \norm{\hat{\omega}}_{\infty} \bigr\rvert \leq \norm{\omega - \hat{\omega}}_{\infty},
\end{displaymath}
and is readily confirmed by the results shown in Table \ref{tab_res_vort} and Table \ref{tab_res_glob}. In addition, note that global errors such as the error of the enstrophy $\mathcal{E}$ can \emph{significantly underestimate} the pointwise error of
the vorticity vector $\omega$. This confirms the inadequacy of the ``conventional'' error indicators in the context of singularity detection.
\begin{table}[h]
  \centering
  \caption{Maximum (relative) change of kinetic energy $E$, minimum circulation $\Gamma_{1}$, and maximum circulation $\Gamma_{2}$ over the interval $[0,0.003505]$. The initial value of each quantity, measured on the finest $2048 \times 2048$ mesh, is
  indicated in the last row ``Init. value'' of the table.}
  \label{tab_res_consv}
  \begin{tabular}{*{4}{>{$}c<{$}}}
    \toprule
     & \multicolumn{3}{c}{$t = 0.003505$} \\
    \cmidrule{2-4}
    \rbs{1.75ex}{Mesh size} & \norm{\delta E}_{\infty,t} & \norm{\delta \Gamma_{1}}_{\infty,t} & \norm{\delta \Gamma_{2}}_{\infty,t} \\
    \midrule
    1024 \times 1024 & 1.5259 \times 10^{-11} & 4.3525 \times 10^{-17} & 1.2485 \times 10^{-14} \\
    1280 \times 1280 & 4.1730 \times 10^{-12} & 3.3033 \times 10^{-17} & 7.7803 \times 10^{-15} \\
    1536 \times 1536 & 2.0787 \times 10^{-12} & 3.1308 \times 10^{-17} & 9.9516 \times 10^{-15} \\
    1792 \times 1792 & 6.4739 \times 10^{-13} & 2.7693 \times 10^{-17} & 2.1351 \times 10^{-14} \\
    2048 \times 2048 & 6.6594 \times 10^{-13} & 2.5308 \times 10^{-17} & 3.4921 \times 10^{-14} \\
    \midrule
    \text{Init. value} & 55.9309 & 0.0000 & 6.2832 \times 10^{2} \\
    \bottomrule
  \end{tabular}
\end{table}%
\begin{table}[h]
  \centering
  \caption{Relative error of enstrophy $\mathcal{E}$, enstrophy production rate $\mathcal{E}_{p}$, and maximum vorticity $\norm{\omega}_{\infty}$ at $t = 0.003505$. The absolute size of each quantity, measured on the finest $2048 \times 2048$ mesh, is
  indicated in the last row ``Ref. value'' of the table.}
  \label{tab_res_glob}
  \begin{tabular}{*{7}{>{$}c<{$}}}
    \toprule
     & \multicolumn{6}{c}{Relative error at $t = 0.003505$} \\
    \cmidrule{2-7}
    \rbs{1.75ex}{Mesh size} & \mathcal{E} & \text{Order} & \mathcal{E}_{p} & \text{Order} & \norm{\omega}_{\infty} & \text{Order} \\
    \midrule
    1024 \times 1024 & 4.6075 \times 10^{-6} & - & 4.6565 \times 10^{-5} & - & 7.7593 \times 10^{-4} & - \\
    1280 \times 1280 & 1.4946 \times 10^{-6} & 5.05 & 1.4488 \times 10^{-5} & 5.23 & 3.0099 \times 10^{-4} & 4.24 \\
    1536 \times 1536 & 5.6161 \times 10^{-7} & 5.37 & 5.3275 \times 10^{-6} & 5.49 & 1.2927 \times 10^{-4} & 4.64 \\
    1792 \times 1792 & 2.3385 \times 10^{-7} & 5.68 & 2.0314 \times 10^{-6} & 6.25 & 6.1010 \times 10^{-5} & 4.87 \\
    \midrule
    \text{Ref. value} & 7.0254 \times 10^{5} & - & 1.4270 \times 10^{10} & - & 1.2401 \times 10^{12} & - \\
    \bottomrule
  \end{tabular}
\end{table}%

\subsubsection{Resolution Study in Time}\label{ssec2_res_time}
\begin{figure}[h]
  \centering
  \subfigure[sup-norm relative error of $u_{1}$]{
    \includegraphics[scale=0.415]{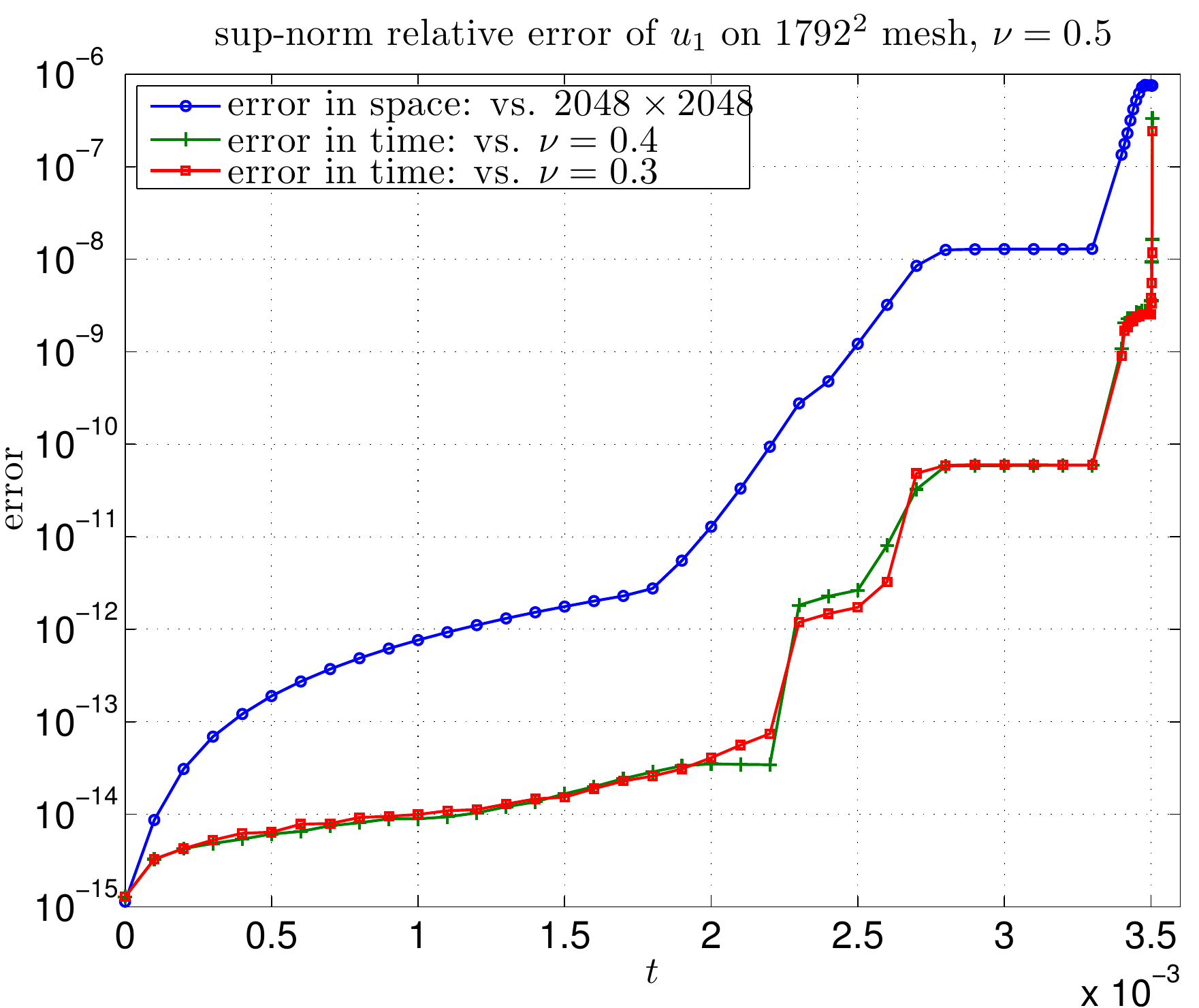}
  }\quad
  \subfigure[sup-norm relative error of $\omega_{1}$]{
    \includegraphics[scale=0.415]{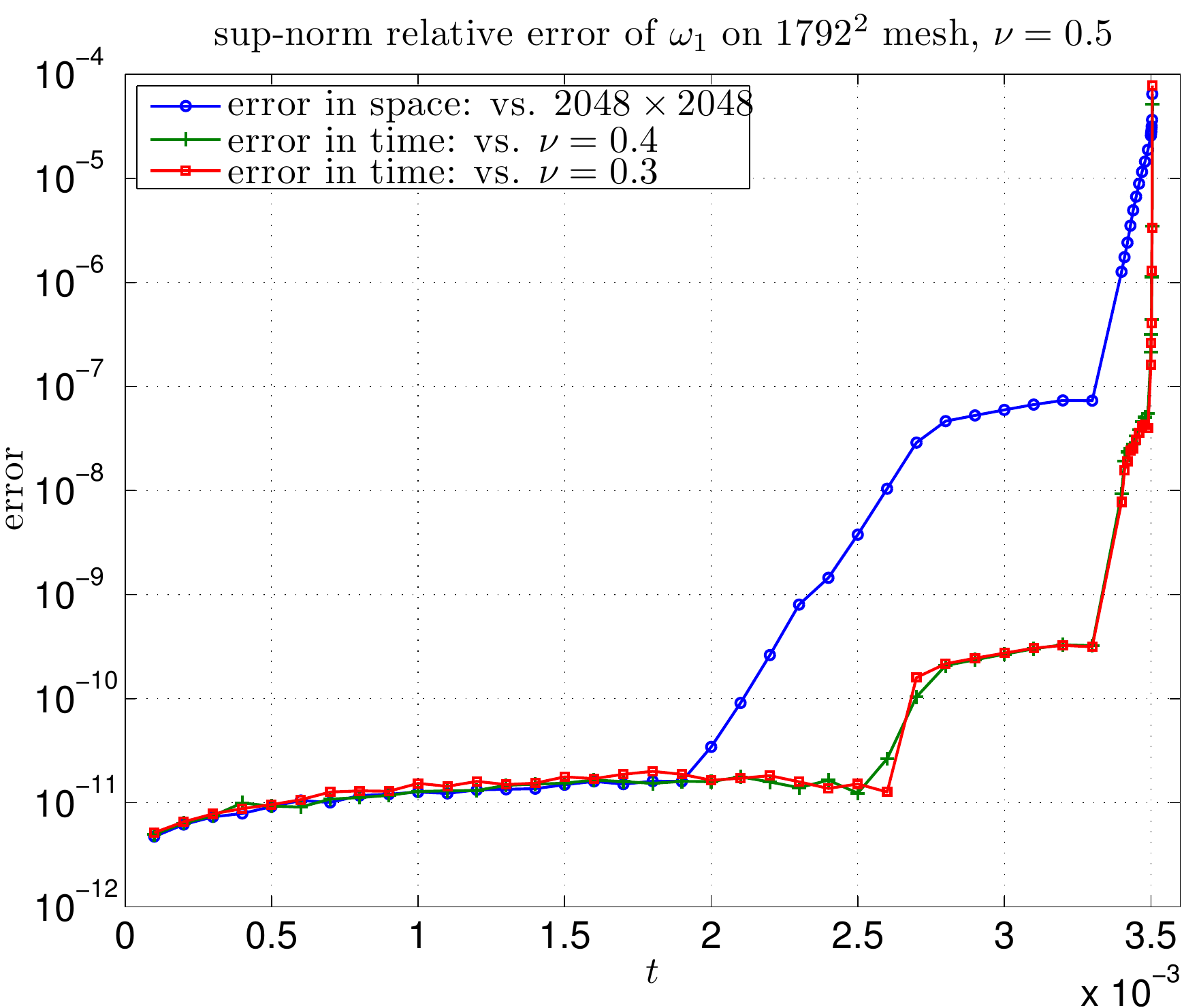}
  }
  \subfigure[sup-norm relative error of $\psi_{1}$]{
    \includegraphics[scale=0.415]{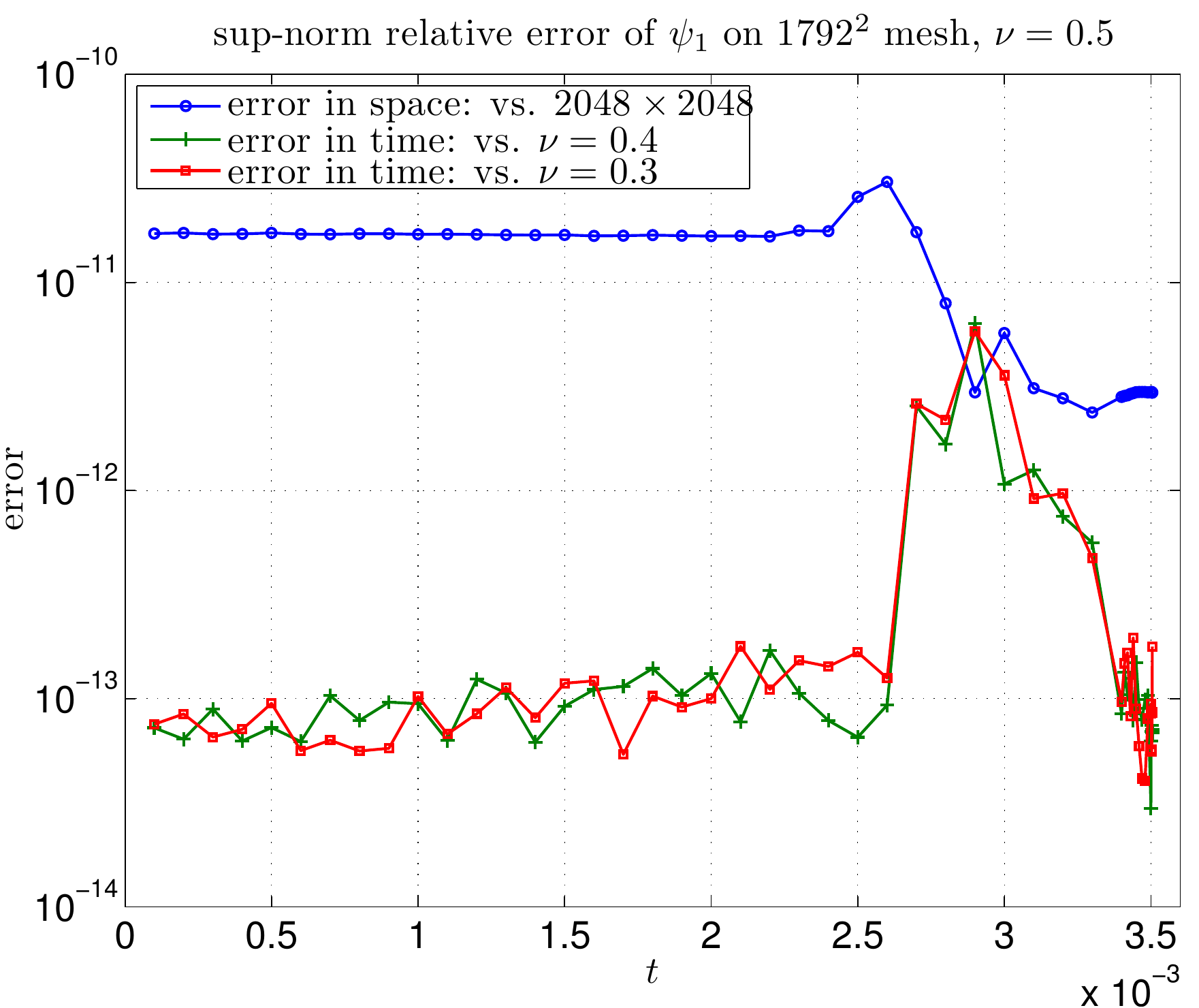}
  }\quad
  \subfigure[sup-norm relative error of $\omega$]{
    \includegraphics[scale=0.415]{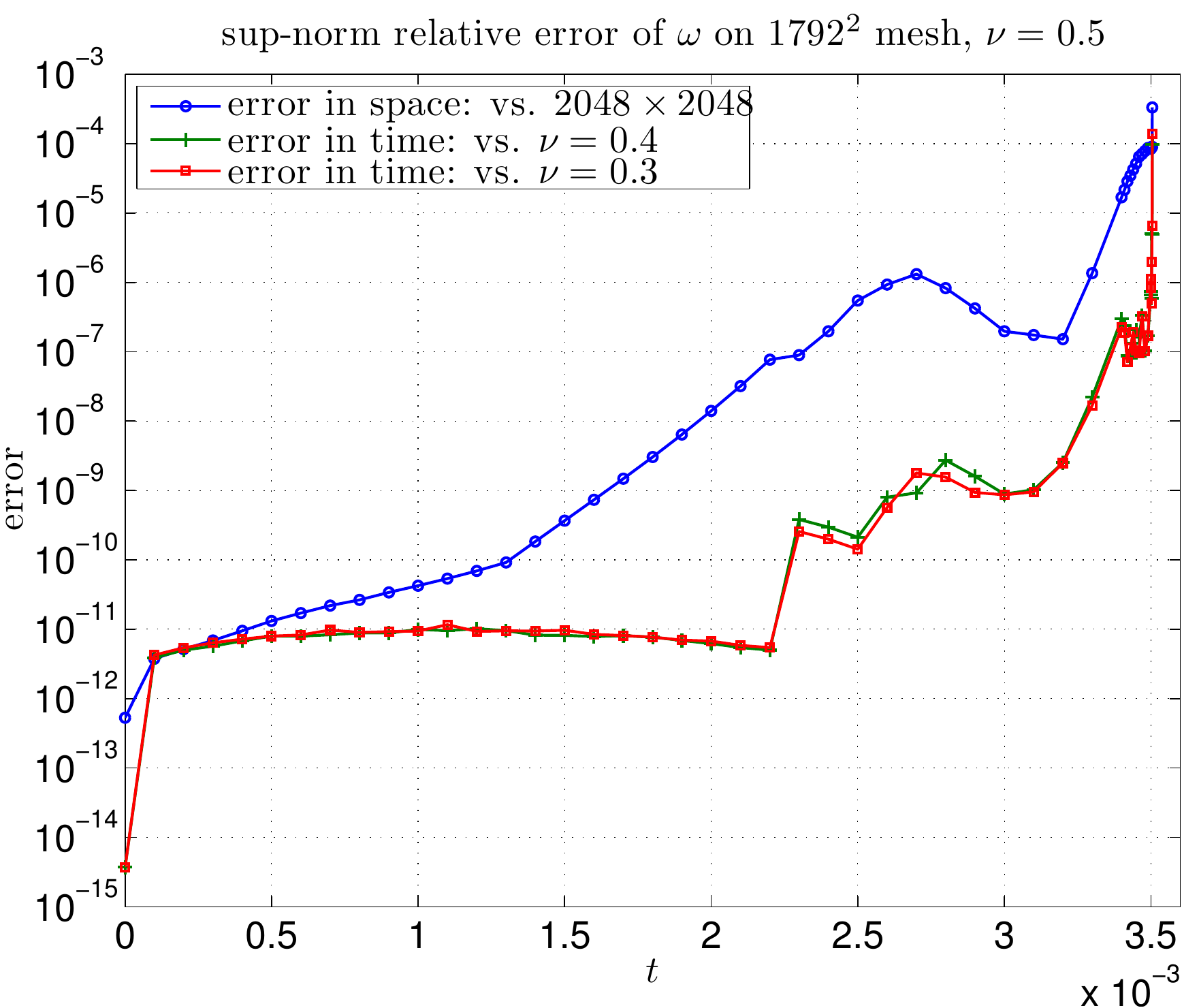}
  }
  \caption{Resolution study in time: sup-norm relative error of (a)--(c) the transformed primitive variables $(u_{1},\omega_{1},\psi_{1})$ and (d) the vorticity vector $\omega$, computed on the $1792 \times 1792$ mesh. The last time instant shown in the
  figure is $t = 0.003505$.}
  \label{fig_res_time}
\end{figure}

Finally, we perform a resolution study in time by repeating the $1792 \times 1792$ mesh computation using smaller time steps $\delta_{t}$. This is achieved by reducing the CFL number from $\nu = 0.5$ to $0.4,\ 0.3$, and the relative growth threshold from
$\epsilon_{t} = 5\%$ to $4\%,\ 3\%$ (Section \ref{ssec_alg}). For each reduced time step computation, the resulting solution $(\hat{u}_{1},\hat{\omega}_{1},\hat{\psi}_{1},\hat{\omega})$ is taken as the reference solution and is compared with the original
solution $(u_{1},\omega_{1},\psi_{1},\omega)$ computed using $(\nu,\epsilon_{t}) = (0.5,5\%)$. The corresponding sup-norm errors are summarized in Figure \ref{fig_res_time} and Table \ref{tab_res_time}. Note that the error between the computations
$(\nu,\epsilon_{t}) = \{(0.3,3\%),(0.5,5\%)\}$ is roughly the same as that between the computations $(\nu,\epsilon_{t}) = \{(0.4,4\%),(0.5,5\%)\}$, which is smaller than the error between the $1792 \times 1792$ and the $2048 \times 2048$ mesh
computations. This indicates that the solutions computed on the $1792 \times 1792$ and all the coarser meshes with $(\nu,\epsilon_{t}) = (0.5,5\%)$ are well resolved in time up to $t = 0.003505$.
\begin{table}[h]
  \centering
  \caption{Sup-norm relative error of the transformed primitive variables $(u_{1},\omega_{1},\psi_{1})$ and the vorticity vector $\omega$, computed on the $1792 \times 1792$ mesh with $(\nu,\epsilon_{t}) = (0.5,5\%)$, and compared with different
  reference solutions. The absolute size of each variable, measured on the finest $2048 \times 2048$ mesh, is indicated in the last row ``Sup-norm'' of the table.}
  \label{tab_res_time}
  \begin{tabular}{*{5}{>{$}c<{$}}}
    \toprule
     & \multicolumn{4}{c}{Sup-norm relative error at $t = 0.003505$} \\
    \cmidrule{2-5}
    \rbs{1.75ex}{Ref. solution} & u_{1} & \omega_{1} & \psi_{1} & \omega \\
    \midrule
    2048 \times 2048 & 7.5561 \times 10^{-7} & 6.4388 \times 10^{-5} & 2.9518 \times 10^{-12} & 3.3212 \times 10^{-4} \\
    (\nu,\epsilon_{t}) = (0.4,4\%) & 3.3350 \times 10^{-7} & 5.1609 \times 10^{-5} & 6.8713 \times 10^{-14} & 9.7514 \times 10^{-5} \\
    (\nu,\epsilon_{t}) = (0.3,3\%) & 2.4197 \times 10^{-7} & 7.7720 \times 10^{-5} & 1.7776 \times 10^{-13} & 1.3800 \times 10^{-4} \\
    \midrule
    \text{Sup-norm} & 1.0000 \times 10^{2} & 1.0877 \times 10^{6} & 2.1610 \times 10^{-1} & 1.2401 \times 10^{12} \\
    \bottomrule
  \end{tabular}
\end{table}%

\subsection{Asymptotic Scaling Analysis I: Maximum Vorticity}\label{ssec_vfit}
With the pointwise error bounds established in the previous section, we are ready to examine the numerical data in greater detail and apply the mathematical criteria reviewed in Section \ref{sec_intro} to assess the likelihood of a finite-time
singularity.

The basic tool that we shall use is the well-known Beale-Kato-Majda (BKM) criterion \citep{bkm1984}. According to this criterion, a smooth solution of the 3D Euler equations blows up at time $t_{s}$ if and only if
\begin{displaymath}
  \int_{0}^{t_{s}} \norm{\omega(\cdot,t)}_{\infty}\,dt = \infty,
\end{displaymath}
where $\norm{\omega(\cdot,t)}_{\infty}$ is the maximum vorticity at time $t$. The BKM criterion was originally proved in \citet{bkm1984} for flows in free space $\mathbb{R}^{3}$, and was later generalized by \citet{ferrari1993} and \citet{sy1993} to
flows in smooth bounded domains subject to no-flow boundary conditions. In view of this criterion, a ``standard'' approach to singularity detection in Euler computations is to \emph{assume} the existence of an appropriate asymptotic scaling for
$\norm{\omega}_{\infty}$, typically in the form of an \emph{inverse power-law}
\begin{equation}
  \norm{\omega(\cdot,t)}_{\infty} \sim c (t_{s}-t)^{-\gamma},\qquad c,\, \gamma > 0.
  \label{eqn_vfit}
\end{equation}
Then an estimate of the (unknown) singularity time $t_{s}$ and the scaling parameters $(c,\gamma)$ is obtained using a line fitting procedure. Normally, the line fitting is computed on some interval $[\tau_{1},\tau_{2}]$ prior to the predicted
singularity time $t_{s}$, and the results are extrapolated forward in time to yield the desired estimates.

Although seemingly straightforward, the above procedure must be used with caution. Indeed, there are examples where inadvertent line fitting has led to false predictions of finite-time singularities. As we shall demonstrate below, the key to the
successful application of the line fitting procedure lies in the choice of the fitting interval $[\tau_{1},\tau_{2}]$. One must realize, upon the invocation of \eqref{eqn_vfit}, that the applicability of this form fit is \emph{not known a priori} and
must be determined from the line fitting itself. In order for the line fitting to work, the interval $[\tau_{1},\tau_{2}]$ must be placed \emph{within the asymptotic regime of \eqref{eqn_vfit}} if scalings of that form do exist. If such an asymptotic
regime cannot be identified, then the validity of \eqref{eqn_vfit} is questionable and any conclusions drawn from the line fitting are likely to be false.

In most existing studies, the choice of the fitting interval $[\tau_{1},\tau_{2}]$ is based on discretionary manual selections, which tend to generate results that lack clear interpretations and are difficult to reproduce. To overcome these difficulties,
we propose to choose $\tau_{1},\ \tau_{2}$ using an automatic procedure which in ideal situations should place $\tau_{2}$ at $t_{s}$ and $\tau_{1}$ at a point ``close enough'' to $t_{s}$, in such a way that $[\tau_{1},\tau_{2}]$ is enclosed in the
asymptotic regime of \eqref{eqn_vfit}. In reality, such a choice can never be made because a singularity time $t_{s}$, if it exists, can never be attained by a numerical simulation. Thus we propose to place $\tau_{1},\ \tau_{2}$ close enough to the
\emph{stopping time $t_{e}$} such that the computed solutions are still ``well resolved'' on $[\tau_{1},\tau_{2}]$ and an asymptotic scaling of the form \eqref{eqn_vfit} exists and dominates in $[\tau_{1},\tau_{2}]$. To this end, we shall choose
$\tau_{2}$ to be the first time instant at which the sup-norm relative error of the vorticity vector $\omega$ exceeds a certain threshold $\epsilon_{2}$, and choose $\tau_{1}$ so that $[\tau_{1},\tau_{2}]$ is the interval on which the line fitting yields
the ``best'' results (in a sense to be made precise below). Note that the accuracy of the computed solutions is measured in terms of the error of $\omega$, not that of $(u_{1},\omega_{1},\psi_{1})$, because $\omega$ is the quantity that controls the
blowup.

We consider a line fitting ``successful'' if both $\tau_{2}$ and the line-fitting predicted singularity time $\hat{t}_{s}$ converge to the \emph{same} finite value as the mesh is refined. The convergence should be \emph{monotone}, i.e. $\tau_{2} \uparrow
t_{s},\ \hat{t}_{s} \downarrow t_{s}$ where $t_{s}$ is the common limit, the true singularity time. In addition, $\tau_{1}$ should converge to a finite value that is strictly less than $t_{s}$ as the mesh is refined. The reason that the convergence of
$\tau_{2},\ \hat{t}_{s}$ to the singularity time $t_{s}$ should be monotone is two-fold: first, the finer the mesh, the longer it takes the error to grow to a given tolerance and hence the larger the $\tau_{2}$ is; second, as $\tau_{2}$ gets increasingly
closer to $t_{s}$, the strong, singular growth of the blowing-up solution is better captured on $[\tau_{1},\tau_{2}]$, which then translates into an earlier estimate $\hat{t}_{s}$ of the blowup time.

If the interval $[\tau_{1},\tau_{2}]$ can be chosen to satisfy all the above criteria, and the scaling parameters $(c,\gamma)$ estimated on this interval converge to some finite values $c_{s} > 0,\ \gamma_{s} \geq 1$ as the mesh is refined, then the
existence of a finite-time singularity is confirmed.

Let's now apply these ideas to our numerical data.

\subsubsection{The Line Fitting Procedure}\label{ssec2_vfit_line}
We first describe a line fitting procedure that will be needed in both the choice of the fitting interval $[\tau_{1},\tau_{2}]$ (Section \ref{ssec2_vfit_tau}) and the computation of the asymptotic scaling \eqref{eqn_vfit}. Under the assumption that the
maximum vorticity $\norm{\omega}_{\infty}$ is approximated sufficiently well by the inverse power-law \eqref{eqn_vfit} on the interval $[\tau_{1},\tau_{2}]$, the \emph{logarithmic time derivative}, or simply the log $t$-derivative, of
$\norm{\omega}_{\infty}$ is easily found to satisfy
\begin{displaymath}
  \frac{d}{dt} \log \norm{\omega(\cdot,t)}_{\infty} = \norm{\omega(\cdot,t)}_{\infty}^{-1} \frac{d}{dt} \norm{\omega(\cdot,t)}_{\infty} \sim \frac{\gamma}{t_{s}-t}.
\end{displaymath}
This leads to the simple linear regression model
\begin{equation}
  y(t) := \biggl[ \frac{d}{dt} \log \norm{\omega(\cdot,t)}_{\infty} \biggr]^{-1} \sim -\frac{1}{\gamma} (t-t_{s}) =: at + b,
  \label{eqn_vfit_ts}
\end{equation}
with response variable $y$, explanatory variable $t$, and model parameters $a = -1/\gamma,\ b = t_{s}/\gamma$. The model parameters in \eqref{eqn_vfit_ts} can be estimated from a standard least-squares procedure. The fitness of the model can be measured
using either the \emph{coefficient of determination} (the $R^{2}$):
\begin{displaymath}
  R^{2} = 1 - \frac{SS_{\text{err}}}{SS_{\text{tot}}},
\end{displaymath}
where a value close to 1 indicates good fitness, or the \emph{fraction of variance unexplained} (FVU):
\begin{displaymath}
  \text{FVU} = 1 - R^{2} = \frac{SS_{\text{err}}}{SS_{\text{tot}}},
\end{displaymath}
where a value close to 0 indicates good fitness. Here
\begin{displaymath}
  SS_{\text{tot}} = \sum_{i} (y_{i} - \bar{y})^{2}
\end{displaymath}
is the \emph{total sum of squares} and
\begin{displaymath}
  SS_{\text{err}} = \sum_{i} (y_{i} - \hat{y}_{i})^{2}
\end{displaymath}
is the \emph{residual sum of squares}, where $y_{i},\ \hat{y}_{i}$ denote the observed and predicted values of the response variable $y$, respectively, and $\bar{y}$ denotes the mean of the observed data $y_{i}$.

To apply the above line fitting procedure to our numerical data, we need the time derivative of the maximum vorticity, $\frac{d}{dt} \norm{\omega}_{\infty}$. For the specific initial data \eqref{eqn_eat_ic}, the maximum vorticity is always attained at
the corner $\tilde{q}_{0} = (1,0)^{T}$. Due to the special symmetry properties of the solution (Section \ref{sec_eqn}) and the no-flow boundary condition \eqref{eqn_eat_bc_r}, the vorticity vector $\omega$ at $\tilde{q}_{0}$ has a particularly simple
form:
\begin{subequations}\label{eqn_vort_dyn0}
\begin{equation}
  \omega(\tilde{q}_{0}) = (-ru_{1,z}, r\omega_{1}, 2u_{1}+ru_{1,r})^{T}|_{\tilde{q}_{0}} = (-u_{1,z}(\tilde{q}_{0}), 0, 0)^{T}.
  \label{eqn_vort_dyn0_v}
\end{equation}
Consequently, the time derivative and the log $t$-derivative of the maximum vorticity can be readily evaluated:
\begin{equation}
  \frac{d}{dt} \norm{\omega(\cdot,t)}_{\infty} = \frac{d}{dt} \abs{\tilde{u}_{1,z}} = -\tilde{\psi}_{1,rz} \abs{\tilde{u}_{1,z}},\qquad \frac{d}{dt} \log \norm{\omega(\cdot,t)}_{\infty}^{-1} = \tilde{\psi}_{1,rz},
  \label{eqn_vort_dyn0_a}
\end{equation}
\end{subequations}
where for simplicity we have written $\tilde{u}_{1,z} = u_{1,z}(\tilde{q}_{0})$ and $\tilde{\psi}_{1,rz} = \psi_{1,rz}(\tilde{q}_{0})$.

Once an estimate $\hat{t}_{s}$ of the singularity time $t_{s}$ is obtained, the scaling parameter $c$ in \eqref{eqn_vfit} can be determined from another linear regression problem:
\begin{equation}
  \tilde{y}(\tilde{t}) := \log \norm{\omega(\cdot,t)}_{\infty} \sim -\gamma \log(\hat{t}_{s}-t) + \log c =: \tilde{a} \tilde{t} + \tilde{b},
  \label{eqn_vfit_c}
\end{equation}
where $\tilde{y}$ is the response variable, $\tilde{t} = \log(\hat{t}_{s}-t)$ is the explanatory variable, and $\tilde{a} = -\gamma,\ \tilde{b} = \log c$ are model parameters. As before, the model parameters in \eqref{eqn_vfit_c} can be estimated from a
standard least-squares procedure, and the fitness of the model can be measured using either the $R^{2}$ or the FVU.

\subsubsection{Determination of $\tau_{1}$ and $\tau_{2}$}\label{ssec2_vfit_tau}
With the above line fitting procedure, we are now ready to describe the algorithm for choosing the fitting interval $[\tau_{1},\tau_{2}]$.

The first step of the algorithm is to determine $\tau_{2}$, which is formally defined to be the first time instant at which the sup-norm relative error of the vorticity vector $\omega$ exceeds a certain threshold $\epsilon_{2}$. Note that this definition
of $\tau_{2}$ needs to be modified on the finest $2048 \times 2048$ mesh because the error of $\omega$ is not available there. In what follows, we shall define the value of $\tau_{2}$ on the $2048 \times 2048$ mesh to be the same as the one computed on
the $1792 \times 1792$ mesh. This is reasonable given that the error computed on the $1792 \times 1792$ mesh is likely an overestimate of the error computed on the $2048 \times 2048$ mesh, as indicated by the resolution study in Section
\ref{ssec2_res_vort} where convergence of $\omega$ under mesh refinement is observed.

Once $\tau_{2}$ is known, the next step of the algorithm is to determine $\tau_{1}$, which is formally defined to be the time instant at which the FVU of the line fitting computed on $[\tau_{1},\tau_{2}]$ attains its minimum. To avoid placing too few or
too many points in $[\tau_{1},\tau_{2}]$, which may lead to line fittings with too much noise or too much bias, we choose $\tau_{1}$ in such a way that $\tau_{1} \leq \tau_{2}-\epsilon_{1}$ for some $\epsilon_{1} > 0$ and the FVU of the line fitting
computed on $[t,\tau_{2}]$, when viewed as a function of $t$, attains a \emph{local} (instead of global) minimum in a neighborhood of $\tau_{1}$.

\subsubsection{Evidence for Finite-Time Blowup}\label{ssec2_vfit_est}
\begin{figure}[h]
  \centering
  \includegraphics[scale=0.42]{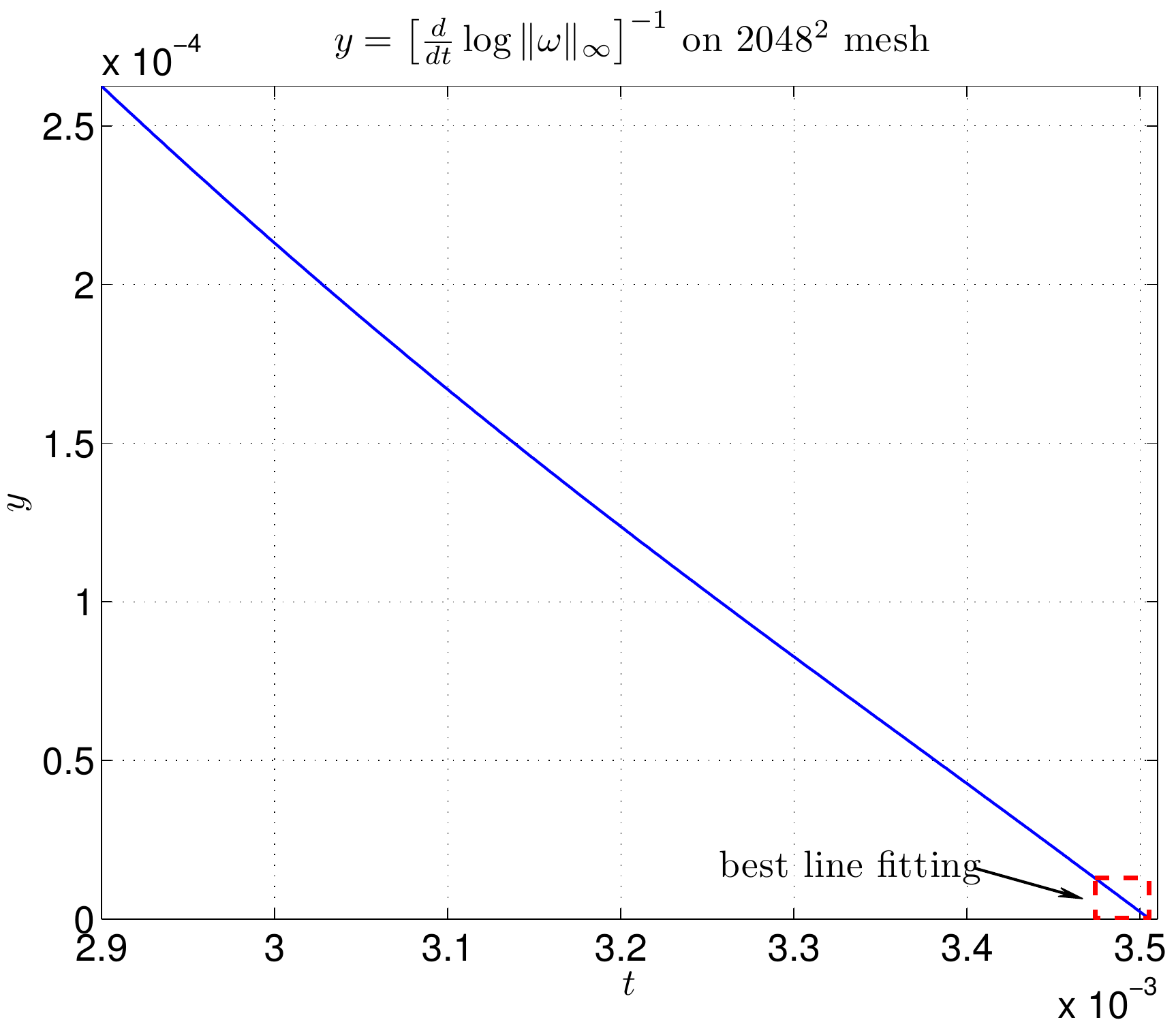}
  \caption{Inverse log $t$-derivative of the maximum vorticity computed on the $2048 \times 2048$ mesh. The dashed line box represents the best fitting interval $[\tau_{1},\tau_{2}]$.}
  \label{fig_vort_log_td}
\end{figure}

We now apply the line fitting procedure described in Section \ref{ssec2_vfit_line}--\ref{ssec2_vfit_tau} to our numerical data to assess the likelihood of a finite-time singularity. As demonstrated earlier in Section \ref{ssec_1st_sign}, the maximum
vorticity $\norm{\omega}_{\infty}$ computed from \eqref{eqn_eat}--\eqref{eqn_eat_ibc} has a growth rate faster than double-exponential (Figure \ref{fig_loglog_vort}). To see whether $\norm{\omega}_{\infty}$ blows up in finite time, we plot in Figure
\ref{fig_vort_log_td} the inverse log $t$-derivative of the maximum vorticity (see \eqref{eqn_vfit_ts})
\begin{displaymath}
  y(t) = \biggl[ \frac{d}{dt} \log \norm{\omega(\cdot,t)}_{\infty} \biggr]^{-1}
\end{displaymath}
computed on the $2048 \times 2048$ mesh. Intuitively, the inverse log $t$-derivative approaches a straight line after $t \approx 0.0032$, which suggests that the maximum vorticity indeed admits an inverse power-law of the form \eqref{eqn_vfit}.
Motivated by this observation, we apply the line fitting to the data $y$ and report the resulting estimates in Table \ref{tab_vfit}. It can be observed from this table that all estimated parameters converge to a finite limit as the mesh is refined, where
in particular both $\tau_{2}$ and $\hat{t}_{s}$ tend to a common limit in a monotonic fashion\footnote{The small discrepancy between the limits of $\tau_{2}$ and $\hat{t}_{s}$ is due to the fact that the sup-norm errors of $\omega$ are computed only at a
\emph{discrete} set of time instants. This restricts the definition of $\tau_{2}$ to a discrete set of values.}. Note also that the limit of $\tau_{1}$ is strictly less than the common limit of $\tau_{2}$ and $\hat{t}_{s}$, indicating the existence of an
asymptotic regime. In addition, both estimates $\hat{\gamma}_{1},\ \hat{\gamma}_{2}$ of $\gamma$ (computed from \eqref{eqn_vfit_ts} and \eqref{eqn_vfit_c} respectively) approach a common limit with a value close to $\frac{5}{2} \geq 1$, and the limit of
$\hat{c}$ is strictly positive. Based on these observations and the BKM criterion, we conclude that the solution of problem \eqref{eqn_eat}--\eqref{eqn_eat_ibc} develops a singularity at $t_{s} \approx 0.0035056$.
\begin{table}[h]
  \centering
  \caption{The best line fittings \eqref{eqn_vfit_ts} and \eqref{eqn_vfit_c} computed on the interval $[\tau_{1},\tau_{2}]$ with $n$ data points.}
  \label{tab_vfit}
  \begin{tabular}{*{8}{>{$}c<{$}}}
    \toprule
    \text{Mesh size} & n & \tau_{1} & \tau_{2} & \hat{t}_{s} & \hat{\gamma}_{1} & \hat{\gamma}_{2} & \hat{c} \\
    \midrule
    1024 \times 1024 & 58 & 0.003306 & 0.003410 & 0.0035070 & 2.5041 & 2.5062 & 4.8293 \times 10^{-4} \\
    1280 \times 1280 & 47 & 0.003407 & 0.003453 & 0.0035063 & 2.4866 & 2.4894 & 5.5362 \times 10^{-4} \\
    1536 \times 1536 & 20 & 0.003486 & 0.003505 & 0.0035056 & 2.4544 & 2.4559 & 7.4912 \times 10^{-4} \\
    1792 \times 1792 & 27 & 0.003479 & 0.003505 & 0.0035056 & 2.4557 & 2.4566 & 7.4333 \times 10^{-4} \\
    2048 \times 2048 & 32 & 0.003474 & 0.003505 & 0.0035056 & 2.4568 & 2.4579 & 7.3273 \times 10^{-4} \\
    \bottomrule
  \end{tabular}
\end{table}%

It is interesting to compare at this point the two estimates $\hat{\gamma}_{1},\ \hat{\gamma}_{2}$ of the scaling exponent $\gamma$ computed from the line fitting problems \eqref{eqn_vfit_ts} and \eqref{eqn_vfit_c}. As can be observed from Table
\ref{tab_vfit}, the estimate $\hat{\gamma}_{2}$ computed from \eqref{eqn_vfit_c} is always slightly larger than the one $\hat{\gamma}_{1}$ computed from \eqref{eqn_vfit_ts}. This is expected, because the singularity time $\hat{t}_s$ estimated from
\eqref{eqn_vfit_ts} decreases monotonically as the mesh is refined, indicating that $\hat{t}_{s}$ is always an \emph{overestimate} of the true singularity time $t_{s}$. Consequently, the inverse power-law $(\hat{t}_{s}-t)^{-\gamma}$ necessarily
underestimates the maximum vorticity $\norm{\omega}_{\infty} \sim (t_{s}-t)^{-\gamma}$ when $t$ is sufficiently close to $t_{s}$, and the scaling exponent $\hat{\gamma}_{2}$ estimated from \eqref{eqn_vfit_c} has to be artificially magnified to compensate
for this discrepancy. This explains the larger value of $\hat{\gamma}_{2}$ compared with $\hat{\gamma}_{1}$.

The computation of $\hat{\gamma}_{1}$ from \eqref{eqn_vfit_ts}, on the other hand, does not suffer from this problem and is expected to yield a more accurate result. Thus in what follows we shall always choose $\hat{\gamma}_{1}$ as the estimate of
$\gamma$.

\begin{figure}[h]
  \centering
  \subfigure[line fitting \eqref{eqn_vfit_ts}]{
    \includegraphics[scale=0.415]{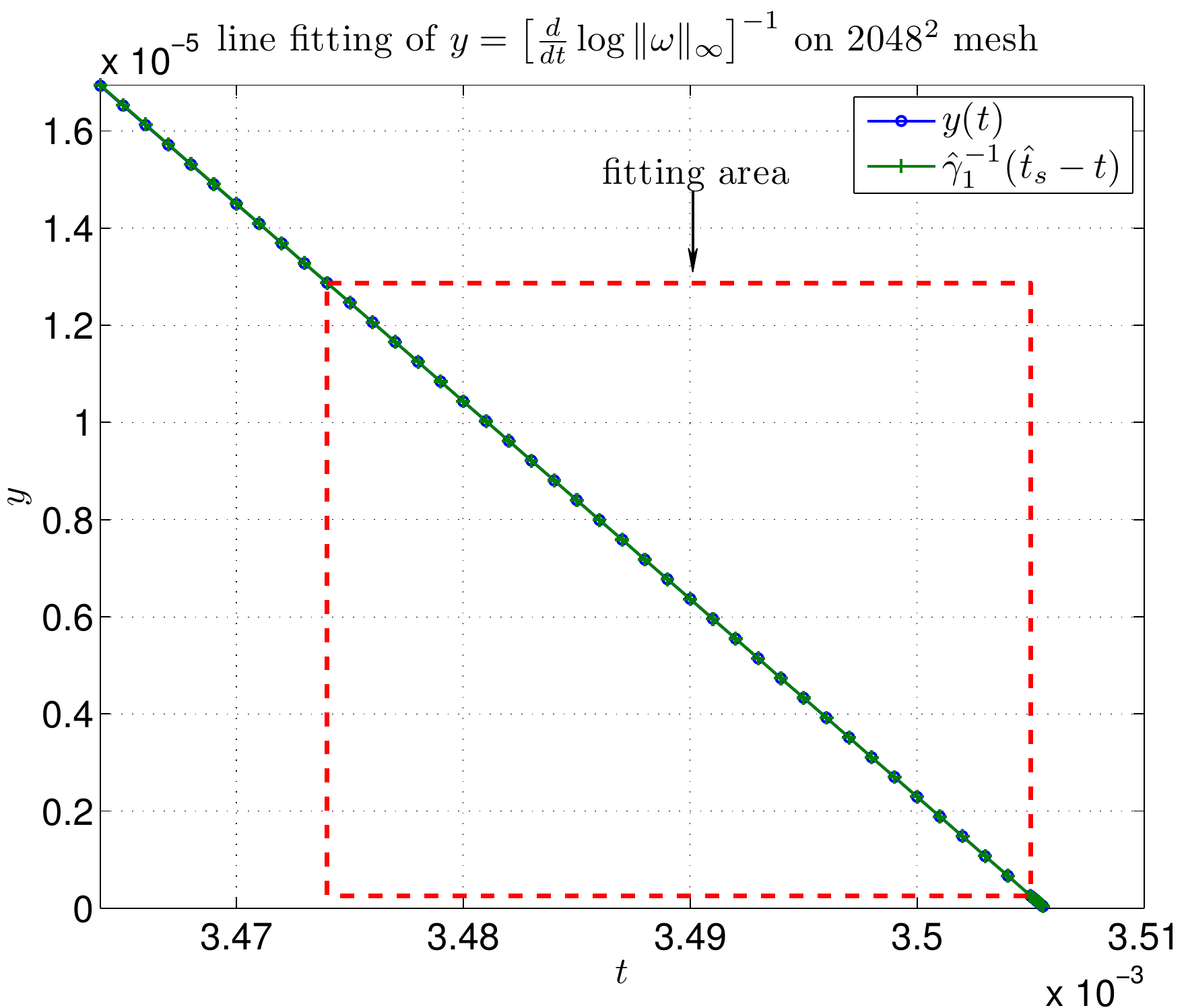}
  }\quad
  \subfigure[line fitting \eqref{eqn_vfit_ts} (zoom-in)]{
    \includegraphics[scale=0.415]{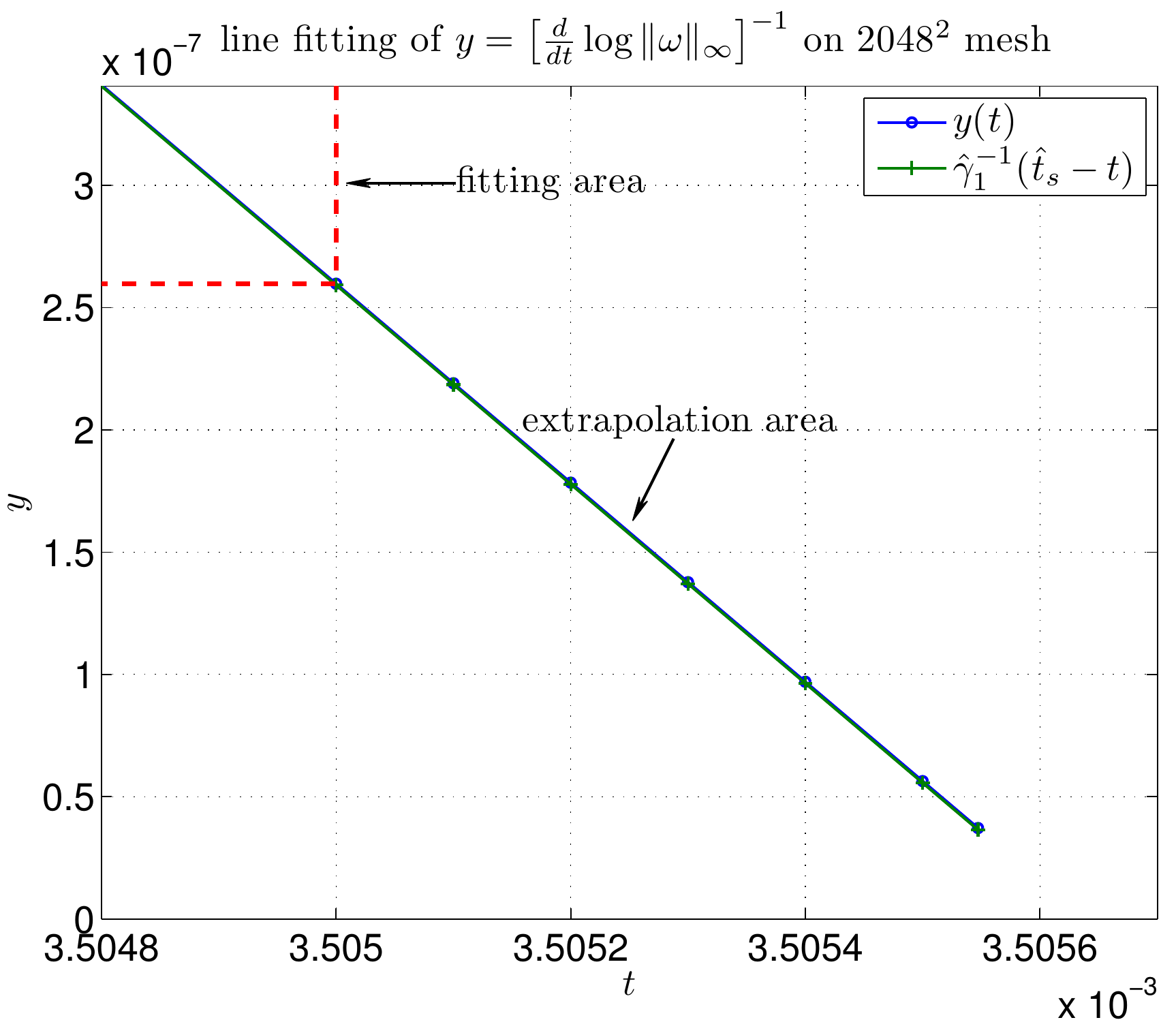}
  }
  \subfigure[line fitting \eqref{eqn_vfit_c}]{
    \includegraphics[scale=0.415]{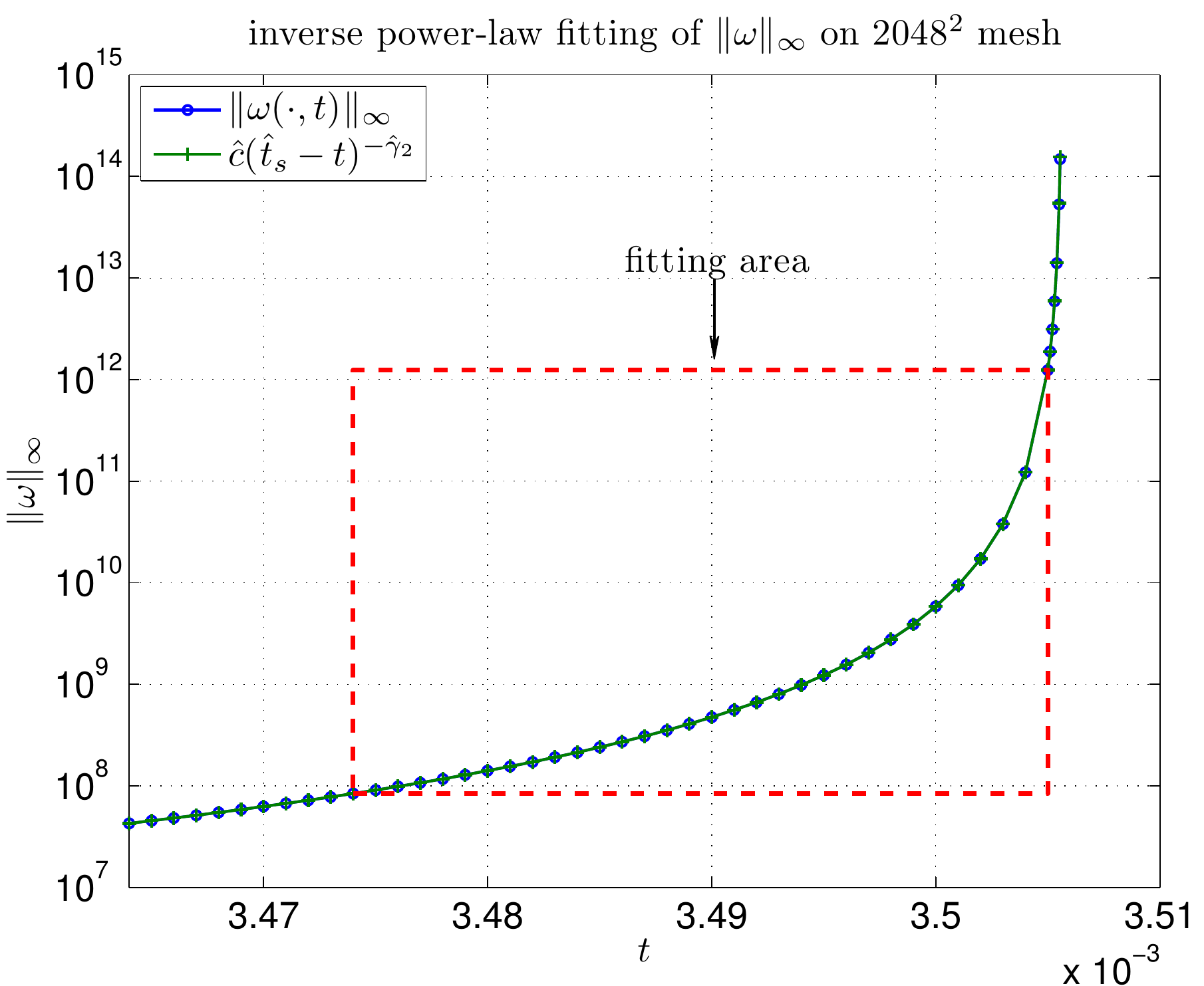}
  }\quad
  \subfigure[line fitting \eqref{eqn_vfit_c} (zoom-in)]{
    \includegraphics[scale=0.415]{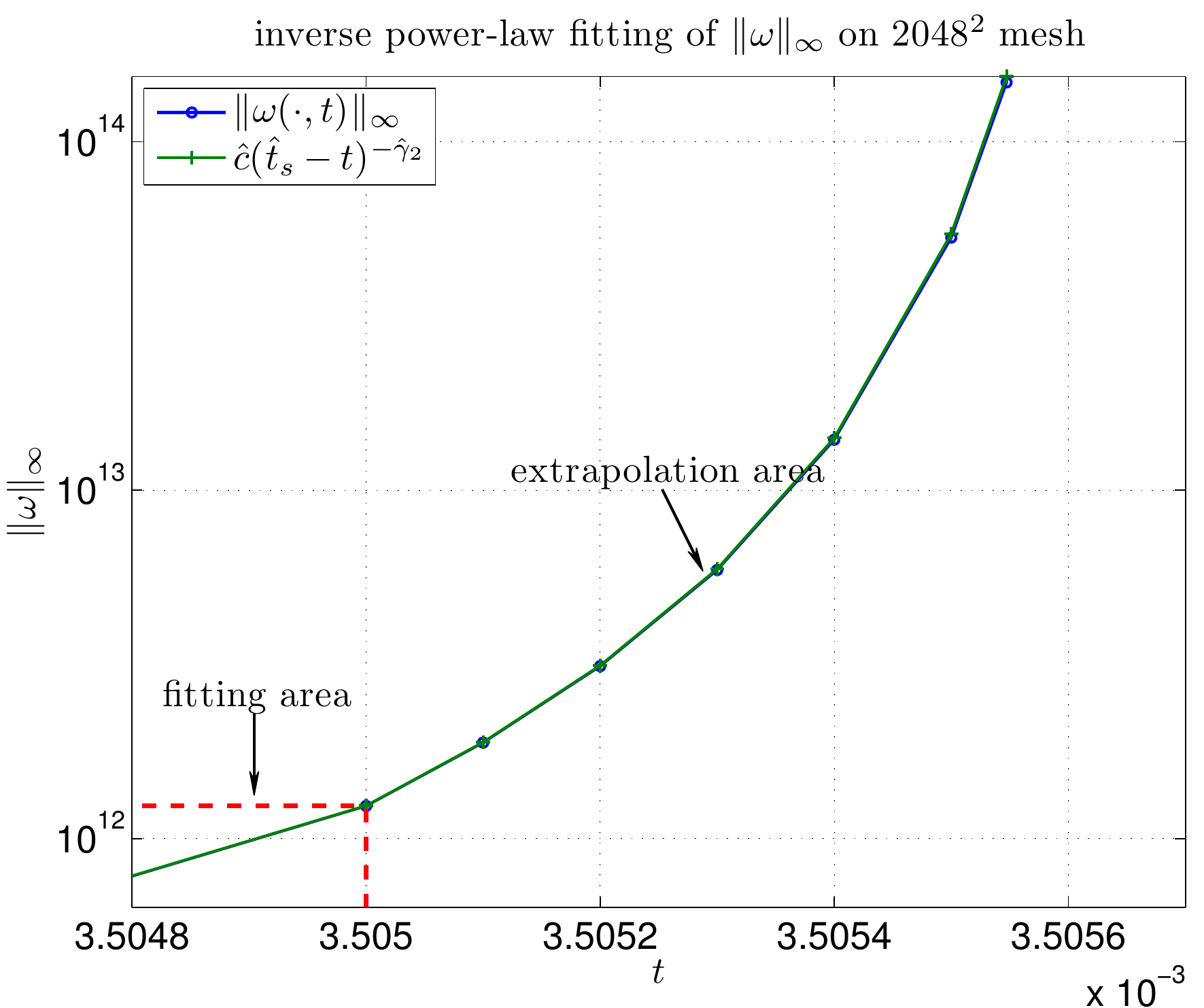}
  }
  \caption{(a) Inverse log $t$-derivative of $\norm{\omega}_{\infty}$ and its line fitting $\hat{\gamma}_{1}^{-1} (\hat{t}_{s}-t)$. (b) A zoom-in view of (a) in the extrapolation interval. (c) Maximum vorticity $\norm{\omega}_{\infty}$ and its inverse
  power-law fitting $\hat{c} (\hat{t}_{s}-t)^{-\hat{\gamma}_{2}}$. (d) A zoom-in view of (c) in the extrapolation interval. All results shown in this figure are computed on the $2048 \times 2048$ mesh.}
  \label{fig_vfit}
\end{figure}

To measure the quality of the line fittings computed in Table \ref{tab_vfit}, we introduce the ``extrapolated FVU'',
\begin{displaymath}
  \text{FVU}_{\text{e}} = \frac{SS_{\text{e,err}}}{SS_{\text{e,tot}}},
\end{displaymath}
where $SS_{\text{e,tot}}$ and $SS_{\text{e,err}}$ are the total sum of squares and residual sum of squares defined on the extrapolation interval $[\tau_{2},t_{e}]$, respectively. These extrapolated FVU, together with the FVU computed on
$[\tau_{1},\tau_{2}]$, are summarized below in Table \ref{tab_vfit_fvu}. We also plot in Figure \ref{fig_vfit} the maximum vorticity $\norm{\omega}_{\infty}$, the inverse log $t$-derivative of $\norm{\omega}_{\infty}$, and their corresponding form fit
computed on the $2048 \times 2048$ mesh.
\begin{table}[h]
  \centering
  \caption{The FVU and FVU$_{\text{e}}$ of the line fitting \eqref{eqn_vfit_ts} and \eqref{eqn_vfit_c}.}
  \label{tab_vfit_fvu}
  \begin{tabular}{*{5}{>{$}c<{$}}}
    \toprule
    \text{Mesh size} & \text{FVU of \eqref{eqn_vfit_ts}} & \text{FVU$_{\text{e}}$ of \eqref{eqn_vfit_ts}} & \text{FVU of \eqref{eqn_vfit_c}} & \text{FVU$_{\text{e}}$ of \eqref{eqn_vfit_c}} \\
    \midrule
    1024 \times 1024 & 8.7255 \times 10^{-7} & 6.1426 \times 10^{-4} & 1.9901 \times 10^{-8} & 1.0657 \times 10^{-1} \\
    1280 \times 1280 & 3.3648 \times 10^{-6} & 6.2433 \times 10^{-4} & 3.0463 \times 10^{-8} & 7.9442 \times 10^{-2} \\
    1536 \times 1536 & 2.4372 \times 10^{-7} & 6.0014 \times 10^{-4} & 4.1369 \times 10^{-7} & 1.0409 \times 10^{-3} \\
    1792 \times 1792 & 1.0127 \times 10^{-7} & 4.5958 \times 10^{-4} & 2.4588 \times 10^{-7} & 8.0410 \times 10^{-4} \\
    2048 \times 2048 & 9.3767 \times 10^{-8} & 1.0956 \times 10^{-4} & 2.8074 \times 10^{-8} & 1.6966 \times 10^{-4} \\
    \bottomrule
  \end{tabular}
\end{table}%
It can be observed from these results that both linear models \eqref{eqn_vfit_ts} and \eqref{eqn_vfit_c} fit the data very well, as clearly indicated by the very small values of FVU. In addition, the line fittings provide an excellent approximation to
the data even in the extrapolation interval, as the small values of FVU$_{\text{e}}$ show. Based on these observations, we conclude that the estimates obtained in Table \ref{tab_vfit} are trustworthy.

\subsubsection{A Comparison}\label{ssec2_vfit_cmp}
We conclude this section with a brief comparison of our results with other representative numerical studies (Table \ref{tab_vfit_cmp}). As is clear from the table, our computation offers a much higher effective resolution and advances the solution to a
point that is asymptotically close to the predicted singularity time. It also produces a much stronger vorticity amplification. In short, our computation gives much more convincing evidence for the existence of a finite-time singularity compared with
other simulations.
\begin{table}[h]
  \centering
  \caption{Comparison of our results with other representative numerical studies. K: \citet{kerr1993}; BP: \citet{bp1994}; GMG: \citet{gmg1998}; OC: \citet{oc2007}; $\tau_{2}$: the last time at which the solution is deemed ``well resolved''.}
  \label{tab_vfit_cmp}
  \begin{tabular}{*{5}{>{$}c<{$}}}
    \toprule
    \text{Studies} & \tau_{2} & t_{s} & \text{Effec. res.} & \text{Vort. amp.} \\
    \midrule
    \text{K} & 17 & 18.7 & \leq 512^{3} & 23 \\
    \text{BP} & 1.6^{\dag} & 2.06 & 1024^{3} & 180 \\
    \text{GMG} & 1.32 & 1.355 & 2048^{3} & 21 \\
    \text{OC} & 2.72 & 2.75 & 1024^{3} & 55 \\
    \text{Ours} & 0.003505 & 0.0035056 & (3 \times 10^{12})^{2} & 3 \times 10^{8} \\
    \bottomrule
    \multicolumn{5}{l}{\footnotesize{$\dag$: According to \citet{hl2008}.}}
  \end{tabular}
\end{table}

\subsection{Asymptotic Scaling Analysis II: Vorticity Moments}\label{ssec_efit}
Given the existence of a finite-time singularity as indicated by the blowing-up maximum vorticity $\norm{\omega}_{\infty}$, we next turn to the interesting question whether the \emph{vorticity moment} integrals
\begin{displaymath}
  \Omega_{2m} = \biggl[ \int_{D(1,L)} \abs{\omega}^{2m}\,dx \biggr]^{1/2m},\qquad m = 1,2,\dotsc,
\end{displaymath}
blow up at the same time as $\norm{\omega}_{\infty}$ does, and if yes, what type of asymptotic scalings they satisfy. According to H\"{o}lder's inequality, higher vorticity moments ``control'' the growth of lower vorticity moments, in the sense that
\begin{displaymath}
  \Omega_{2m} \leq \Omega_{2n} \abs{D(1,L)}^{(n-m)/(2mn)},\qquad 1 \leq m < n.
\end{displaymath}
Thus the blowup of any vorticity moment $\Omega_{2m}$ implies the blowup of all higher moments $\Omega_{2n}\ (n \geq m)$. In particular, since $\norm{\omega}_{\infty} = \Omega_{\infty}$, the blowup of any finite-order vorticity moment provides additional
supporting evidence for the existence of a finite-time singularity.

We have carried out a detailed analysis of the vorticity moments and discovered that all moments of order higher than 2 blow up at a finite time. For the purpose of illustration, we report in Table \ref{tab_vmfit} the singularity time $\hat{t}_{2m,s}$
and the scaling exponent $\hat{\gamma}_{2m,1}$ estimated from the line fitting
\begin{equation}
  y(t) := \biggl[ \frac{d}{dt} \log \Omega_{2m} \biggr]^{-1} \sim -\frac{1}{\gamma_{2m}} (t-t_{s}) =: at + b,
  \label{eqn_vmfit_ts}
\end{equation}
for $m = 2,3,4$, where $\Omega_{2m}$ is assumed to satisfy the scaling law $\Omega_{2m}(t) \sim c_{2m} (t_{s}-t)^{-\gamma_{2m}}$.
\begin{table}[h]
  \centering
  \caption{The line fitting \eqref{eqn_vmfit_ts} of the $2m$-th vorticity moment $\Omega_{2m},\ m = 2,3,4$, computed on the interval $[\tau_{1},\tau_{2}]$. For comparison, the singularity time $\hat{t}_{s}$ estimated from \eqref{eqn_vfit_ts} is also
  included.}
  \label{tab_vmfit}
  \begin{tabular}{*{8}{>{$}c<{$}}}
    \toprule
     &  & \multicolumn{3}{c}{$\hat{t}_{2m,s}$} & \multicolumn{3}{c}{$\hat{\gamma}_{2m,1}$} \\
    \cmidrule{3-8}
    \rbs{1.75ex}{Mesh size} & \rbs{1.75ex}{$\hat{t}_{s}$ from \eqref{eqn_vfit_ts}} & m = 2 & m = 3 & m = 4 & m = 2 & m = 3 & m = 4 \\
    \midrule
    1024 \times 1024 & 0.0035070 & 0.0035231 & 0.0035124 & 0.0035097 & 1.2542 & 1.6129 & 1.8176 \\
    1280 \times 1280 & 0.0035063 & 0.0035115 & 0.0035074 & 0.0035067 & 1.1306 & 1.5383 & 1.7658 \\
    1536 \times 1536 & 0.0035056 & 0.0035056 & 0.0035056 & 0.0035056 & 1.0019 & 1.4857 & 1.7289 \\
    1792 \times 1792 & 0.0035056 & 0.0035057 & 0.0035056 & 0.0035056 & 1.0039 & 1.4855 & 1.7285 \\
    2048 \times 2048 & 0.0035056 & 0.0035057 & 0.0035056 & 0.0035056 & 1.0062 & 1.4857 & 1.7285 \\
    \bottomrule
  \end{tabular}
\end{table}%
It can be observed from this table that all $\Omega_{2m}$ with $m > 1$ satisfy an inverse power-law with an exponent monotonically approaching $\hat{\gamma} \approx \frac{5}{2}$, and they all blow up at a finite time $\hat{t}_{2m,s}$ approximately equal
to the singularity time $\hat{t}_{s}$ estimated from \eqref{eqn_vfit_ts}. This confirms the blowup of $\norm{\omega}_{\infty}$ at the predicted singularity time $t_{s}$ and hence the existence of a finite-time singularity.

\subsection{Vorticity Directions and Spectral Dynamics}\label{ssec_geosd}
The BKM criterion characterizes the finite-time blowup of the 3D Euler equations in terms of the sup-norm of the vorticity \emph{magnitude} $\abs{\omega}$ but makes no assumption on the vorticity \emph{direction} $\xi = \omega/\abs{\omega}$. When less
regularity is required on the vorticity magnitude, say boundedness in $L^{p}$ ($p < \infty$) instead of boundedness in $L^{\infty}$, the regularity of the vorticity direction can also play a role in controlling the blowup of the Euler solutions
\citep{constantin1994}. To see more precisely how the direction vector $\xi$ enters the analysis, recall the vorticity amplification equation
\begin{subequations}\label{eqn_vort_dyn}
\begin{equation}
  \abs{\omega}_{t} + u \cdot \nabla \abs{\omega} = \alpha \abs{\omega},
  \label{eqn_vort_dyn_v}
\end{equation}
where $\alpha$ is the vorticity amplification factor:
\begin{equation}
  \alpha = \xi \cdot \nabla u \cdot \xi = \xi \cdot S \xi,\qquad S = \frac{1}{2} \bigl( \nabla u + \nabla u^{T} \bigr).
  \label{eqn_vort_dyn_a}
\end{equation}
It can be shown that \citep{constantin1994}
\begin{equation}
  \alpha(x) = \frac{3}{4\pi} \PV\int_{\mathbb{R}^{3}} D(\hat{y},\xi(x+y),\xi(x)) \abs{\omega(x+y)}\,\frac{dy}{\abs{y}^{3}},
  \label{eqn_vort_dyn_ai}
\end{equation}
\end{subequations}
where $\hat{y} = y/\abs{y}$ is the unit vector pointing in the direction of $y$ and
\begin{displaymath}
  D(e_{1},e_{2},e_{3}) = (e_{1} \cdot e_{3}) \det(e_{1},e_{2},e_{3}).
\end{displaymath}
Note that the quantity $D(e_{1},e_{2},e_{3})$ is small when $e_{2}$ and $e_{3}$ are nearly aligned or anti-aligned, so a smoothly-varying vorticity direction field $\xi$ near a spatial point $x$ can induce strong cancellation in the vorticity
amplification $\alpha(x)$, thus preventing the vorticity $\abs{\omega(x)}$ at $x$ from growing unboundedly. The most well-known (non)blowup criteria in this direction are those of Constantin-Fefferman-Majda \citep{cfm1996} and Deng-Hou-Yu
\citep{dhy2005}. Under the assumption that the vorticity direction $\xi$ is ``not too twisted'' near the location of the maximum vorticity, they show that a suitable upper bound can be obtained for $\alpha$ and hence for $\norm{\omega}_{\infty}$,
establishing the regularity of the solutions to the 3D Euler equations.

The non-blowup criteria of Constantin-Fefferman-Majda (CFM) and Deng-Hou-Yu (DHY) are useful for excluding false blowup candidates, but cannot be used directly to verify a finite-time singularity. The reason is that these criteria provide only upper
bounds for the amplification factor $\alpha$ while a blowup estimate requires a lower bound. Nevertheless, a careful examination of our numerical data against these criteria provides additional evidence for a finite-time singularity. It also offers
additional insights into the nature of the blowup.

In what follows, we shall state the non-blowup criteria of CFM and DHY and apply them to our numerical data (Section \ref{ssec2_geosd_cfm}--\ref{ssec2_geosd_geo}). We shall also investigate the vorticity amplification factor $\alpha$ directly at the
location of the maximum vorticity and establish a connection between $\alpha$ and the eigenstructure of the symmetric strain tensor $S$ (Section \ref{ssec2_geosd_sd}). Before proceeding, however, we shall point out that the representation formula
\eqref{eqn_vort_dyn_ai} for the vorticity amplification factor $\alpha$ is valid only in free space $\mathbb{R}^{3}$ and does not hold true for periodic-axisymmetric flows bounded by solid walls. In principle, formulas similar to \eqref{eqn_vort_dyn_ai}
can be derived in bounded and/or periodic domains; for example, in our case the vorticity amplification equation at the location of the maximum vorticity can be shown to take the form (see \eqref{eqn_vort_dyn0_a})
\begin{displaymath}
  \frac{d}{dt} \norm{\omega(\cdot,t)}_{\infty} = -\tilde{\psi}_{1,rz} \norm{\omega(\cdot,t)}_{\infty},
\end{displaymath}
where
\begin{equation}
  \tilde{\psi}_{1,rz} = \psi_{1,rz}(\tilde{q}_{0}) = \frac{1}{L} \int_{0}^{1} r^{3} \int_{0}^{L} \omega_{1}(r,z) G_{1,z}(r,z)\,dz\,dr,
  \label{eqn_vort_dyn_aa}
\end{equation}
and $G_{1}$ is certain ``fundamental solution'' of the five-dimensional Laplace operator. On the other hand, these representation formulas are often considerably more complicated than \eqref{eqn_vort_dyn_ai}, and in the presence of axial symmetry they
may even obscure the connection between the vorticity amplification factor $\alpha$ and the vorticity direction $\xi$, as the formula \eqref{eqn_vort_dyn_aa} shows. Hence, instead of deriving and using a formula of the form \eqref{eqn_vort_dyn_aa}, we
shall apply in what follows the elegant formula \eqref{eqn_vort_dyn_ai} directly to our numerical data. Although the analysis that results is not strictly rigorous, it reveals more clearly the role played by the vorticity direction $\xi$, hence leading
to a better understanding of the interplay between the geometry of $\xi$ and the dynamics of the vorticity amplification $\alpha$.

\subsubsection{The Constantin-Fefferman-Majda Criterion}\label{ssec2_geosd_cfm}
The CFM criterion consists of two parts. To state the results, we first recall the notion of \emph{smoothly directed} and \emph{regularly directed} sets.

Let $u = u(x,t)$ be the velocity field for the 3D incompressible Euler equations \eqref{eqn_eu} and $X(q,t)$ be the corresponding flow map, defined by
\begin{displaymath}
  \frac{dX}{dt} = u(X,t),\qquad X(q,0) = q.
\end{displaymath}
Denote by $W_{t} = X(W_{0},t)$ the image of a set $W_{0}$ at time $t$ and by $B_{r}(W)$ the neighborhood of $W$ formed with points situated at Euclidean distance not larger than $r$ from $W$. A set $W_{0}$ is said to be \emph{smoothly directed} if there
exists $\rho > 0$ and $r \in (0,\frac{1}{2} \rho]$ such that the following three conditions are satisfied: first, for every $q \in W_{0}^{*}$ where
\begin{displaymath}
  W_{0}^{*} = \Bigl\{ q \in W_{0}\colon \abs{\omega_{0}(q)} \neq 0 \Bigr\}
\end{displaymath}
and for all $t \in [0,T)$, the vorticity direction $\xi = \omega/\abs{\omega}$ has a Lipschitz extension to the Euclidean ball of radius $4\rho$ centered at $X(q,t)$ and
\begin{subequations}\label{eqn_cfm_sds}
\begin{equation}
  M = \lim_{t \to T} \sup_{q \in W_{0}^{*}} \int_{0}^{t} \norm{\nabla \xi(\cdot,\tau)}_{L^{\infty}(B_{4\rho}(X(q,\tau)))}^{2}\,d\tau < \infty;
  \label{eqn_cfm_sds_xi}
\end{equation}
second,
\begin{equation}
  \sup_{B_{3r}(W_{t})} \abs{\omega(x,t)} \leq m \sup_{B_{r}(W_{t})} \abs{\omega(x,t)}
  \label{eqn_cfm_sds_b}
\end{equation}
holds for all $t \in [0,T)$ with $m \geq 0$ constant; and finally,
\begin{equation}
  \sup_{B_{4\rho}(W_{t})} \abs{u(x,t)} \leq U
  \label{eqn_cfm_sds_u}
\end{equation}
\end{subequations}
holds for all $t \in [0,T)$. A set $W_{0}$ is said to be \emph{regularly directed} if there exists $\rho > 0$ such that
\begin{subequations}\label{eqn_cfm_rds}
\begin{equation}
  \sup_{q \in W_{0}^{*}} \int_{0}^{T} K_{\rho}(X(q,t),t)\,dt < \infty,
  \label{eqn_cfm_rds_i}
\end{equation}
where
\begin{equation}
  K_{\rho}(x,t) = \int_{\abs{y} \leq \rho} \abs{D(\hat{y},\xi(x+y,t),\xi(x,t))} \cdot \abs{\omega(x+y,t)}\,\frac{dy}{\abs{y}^{3}}
  \label{eqn_cfm_rds_k}
\end{equation}
and
\begin{equation}
  D(\hat{y},\xi(x+y),\xi(x)) = (\hat{y} \cdot \xi(x)) \det(\hat{y},\xi(x+y),\xi(x)).
  \label{eqn_cfm_rds_d}
\end{equation}
\end{subequations}

The CFM criterion asserts that \citep{cfm1996}
\begin{theorem}
Assume $W_{0}$ is smoothly directed. Then there exists $\tau > 0$ and $\Gamma > 0$ such that
\begin{displaymath}
  \sup_{B_{r}(W_{t})} \abs{\omega(x,t)} \leq \Gamma \sup_{B_{\rho}(W_{t_{0}})} \abs{\omega(x,t_{0})}
\end{displaymath}
holds for any $0 \leq t_{0} < T$ and $0 \leq t-t_{0} \leq \tau$.
\label{thm_cfm_1}
\end{theorem}
\begin{theorem}
Assume $W_{0}$ is regularly directed. Then there exists $\Gamma > 0$ such that
\begin{displaymath}
  \sup_{q \in W_{0}} \abs{\omega(X(q,t),t)} \leq \Gamma \sup_{q \in W_{0}} \abs{\omega_{0}(q)}
\end{displaymath}
holds for all $t \in [0,T]$.
\label{thm_cfm_2}
\end{theorem}

Both Theorems  \ref{thm_cfm_1} and \ref{thm_cfm_2} can be reformulated in cylindrical coordinates. To fix the notations in the rest of this section, we shall denote by $x = (x_{1},x_{2},x_{3})^{T}$ a point in $\mathbb{R}^{3}$ and by $\tilde{x} =
(r,x_{3})^{T}$ its projection onto the $rz$-plane, where $r = \sqrt{x_{1}^{2}+x_{2}^{2}}$. For any radially symmetric function $f$, we shall write $f(x)$ and $f(\tilde{x})$ interchangeably depending on the context. The notation $B_{\rho}(q)$ can denote a
3D Euclidean ball if its center $q$ is a point in $\mathbb{R}^{3}$, or a 2D Euclidean ball if $q$ is a point in the 2D $rz$-plane.

To check our numerical data against the CFM criterion, we define, for each fixed time instant $t$, the neighborhood of the maximum vorticity:
\begin{equation}
  D_{\infty}(t) = \Bigl\{ (r,z) \in D(1,\tfrac{1}{4} L)\colon \abs{\omega(r,z,t)} \geq \tfrac{1}{2} \norm{\omega(\cdot,t)}_{\infty} \Bigr\}.
  \label{eqn_d_inf}
\end{equation}
As will be demonstrated below in Section \ref{ssec_selfsim}, the diameter of $D_{\infty}(t)$ shrinks rapidly to 0 as the predicted singularity time $t_{s}$ is approached (see Figure \ref{fig_ssim_contv}\subref{fig_ssim_contv_lin}). Since the maximum
vorticity is always attained at $\tilde{q}_{0} = (1,0)^{T}$, i.e. $\tilde{q}_{0} \in D_{\infty}(t)$ for all $t$, it follows that
\begin{displaymath}
  D_{\infty}(t) \subseteq B_{\delta}(\tilde{q}_{0}) := \Bigl\{ (r,z)\colon (r-1)^{2} + z^{2} < \delta^{2} \Bigr\},
\end{displaymath}
for any fixed $\delta > 0$ provided that $t$ is sufficiently close to $t_{s}$. On the other hand, $\tilde{q}_{0}$ is a stagnation point of the flow field:
\begin{displaymath}
  u^{r}(\tilde{q}_{0}) = -\psi_{1,z}(\tilde{q}_{0}) = 0,\qquad u^{\theta}(\tilde{q}_{0}) = u_{1}(\tilde{q}_{0}) = 0,\qquad u^{z}(\tilde{q}_{0}) = 2\psi_{1}(\tilde{q}_{0}) + \psi_{1,r}(\tilde{q}_{0}) = 0,
\end{displaymath}
in view of the no-flow boundary condition $\psi_{1}(1,z) = 0$ (see \eqref{eqn_eat_bc_r}) and the odd symmetry of $u_{1},\ \psi_{1}$ at $z = 0$ (see Section \ref{sec_eqn}). This means that
\begin{displaymath}
  X(q_{0},t) \equiv q_{0},\qquad q_{0} = (1,0,0)^{T},\ \forall t > 0,
\end{displaymath}
and thus for any fixed $\rho > 0$ and $t$ sufficiently close to $t_{s}$, the projection of the 3D Euclidean ball $B_{4\rho}(X(q_{0},t)) \equiv B_{4\rho}(q_{0})$ onto the $rz$-plane will always contain the set $D_{\infty}(t)$.

\begin{figure}[h]
  \centering
  \includegraphics[scale=0.415]{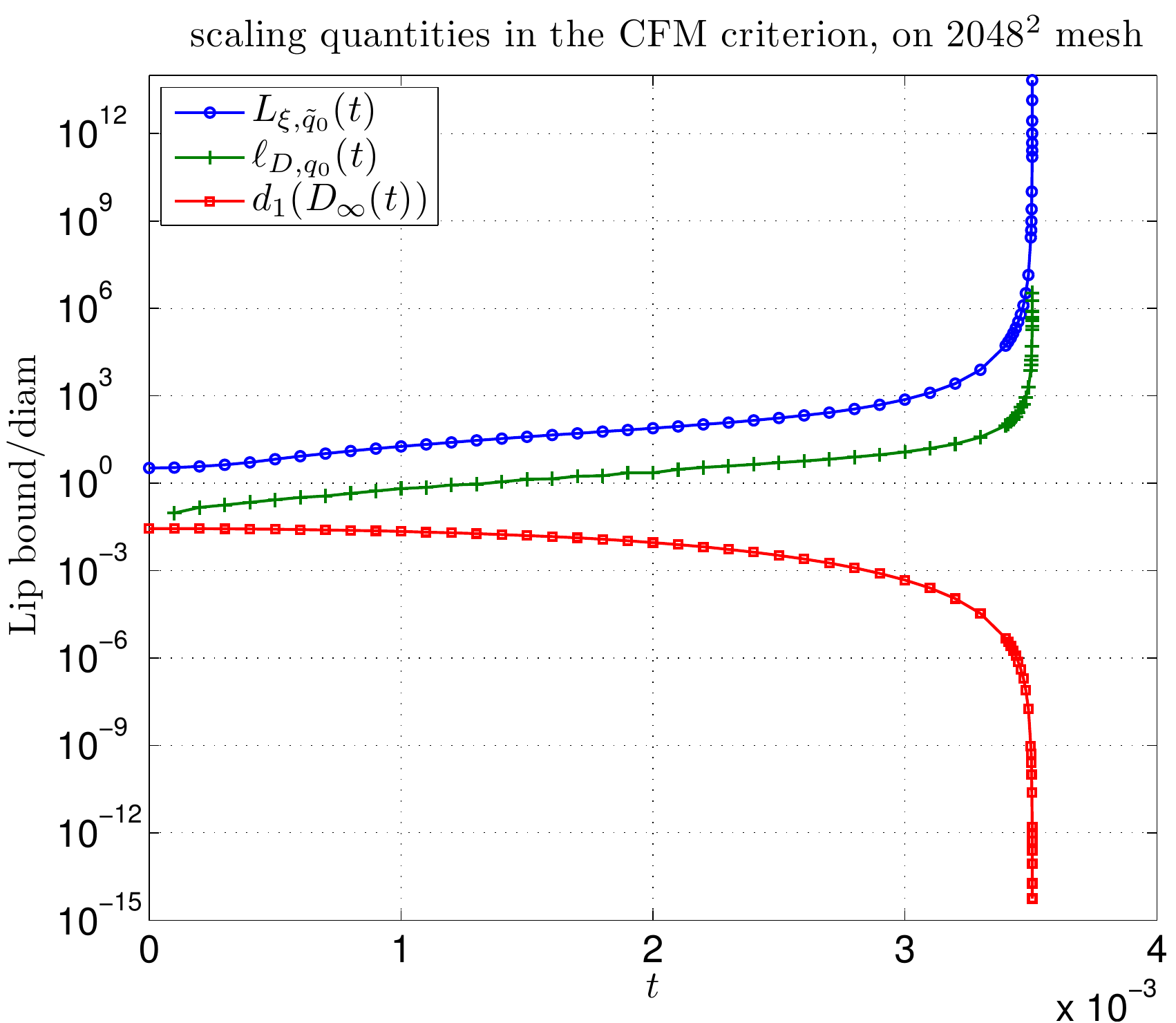}
  \caption{The local Lipschitz constants $L_{\xi,\tilde{q}_{0}},\ \ell_{D,q_{0}}$ and the length scale $d_{1}$ considered in the CFM criterion.}
  \label{fig_vdir_lip_bds}
\end{figure}

We are now ready to show that Theorem \ref{thm_cfm_1}, when applied to our numerical data, does not exclude the possibility of a finite-time singularity. More specifically, we shall show that the condition \eqref{eqn_cfm_sds_xi} that is required to
define a smoothly directed set is not met by our numerical data. For this purpose, we take
\begin{displaymath}
  W_{0} = W_{0}^{*} = \Bigl\{ (x_{1},x_{2},x_{3}) \in \mathbb{R}^{3}\colon ({\textstyle \sqrt{x_{1}^{2}+x_{2}^{2}}},x_{3}) \in D_{\infty}(0) \Bigr\},
\end{displaymath}
and note that
\begin{align*}
  \sup_{q \in W_{0}^{*}} \int_{0}^{t} \norm{\nabla \xi(\cdot,\tau)}_{L^{\infty}(B_{4\rho}(X(q,\tau)))}^{2}\,d\tau & \geq \int_{t_{0}}^{t} \norm{\nabla \xi(\cdot,\tau)}_{L^{\infty}(B_{4\rho}(q_{0}))}^{2}\,d\tau \\
  & \geq \int_{t_{0}}^{t} \norm{\nabla \xi(\cdot,\tau)}_{L^{\infty}(D_{\infty}(\tau))}^{2}\,d\tau,
\end{align*}
for any $t_{0} \in (0,t_{s})$ sufficiently close to $t_{s}$ and any $t \in (t_{0},t_{s})$. This shows that, with $T = t_{s}$,
\begin{displaymath}
  M = \lim_{t \to t_{s}} \sup_{q \in W_{0}^{*}} \int_{0}^{t} \norm{\nabla \xi(\cdot,\tau)}_{L^{\infty}(B_{4\rho}(X(q,\tau)))}^{2}\,d\tau \geq \lim_{t \to t_{s}} \int_{t_{0}}^{t} \norm{\nabla \xi(\cdot,\tau)}_{L^{\infty}(D_{\infty}(\tau))}^{2}\,d\tau.
\end{displaymath}
To obtain a lower bound for the above integral, we consider the quantity
\begin{displaymath}
  L_{\xi,\tilde{q}_{0}}(t) = \sup_{\tilde{y} \in D_{\infty}(t)} \frac{\abs{\xi(\tilde{y},t)-\xi(\tilde{q}_{0},t)}}{\abs{\tilde{y}-\tilde{q}_{0}}},
\end{displaymath}
which defines the (local) Lipschitz constant of the vorticity direction $\xi$ at $\tilde{q}_{0}$ and which gives a lower bound of $\norm{\nabla \xi}_{L^{\infty}(D_{\infty}(\tau))}$ in view of the standard estimate
\begin{displaymath}
  \abs{\xi(\tilde{y},t)-\xi(\tilde{q}_{0},t)} \leq \int_{0}^{1} \abs{\nabla \xi(\tilde{q}_{0} + s(\tilde{y}-\tilde{q}_{0}),t)} \cdot \abs{\tilde{y}-\tilde{q}_{0}}\,ds
  \leq \norm{\nabla \xi(\cdot,t)}_{L^{\infty}(D_{\infty}(t))} \abs{\tilde{y}-\tilde{q}_{0}},
\end{displaymath}
(we note that $D_{\infty}(t)$ is convex; see Figure \ref{fig_ssim_contv}\subref{fig_ssim_contv_s}). Since the quantity $L_{\xi,\tilde{q}_{0}}$ estimated from our numerical data grows rapidly with $t$, as is clear from Figure \ref{fig_vdir_lip_bds}, and a
line fitting similar to \eqref{eqn_vfit_c} yields
\begin{displaymath}
  L_{\xi,\tilde{q}_{0}}(t) \sim c (\hat{t}_{s}-t)^{-2.9165},\qquad c = 1.3497 \times 10^{-7},
\end{displaymath}
where $\hat{t}_{s}$ is the singularity time estimated from \eqref{eqn_vfit_ts}, it follows that the time integral of $\norm{\nabla \xi}_{L^{\infty}(D_{\infty}(\tau))}^{2}$ cannot remain bounded as $t$ approaches $t_{s}$. Hence \eqref{eqn_cfm_sds_xi}
cannot be satisfied by our choice of $W_{0}$. Returning to the statement of Theorem \ref{thm_cfm_1}, we see that
\begin{displaymath}
  \sup_{B_{r}(W_{t})} \abs{\omega(x,t)} = \norm{\omega(\cdot,t)}_{\infty},
\end{displaymath}
since $q_{0}$, the location of the maximum vorticity, lies in $W_{t}$ for all $t$. This shows that no \emph{a priori} bound on the maximum vorticity can be inferred from Theorem \ref{thm_cfm_1}.

Similarly, we can argue that Theorem \ref{thm_cfm_2}, when applied to our numerical data, does not exclude the possibility of a finite-time singularity. To see this, we choose $W_{0}$ as above and note that
\begin{displaymath}
  \sup_{q \in W_{0}^{*}} \int_{0}^{T} K_{\rho}(X(q,t),t)\,dt \geq \int_{0}^{T} K_{\rho}(q_{0},t)\,dt,
\end{displaymath}
where
\begin{displaymath}
  K_{\rho}(q_{0},t) = \int_{\abs{y} \leq \rho} \abs{D(\hat{y},\xi(q_{0}+y,t),\xi(q_{0},t))} \cdot \abs{\omega(q_{0}+y,t)}\,\frac{dy}{\abs{y}^{3}}.
\end{displaymath}
The above integral has a lower bound estimate (Appendix \ref{app_k_lbd})
\begin{subequations}\label{eqn_k_lbd}
\begin{equation}
  K_{\rho}(q_{0},t) \geq \frac{3\pi}{640}\, d_{1}(D_{\infty}(t)) \norm{\omega(\cdot,t)}_{\infty} \ell_{D,q_{0}}(t),
  \label{eqn_k_lbd_m}
\end{equation}
where $\ell_{D,q_{0}}$ is the infimum of $\abs{D}/\abs{y}$ over some neighborhood of $q_{0}$ and $d_{1}$ is (roughly) the diameter of $D_{\infty}(t)$. Thus to complete the analysis, it suffices to estimate the quantities $\norm{\omega}_{\infty},\ d_{1}$,
and $\ell_{D,q_{0}}$ from the numerical data. The estimate of $\norm{\omega}_{\infty}$ is derived in Section \ref{ssec2_vfit_est} and has the form
\begin{align*}
  \norm{\omega(\cdot,t)}_{\infty} \sim c_{1} (t_{s}-t)^{-2.4568},&\qquad c_{1} = 7.3273 \times 10^{-4}. \\
  \intertext{As for the other two quantities, it is observed that $\ell_{D,q_{0}}$ grows rapidly with $t$ while $d_{1}$ decays with $t$ (Figure \ref{fig_vdir_lip_bds}). A line fitting similar to \eqref{eqn_vfit_c} then yields}
  \ell_{D,q_{0}}(t) \sim c_{2} (\hat{t}_{s}-t)^{-1.4597},&\qquad c_{2} = 1.7596 \times 10^{-4}, \\
  d_{1}(D_{\infty}(t)) \sim [\delta^{-1} (\hat{t}_{s}-t)]^{2.9181},&\qquad \delta = 7.0214 \times 10^{-3},
\end{align*}
which, together with the estimate of $\norm{\omega}_{\infty}$, shows that
\begin{equation}
  K_{\rho}(q_{0},t) \geq C (t_{s}-t)^{-0.9984}.
  \label{eqn_k_lbd_t}
\end{equation}
\end{subequations}
Taking into account the effect of numerical errors, we may conclude that $K_{\rho}(q_{0},t) \gtrapprox C (t_{s}-t)^{-1}$ and the time integral of $K_{\rho}(q_{0},t)$ diverges as $t$ approaches $t_{s}$. Thus the condition \eqref{eqn_cfm_rds_i} is not
satisfied by our numerical data.

At the first glance, the estimate \eqref{eqn_k_lbd_t} may look a bit surprising because the growth of the maximum vorticity $\norm{\omega}_{\infty}$ is so strong while the blowup of $K_{\rho}(q_{0})$ implied from \eqref{eqn_k_lbd_t} is so marginal.
Still, we believe this is not unreasonable because \eqref{eqn_k_lbd_t} provides only a \emph{lower bound} for $K_{\rho}(q_{0})$ which does not necessarily capture the rapid growth of $K_{\rho}(q_{0})$. More importantly, both $K_{\rho}(q_{0})$ and the
amplification factor $\alpha(q_{0})$ are roughly of the same order when $D(\hat{y},\xi(q_{0}+y),\xi(q_{0}))$ does not change sign in a neighborhood of $y = 0$ (see \eqref{eqn_vort_dyn_ai}). Since $\alpha(q_{0})$ must grow like $(t_{s}-t)^{-1}$ if the
maximum vorticity obeys an inverse power-law, the ``marginal blowup'' of $K_{\rho}(q_{0})$ as indicated by \eqref{eqn_k_lbd_t} may indeed be what is to be expected.

We also emphasize that the above analysis is purely formal since the representation formula \eqref{eqn_cfm_rds_k} for the quantity $K_{\rho}(x)$ is not valid in bounded and/or periodic domains. On the other hand, the analysis suggests, through the key
estimate \eqref{eqn_k_lbd_m}, that the formation of a singularity in the 3D Euler equations is likely a result of the subtle balance among the three competing ``forces'', namely the growth rate of the maximum vorticity $\norm{\omega}_{\infty}$, the
collapsing rate of the support of the vorticity as measured by $d_{1}$, and the smoothness of the vorticity direction field as measured by $\ell_{D,q_{0}}$. This observation is expected to hold true even in bounded and/or periodic domains where
\eqref{eqn_cfm_rds_k} is not valid, and this is where the significance of the above formal analysis lies.

\subsubsection{The Deng-Hou-Yu Criterion}\label{ssec2_geosd_dhy}
The DHY criterion improves the non-blowup criterion of CFM, in particular the part stated in Theorem \ref{thm_cfm_1}, by relaxing the regularity assumptions made on the velocity field $u$ and the vorticity direction $\xi$. Instead of assuming the
integrability of the gradient of $\xi$ in an $O(1)$ region, the DHY criterion requires only the integrability of the \emph{divergence} of $\xi$ along a \emph{vortex line} segment whose length is allowed to shrink to 0 (Theorem \ref{thm_dhy_1}). In
addition, the velocity field $u$ is allowed to grow unboundedly in time, provided that a mild partial regularity condition on $u$ is satisfied along a vortex line (Theorem \ref{thm_dhy_2}). These improvements make the criterion easier to apply in actual
numerical simulations.

Like the CFM criterion, the DHY criterion consists of two parts, the first of which excludes the possibility of a point singularity under certain regularity assumption on the divergence of the vorticity direction $\nabla \cdot \xi$.
\begin{theorem}
Consider the 3D incompressible Euler equations \eqref{eqn_eu} and let $x(t)$ be a family of points such that
\begin{displaymath}
  \abs{\omega(x(t),t)} \geq c_{0} \norm{\omega(\cdot,t)}_{\infty}
\end{displaymath}
for some absolute constant $c_{0} > 0$. Let $y(t)$ be another family of points such that, for each $t \in [0,T)$, $y(t)$ lies on the same vortex line as $x(t)$ and the vorticity direction $\xi = \omega/\abs{\omega}$ is well-defined along the vortex line
lying between $x(t)$ and $y(t)$. If
\begin{subequations}\label{eqn_dhy_dxi}
\begin{equation}
  \biggl| \int_{x(t)}^{y(t)} (\nabla \cdot \xi)(s,t)\,ds \biggr| \leq C,\qquad \forall t \in [0,T),
  \label{eqn_dhy_dxi_i}
\end{equation}
for some absolute constant $C$ and
\begin{equation}
  \int_{0}^{T} \abs{\omega(y(t),t)}\,dt < \infty,
  \label{eqn_dhy_dxi_y}
\end{equation}
\end{subequations}
then there will be no blowup of $\omega(x(t),t)$ up to time $T$. Moreover,
\begin{displaymath}
  \re^{-C} \leq \frac{\abs{\omega(x(t),t)}}{\abs{\omega(y(t),t)}} \leq \re^{C},\qquad \forall t \in [0,T).
\end{displaymath}
\label{thm_dhy_1}
\end{theorem}

The second part of the DHY criterion concerns the dynamic blowup of the vorticity along a vortex line. More specifically, consider a family of vortex line segments $L_{t}$ along which the vorticity is comparable to $\norm{\omega(\cdot,t)}_{\infty}$.
Denote by $L(t)$ the arc length of $L_{t}$ and define
\begin{displaymath}
  U_{\xi}(t) = \max_{x,y \in L_{t}} \abs{(u \cdot \xi)(x,t) - (u \cdot \xi)(y,t)},\qquad U_{n}(t) = \max_{L_{t}} \abs{u \cdot n},
\end{displaymath}
and
\begin{displaymath}
  M(t) = \max \bigl\{ \norm{\nabla \cdot \xi}_{L^{\infty}(L_{t})}, \norm{\kappa}_{L^{\infty}(L_{t})} \bigr\},
\end{displaymath}
where $\kappa = \abs{\xi \cdot \nabla \xi}$ is the curvature of the vortex line and $n$ is the unit normal vector of $L_{t}$.
\begin{theorem}
Assume that there exists a family of vortex line segments $L_{t}$ and a $T_{0} \in [0,T)$ such that $X(L_{t_{1}},t_{1},t_{2}) \supseteq L_{t_{2}}$ for all $T_{0} < t_{1} < t_{2} < T$. Assume also that $\norm{\omega(\cdot,t)}_{\infty}$ is monotonically
increasing and that
\begin{displaymath}
  \norm{\omega(\cdot,t)}_{L^{\infty}(L_{t})} \geq c_{0} \norm{\omega(\cdot,t)}_{\infty}
\end{displaymath}
for some absolute constant $c_{0} > 0$ when $t$ is sufficiently close to $T$. If
\begin{enumerate}
  \item $U_{\xi}(t) + U_{n}(t) M(t) L(t) \leq c_{A} (T-t)^{-A}$ for some $A \in (0,1)$,
  \item $M(t) L(t) \leq C_{0}$, and
  \item $L(t) \geq c_{B} (T-t)^{B}$ for some $B < 1-A$,
\end{enumerate}
where $c_{A},\ c_{B},\ C_{0}$ are all absolute constants, then there will be no blowup of $\norm{\omega(\cdot,t)}_{\infty}$ up to time $T$.
\label{thm_dhy_2}
\end{theorem}

\begin{figure}[h]
  \centering
  \subfigure[$i_{1}(t)$]{
    \includegraphics[scale=0.415]{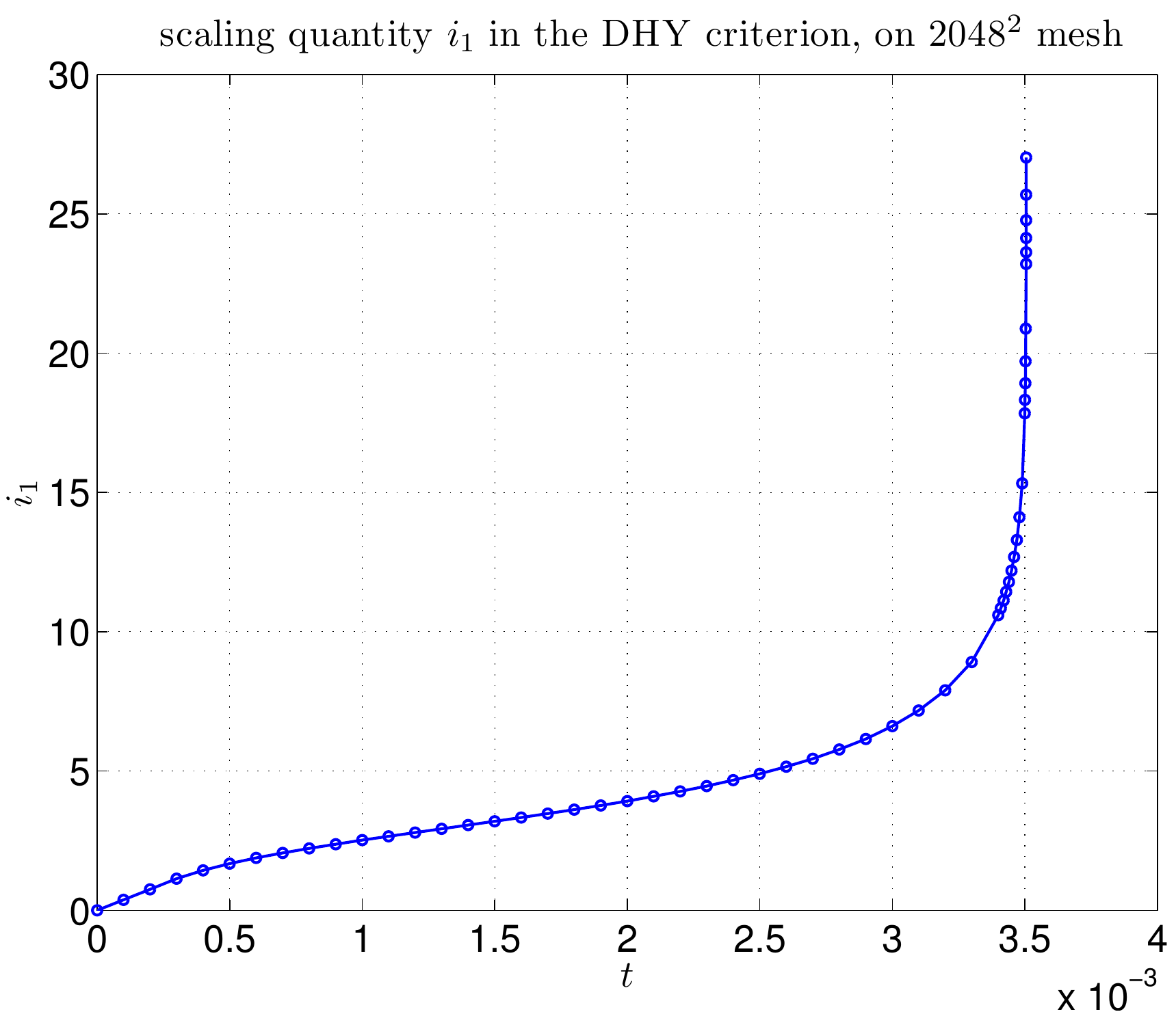}
    \label{fig_vdir_grad_bds_1}
  }\quad
  \subfigure[$M_{1}(t)$]{
    \includegraphics[scale=0.415]{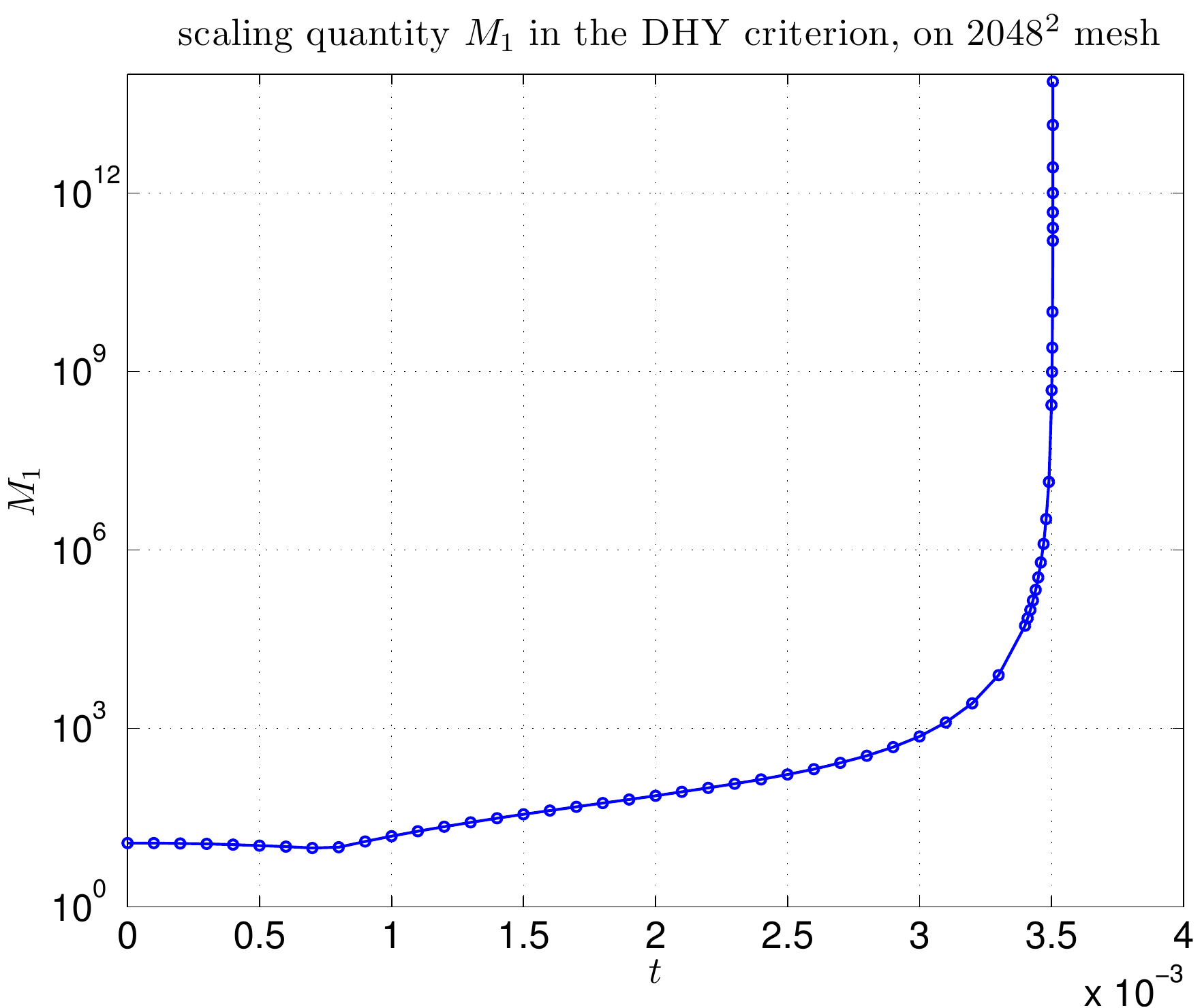}
    \label{fig_vdir_grad_bds_2}
  }
  \caption{The quantities $i_{1}$ and $M_{1}$ considered in the DHY criterion.}
  \label{fig_vdir_grad_bds}
\end{figure}

To check our numerical data against the DHY criterion, we first note that any vortex line segment containing the point $q_{0} = (1,0,0)^{T}$ must lie on the ray
\begin{displaymath}
  [0,q_{0}] := \Bigl\{ (x_{1},0,0) \in \mathbb{R}^{3}\colon x_{1} \in (0,1) \Bigr\}.
\end{displaymath}
This follows directly from the fact that the vorticity direction vectors $\xi(x)$, when restricted to $[0,q_{0}]$, all point in the same direction $(-1,0,0)^{T}$. Now we argue that the conditions of Theorem \ref{thm_dhy_1} cannot be satisfied for the
particular choice $x(t) \equiv q_{0}$. Indeed, if $y(t)$ is a family of points satisfying the conditions of the theorem, then each $y(t)$ must lie on the same vortex line as $q_{0}$ and hence must lie on the ray $[0,q_{0}]$. Now consider the quantity
\begin{displaymath}
  i_{1}(t) = \min_{x \in [0,q_{0}]} \biggl\{ \int_{0}^{t} \abs{\omega(x,s)}\,ds + \int_{x}^{q_{0}} (\nabla \cdot \xi)(y,t)\,dy \biggr\}.
\end{displaymath}
If we define, for each fixed $t \in (0,t_{s})$ and $q \in (0,q_{0})$, the particle trajectory
\begin{displaymath}
  \frac{dX_{q}}{ds} = u^{r}(X_{q},s) = -\abs{X_{q}} \psi_{1,z}(X_{q},s),\qquad X_{q}(t) = q,
\end{displaymath}
then clearly $i_{1}$ gives a lower bound for
\begin{displaymath}
  \int_{0}^{t} \abs{\omega(X_{q}(s),s)}\,ds + \biggl| \int_{q}^{q_{0}} (\nabla \cdot \xi)(y,t)\,dy \biggr|,
\end{displaymath}
since it is numerically observed that $u^{r} < 0$ on $[0,q_{0}]$ and $\abs{\omega}$ is monotonically increasing on $[0,q_{0}]$, which means that
\begin{displaymath}
  \abs{\omega(X_{q}(s),s)} \geq \abs{\omega(X_{q}(t),s)} = \abs{\omega(q,s)},\qquad \forall s \in [0,t].
\end{displaymath}
As is clear from Figure \ref{fig_vdir_grad_bds}\subref{fig_vdir_grad_bds_1}, the quantity $i_{1}$ grows unboundedly as $t$ approaches $t_{s}$, hence the two conditions \eqref{eqn_dhy_dxi_i} and \eqref{eqn_dhy_dxi_y} stated in Theorem \ref{thm_dhy_1}
cannot be satisfied simultaneously.

To apply Theorem \ref{thm_dhy_2} to our data, we consider the quantity
\begin{displaymath}
  M_{1}(t) = \max_{x \in [0,q_{0}]} (\nabla \cdot \xi)(x,t),
\end{displaymath}
which defines the local maximum of the divergence of $\xi$ on $[0,q_{0}]$. As can be seen from Figure \ref{fig_vdir_grad_bds}\subref{fig_vdir_grad_bds_2}, the quantity $M_{1}(t)$ grows rapidly with $t$, and a line fitting similar to \eqref{eqn_vfit_c}
shows that
\begin{equation}
  M_{1}(t) \sim c (\hat{t}_{s}-t)^{-2.9165},\qquad c = 1.3497 \times 10^{-7}.
  \label{eqn_dhy_pow_M1}
\end{equation}
We now argue that the conditions of Theorem \ref{thm_dhy_2} cannot be satisfied for any family of vortex line segments $L_{t}$ containing the point $q_{0}$. Indeed, as our numerical data shows, the maximum of $\nabla \cdot \xi$ on $[0,q_{0}]$ is always
attained at $q_{0}$, i.e.
\begin{displaymath}
  M_{1}(t) = (\nabla \cdot \xi)(q_{0},t) \leq \norm{\nabla \cdot \xi}_{L^{\infty}(L_{t})} \leq M(t).
\end{displaymath}
Thus conditions (b) and (c) in Theorem \ref{thm_dhy_2} cannot be satisfied simultaneously, since condition (b), when combined with \eqref{eqn_dhy_pow_M1}, implies that
\begin{displaymath}
  L(t) \leq C_{0} M^{-1}(t) \leq C_{0} M_{1}^{-1}(t) \sim C (\hat{t}_{s}-t)^{2.9165},
\end{displaymath}
which violates condition (c).

\subsubsection{The Geometry of the Vorticity Direction}\label{ssec2_geosd_geo}
It is illuminating to examine at this point the local geometric structure of the vorticity direction $\xi$ near the location of the maximum vorticity. Figure \ref{fig_vdir_geo} below shows a plot of the 2D vorticity direction $\tilde{\xi} =
(\xi^{r},\xi^{z})^{T}$ and a plot of the $z$-direction component $\xi^{z}$, both defined at $t = 0.003505$ on the set
\begin{displaymath}
  \tilde{D}_{\infty} = [1 - 5.99 \times 10^{-11},1] \times [0,2.09 \times 10^{-12}].
\end{displaymath}
The through-plane ($\theta$) component of $\xi$ has a maximum absolute value of $2.1874 \times 10^{-6}$ in $\tilde{D}_{\infty}$ and hence is negligible there. It can be observed from Figure \ref{fig_vdir_geo} that the $z$-direction component $\xi^{z}$
experiences an $O(1)$ change in $\tilde{D}_{\infty}$ along the $z$-dimension. This corresponds to a set of ``densely packed'' vortex lines near the location of the maximum vorticity, and is responsible for the rapid growth of quantities like
$L_{\xi,\tilde{q}_{0}}$ and $\nabla \cdot \xi$ observed in Figure \ref{fig_vdir_lip_bds}--\ref{fig_vdir_grad_bds}.
\begin{figure}[h]
  \centering
  \subfigure[2D vorticity direction $\tilde{\xi} = (\xi^{r},\xi^{z})^{T}$]{
    \psfrag{r_l}{\tiny $r_{l}$}
    \psfrag{z_r}{\tiny $z_{r}$}
    \includegraphics[scale=0.415]{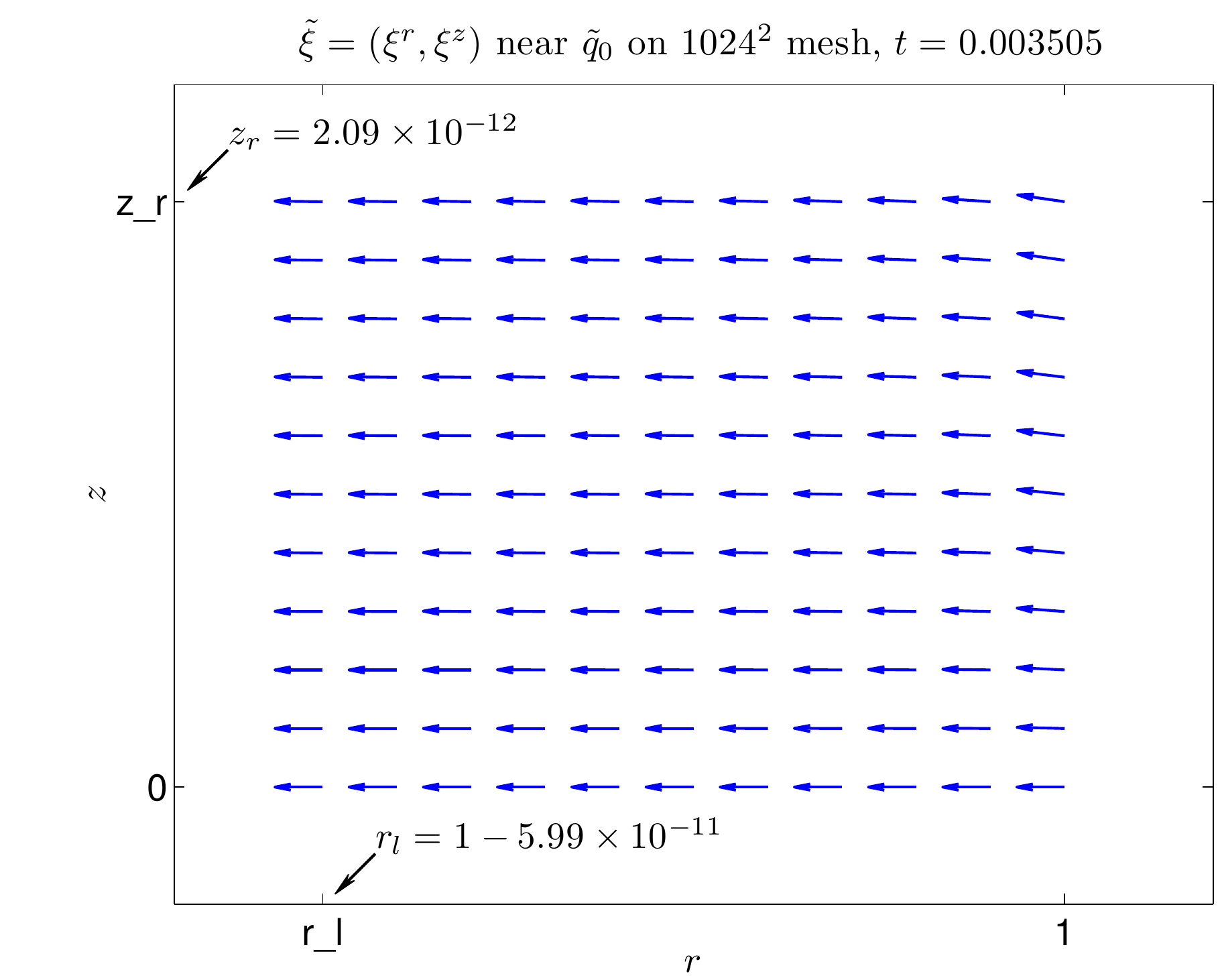}
    \label{fig_vdir_rz}
  }\quad
  \subfigure[$z$-component $\xi^{z}$ of vorticity direction]{
    \psfrag{r_l}{\tiny $r_{l}$}
    \psfrag{z_r}{\tiny $z_{r}$}
    \includegraphics[scale=0.415]{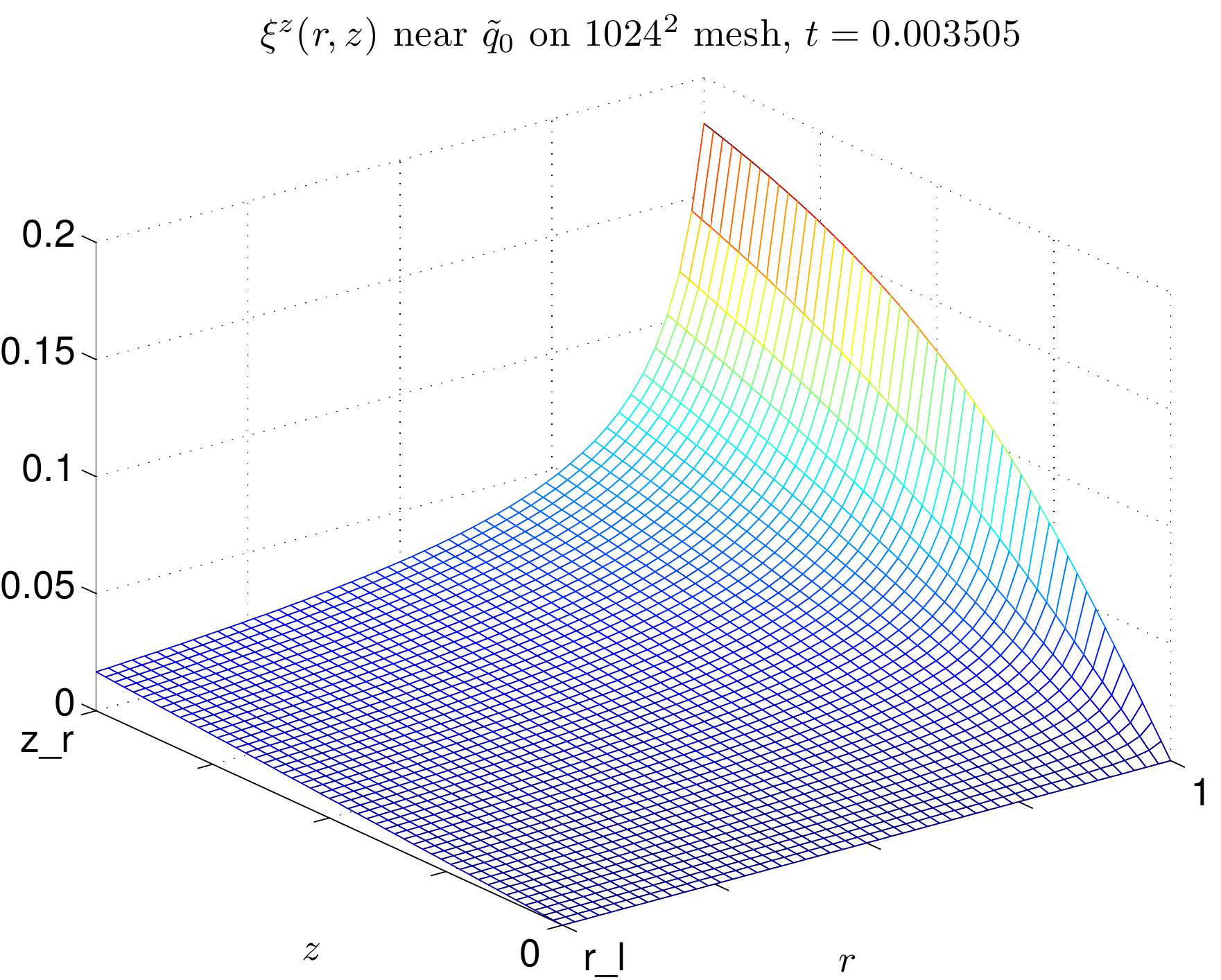}
    \label{fig_vdir_z}
  }
  \caption{(a) the 2D vorticity direction $\tilde{\xi} = (\xi^{r},\xi^{z})^{T}$ and (b) the $z$-direction component $\xi^{z}$ computed on the $1024 \times 1024$ mesh at $t = 0.003505$. All plots in this figure are defined on the region
  $[r_{l},1] \times [0,z_{r}]$ where $r_{l} = 1 - 5.99 \times 10^{-11}$ and $z_{r} = 2.09 \times 10^{-12}$.}
  \label{fig_vdir_geo}
\end{figure}%

\subsubsection{The Spectral Dynamics}\label{ssec2_geosd_sd}
The analysis presented in the previous sections suggests that the growth of the vorticity amplification factor $\alpha$ depends on the local \emph{geometric} structures of the vorticity vector. On the other hand, the dynamics of the vorticity
amplification can also be investigated from an \emph{algebraic} point of view, where the defining relation (see \eqref{eqn_vort_dyn_a})
\begin{displaymath}
  \alpha = \xi \cdot \nabla u \cdot \xi = \xi \cdot S \xi,\qquad S = \frac{1}{2} \bigl( \nabla u + \nabla u^{T} \bigr),
\end{displaymath}
is studied directly and the eigenstructure of the symmetric strain tensor $S$ plays the central role.

In what follows, we shall derive a remarkable connection between the vorticity amplification factor $\alpha$ and the eigenstructure of $S$ at the location of the maximum vorticity. The derivation starts with the representation formula of the velocity
vector in cylindrical coordinates:
\begin{displaymath}
  u = u^{r} e_{r} + u^{\theta} e_{\theta} + u^{z} e_{z},
\end{displaymath}
where the three Cartesian components of $u$ are expressed in terms of the transformed variables $(u_{1},\psi_{1})$:
\begin{align*}
  v_{1} & = -r\psi_{1,z} \cos\theta - ru_{1} \sin\theta, \\
  v_{2} & = -r\psi_{1,z} \sin\theta + ru_{1} \cos\theta, \\
  v_{3} & = 2\psi_{1} + r\psi_{1,r}.
\end{align*}
The entries of the deformation tensor $\nabla u$ can be readily computed, yielding
\begin{align*}
  \partial_{x} v_{1}|_{\theta = 0} & = -\psi_{1,z} - r\psi_{1,rz}, & \partial_{y} v_{1}|_{\theta = 0} & = -u_{1}, & \partial_{z} v_{1}|_{\theta = 0} & = -r\psi_{1,zz}, \\
  \partial_{x} v_{2}|_{\theta = 0} & = u_{1} + ru_{1,r}, & \partial_{y} v_{2}|_{\theta = 0} & = -\psi_{1,z}, & \partial_{z} v_{2}|_{\theta = 0} & = ru_{1,z}, \\
  \partial_{x} v_{3}|_{\theta = 0} & = 3\psi_{1,r} + r\psi_{1,rr}, & \partial_{y} v_{3}|_{\theta = 0} & = 0, & \partial_{z} v_{3}|_{\theta = 0} & = 2\psi_{1,z} + r\psi_{1,rz}.
\end{align*}
Note that due to axial symmetry the evaluation needs only to be done on the meridian plane $\theta = 0$. When further restricted to the point $\tilde{q}_{0} = (1,0)^{T}$, the location of the maximum vorticity, the above expressions reduce to
\begin{displaymath}
  \nabla \tilde{u} =
  \begin{pmatrix*}
    \ \, -\tilde{\psi}_{1,rz} & 0 & 0\ \, \\
    \ \, 0 & 0 & \tilde{u}_{1,z}\ \, \\
    \ \, 0 & 0 & \tilde{\psi}_{1,rz}\ \,
  \end{pmatrix*},\qquad \tilde{S} =
  \begin{pmatrix*}
    \ \, -\tilde{\psi}_{1,rz} & 0 & 0\ \, \\
    \ \, 0 & 0 & \frac{1}{2} \tilde{u}_{1,z}\ \, \\
    \ \, 0 & \frac{1}{2} \tilde{u}_{1,z} & \tilde{\psi}_{1,rz}\ \,
  \end{pmatrix*},
\end{displaymath}
where for simplicity we have written $\nabla \tilde{u} = \nabla u(\tilde{q}_{0}),\ \tilde{S} = S(\tilde{q}_{0})$, etc.

Now the eigenvalues of $\tilde{S}$ can be easily found to be
\begin{displaymath}
  \tilde{\lambda}_{1} = \frac{1}{2} \Bigl\{ \tilde{\psi}_{1,rz} + \bigl[ \tilde{\psi}_{1,rz}^{2} + \tilde{u}_{1,z}^{2} \bigr]^{1/2} \Bigr\},\qquad \tilde{\lambda}_{2} = -\tilde{\psi}_{1,rz},\qquad
  \tilde{\lambda}_{3} = \frac{1}{2} \Bigl\{ \tilde{\psi}_{1,rz} - \bigl[ \tilde{\psi}_{1,rz}^{2} + \tilde{u}_{1,z}^{2} \bigr]^{1/2} \Bigr\},
\end{displaymath}
with corresponding eigenvectors
\begin{displaymath}
  \tilde{w}_{1} =
  \begin{pmatrix*}
    \ \, 0\ \, \\
    \ \, \frac{1}{2} \tilde{u}_{1,z}\ \, \\
    \ \, \tilde{\lambda}_{1}\ \,
  \end{pmatrix*},\qquad \tilde{w}_{2} =
  \begin{pmatrix*}
    \ \, 1\ \, \\
    \ \, 0\ \, \\
    \ \, 0\ \,
  \end{pmatrix*},\qquad \tilde{w}_{3} =
  \begin{pmatrix*}
    \ \, 0\ \, \\
    \ \, \frac{1}{2} \tilde{u}_{1,z}\ \, \\
    \ \, \tilde{\lambda}_{3}\ \,
  \end{pmatrix*}.
\end{displaymath}
On the other hand, the vorticity vector $\omega$ at $\tilde{q}_{0}$ takes the form (see \eqref{eqn_vort_dyn0_v})
\begin{displaymath}
  \tilde{\omega} =
  \begin{pmatrix*}
    \ \, -\tilde{u}_{1,z}\ \, \\
    \ \, 0\ \, \\
    \ \, 0\ \,
  \end{pmatrix*},\qquad \text{with}\qquad \tilde{\xi} = \frac{\tilde{\omega}}{\abs{\tilde{\omega}}} =
  \begin{pmatrix*}
    \ \, -1\ \, \\
    \ \, 0\ \, \\
    \ \, 0\ \,
  \end{pmatrix*}.
\end{displaymath}
Thus the vorticity direction $\tilde{\xi}$ at the location of the maximum vorticity is \emph{perfectly aligned with $\tilde{w}_{2}$}, the second eigenvector of $\tilde{S}$. In addition,
\begin{displaymath}
  \alpha_{\infty} := \tilde{\alpha} = \tilde{\xi} \cdot \tilde{S} \tilde{\xi} = \tilde{\lambda}_{2} = -\tilde{\psi}_{1,rz},
\end{displaymath}
consistent with the result derived earlier in Section \ref{ssec2_vfit_line} (see \eqref{eqn_vort_dyn0_a}).

It is worth noting that, when viewed in $\mathbb{R}^{3}$, the eigenvectors $\{\tilde{w}_{1},\tilde{w}_{2},\tilde{w}_{3}\}$ restricted to the ``singularity ring''
\begin{displaymath}
  C = \Bigl\{ (x,y,z) \in \mathbb{R}^{3}\colon x^{2}+y^{2} = 1,\ z = 0 \Bigr\}
\end{displaymath}
form an orthogonal frame, with $\tilde{w}_{2}$ pointing in the radial direction and $\tilde{w}_{1},\ \tilde{w}_{3}$ pointing in directions tangential to the lateral surface of the cylinder $r = 1$.

Finally, by making use of the relations
\begin{displaymath}
  \alpha_{\infty} = -\tilde{\psi}_{1,rz},\qquad \norm{\omega}_{\infty} = \abs{\tilde{\omega}} = \abs{\tilde{u}_{1,z}},
\end{displaymath}
we may also express the first and third eigenvalues of $\tilde{S}$ in the form
\begin{displaymath}
  \tilde{\lambda}_{1,3} = \frac{1}{2} \Bigl\{ -\alpha_{\infty} \pm \bigl[ \alpha_{\infty}^{2} + \norm{\omega}_{\infty}^{2} \bigr]^{1/2} \Bigr\}.
\end{displaymath}
Since $\alpha_{\infty}$ and $\norm{\omega}_{\infty}$ both satisfy an inverse power-law with an exponent roughly equal to $-1$ (for $\alpha_{\infty}$) and $-\frac{5}{2}$ (for $\norm{\omega}_{\infty}$), it follows that
\begin{displaymath}
  \tilde{\lambda}_{1,3} \sim \pm \frac{1}{2}\, \norm{\omega}_{\infty},\qquad t \to t_{s}^{-}.
\end{displaymath}
This is confirmed by a line fitting similar to \eqref{eqn_vfit_c}, which yields
\begin{align*}
  \tilde{\lambda}_{1} & \sim c_{1} (\hat{t}_{s}-t)^{-2.4582},\qquad c_{1} = 3.6514 \times 10^{-4}, \\
  \tilde{\lambda}_{3} & \sim c_{3} (\hat{t}_{s}-t)^{-2.4576},\qquad c_{3} = -3.6759 \times 10^{-4},
\end{align*}
where $\hat{t}_{s}$ is the singularity time estimated from \eqref{eqn_vfit_ts}.

\subsection{Locally Self-Similar Structure}\label{ssec_selfsim}
It is well known that the 3D Euler equations \eqref{eqn_eu} have the special scaling property that, if $u(x,t)$ is a solution of the equations, then
\begin{displaymath}
  u_{\lambda}(x,t) := \lambda^{\alpha} u(\lambda x,\lambda^{\alpha+1} t),\qquad \forall \lambda > 0,\ \forall \alpha \in \mathbb{R},
\end{displaymath}
is also a solution. A natural question is then whether the 3D Euler equations have self-similar solutions of the form
\begin{subequations}\label{eqn_ssim0}
\begin{equation}
  u(x,t) = \frac{1}{[T-t]^{\gamma}} U \biggl( \frac{x-x_{0}}{[T-t]^{\beta}} \biggr),
  \label{eqn_ssim0_u}
\end{equation}
where $U$ is a self-similar velocity profile and $\beta,\ \gamma$ are scaling exponents. By substituting \eqref{eqn_ssim0_u} into \eqref{eqn_eu}, it is easily shown that
\begin{equation}
  \beta = \frac{1}{\alpha+1},\quad \gamma = \frac{\alpha}{\alpha+1},\qquad \forall \alpha \neq -1,
  \label{eqn_ssim0_exp}
\end{equation}
which in particular implies that
\begin{equation}
  \nabla u(x,t) = \frac{1}{T-t} \nabla U \biggl( \frac{x-x_{0}}{[T-t]^{\beta}} \biggr).
  \label{eqn_ssim0_du}
\end{equation}

In \citet{chae2007,chae2010}, the existence of \emph{global} self-similar solutions of the form \eqref{eqn_ssim0_du} is excluded under one of the following conditions: either \citep{chae2010}
\begin{displaymath}
  \limsup_{t \to T^{-}} (T-t) \norm{\nabla u(\cdot,t)}_{\infty} = \norm{\nabla U}_{\infty} < 1,
\end{displaymath}
or \citep{chae2007}
\begin{displaymath}
  \Omega := \nabla \times U \neq 0\qquad \text{and}\qquad \Omega \in L^{p}(\mathbb{R}^{3}),\quad \forall p \in (0,p_{1}),
\end{displaymath}
for some $p_{1} > 0$. Note that the first condition is not easy to interpret physically while the second is too strong, effectively requiring that $\Omega$ decay exponentially fast at infinity or have compact support. These nonexistence results were
generalized later in \citet{chae2010} to \emph{$\alpha$-asymptotically global} self-similar solutions $\bar{U}$ of the form
\begin{equation}
  \lim_{t \to T^{-}} [T-t]^{1-(3\beta/p)} \biggl\| \nabla u (\cdot,t) - \frac{1}{T-t} \nabla \bar{U} \biggl( \frac{\cdot-x_{0}}{[T-t]^{\beta}} \biggr) \biggr\|_{L^{p}(\mathbb{R}^{3})} = 0,\qquad \beta = \frac{1}{\alpha+1},
  \label{eqn_ssim0_asym}
\end{equation}
\end{subequations}
where the convergence of $u$ to the self-similar profile $\bar{U}$ is understood in the sense of $L^{p},\ p \in (0,\infty]$. Similar nonexistence results for \emph{local} self-similar solutions were also obtained in \citet{chae2011}.

In axisymmetric flows, self-similar solutions\footnote{In what follows, whenever we say ``self-similar solutions'' for an axisymmetric flow we always mean ``self-similar solutions in the meridian plane''.} naturally take the form
\begin{align*}
  u_{1}(\tilde{x},t) & \sim [T-t]^{\gamma_{u}} U \biggl( \frac{\tilde{x}-\tilde{x}_{0}}{[T-t]^{\gamma_{l}}} \biggr), \\
  \omega_{1}(\tilde{x},t) & \sim [T-t]^{\gamma_{\omega}} \Omega \biggl( \frac{\tilde{x}-\tilde{x}_{0}}{[T-t]^{\gamma_{l}}} \biggr), \\
  \psi_{1}(\tilde{x},t) & \sim [T-t]^{\gamma_{\psi}} \Psi \biggl( \frac{\tilde{x}-\tilde{x}_{0}}{[T-t]^{\gamma_{l}}} \biggr),\qquad \tilde{x} \to \tilde{x}_{0},\ t \to T^{-},
\end{align*}
where $\tilde{x} = (r,z)^{T}$ is a point on the $rz$-plane and $(U,\Omega,\Psi)$ are self-similar profiles. Note that this ansatz is \emph{not} of the Leray type and does \emph{not} correspond to a ``conventional'' self-similar solution when viewed in
$\mathbb{R}^{3}$. Rather, it gives a tube-like anisotropic singularity due to the presence of axial symmetry. In addition, the ansatz induces a scaling law (see Section \ref{ssec2_ssim_fit})
\begin{displaymath}
  \norm{\nabla u(\cdot,t)}_{\infty} = O(T-t)^{\min\{\gamma_{u}-\gamma_{\ell},-1\}}
\end{displaymath}
that is \emph{very different} from the ``standard'' law $\norm{\nabla u(\cdot,t)}_{\infty} = O(T-t)^{-1}$ assumed by \citet{chae2007,chae2010,chae2011}. Hence it gives new hope for the existence of a (meridian-plane) self-similar solution.

In what follows, we shall carry out an extensive study of the numerical solution near the location of the maximum vorticity $\tilde{q}_{0} = (1,0)^{T}$ and demonstrate the existence of a locally self-similar blowup. By applying a line fitting similar to
\eqref{eqn_vfit_c}, we also confirm the scaling law $\norm{\nabla u(\cdot,t)}_{\infty} = O(T-t)^{-2.4529}$ satisfied by the self-similar solution, hence providing another supporting evidence for the finite-time blowup of the 3D Euler solution.

\subsubsection{Existence of Self-Similar Neighborhood}\label{ssec2_ssim_contv}
\begin{figure}[h]
  \centering
  \subfigure[linear-linear]{
    \includegraphics[scale=0.415]{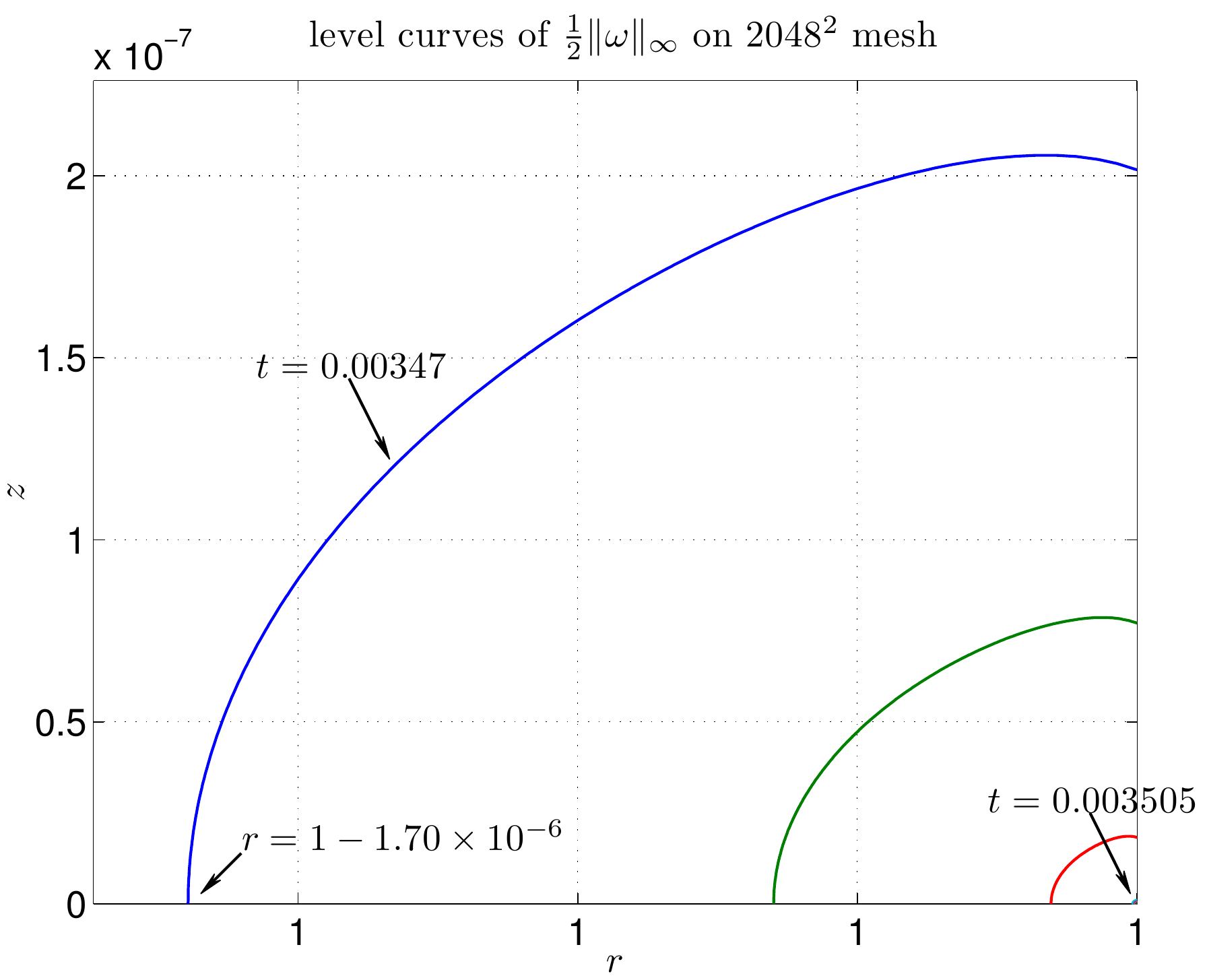}
    \label{fig_ssim_contv_lin}
  }\quad
  \subfigure[log-log]{
    \includegraphics[scale=0.415]{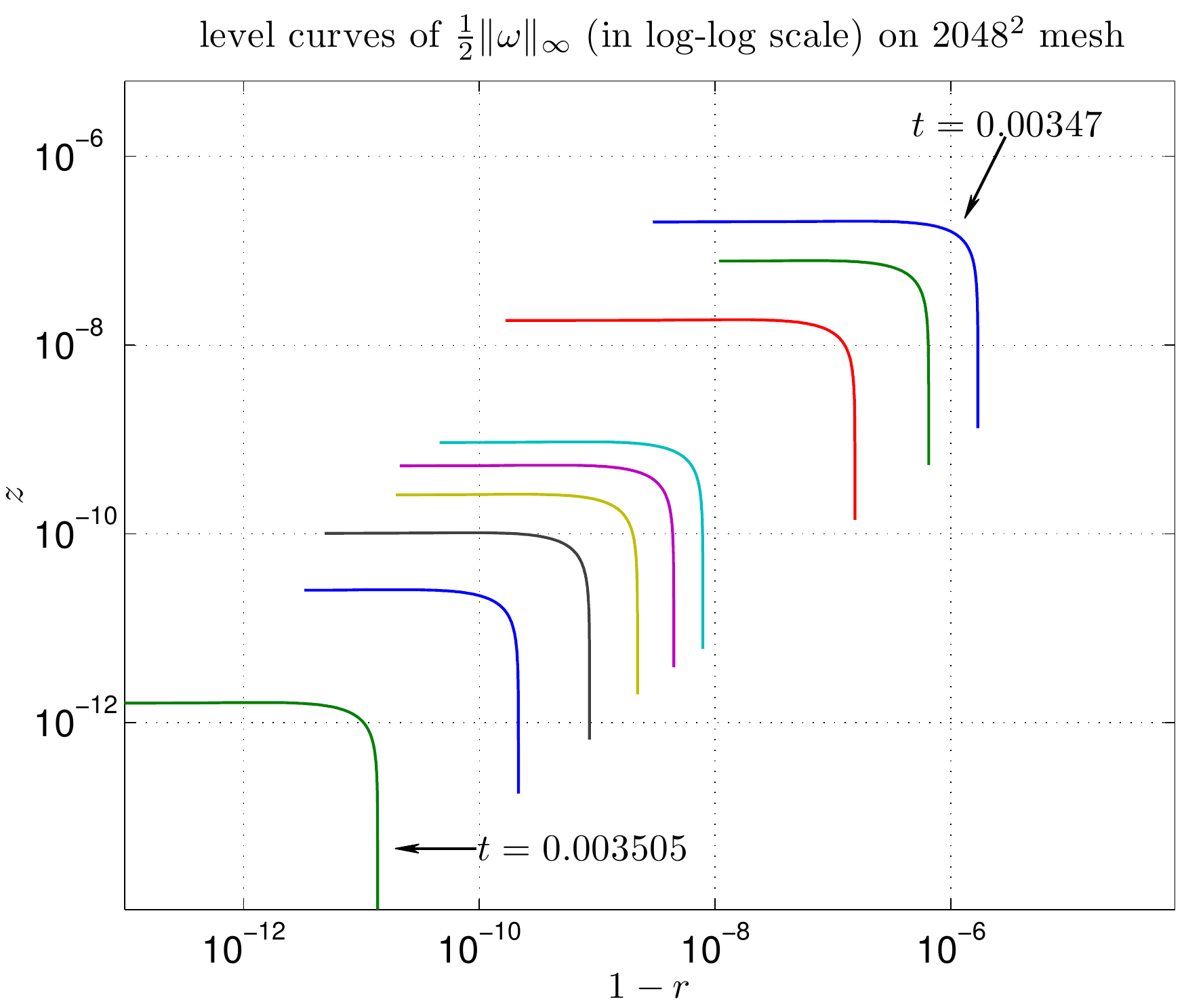}
    \label{fig_ssim_contv_log}
  }
  \subfigure[rescaled]{
    \includegraphics[scale=0.415]{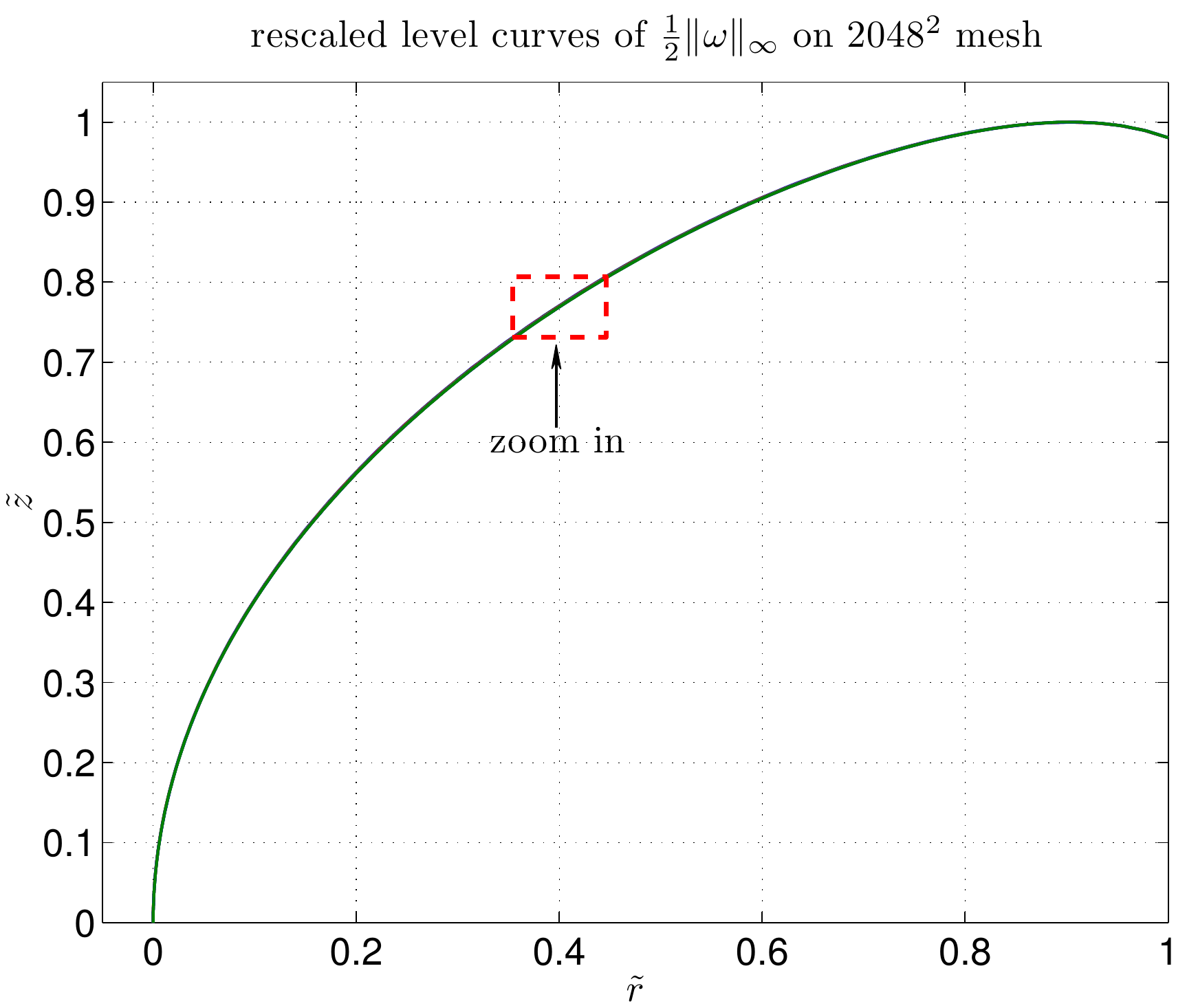}
    \label{fig_ssim_contv_s}
  }\quad
  \subfigure[rescaled (zoom-in)]{
    \includegraphics[scale=0.415]{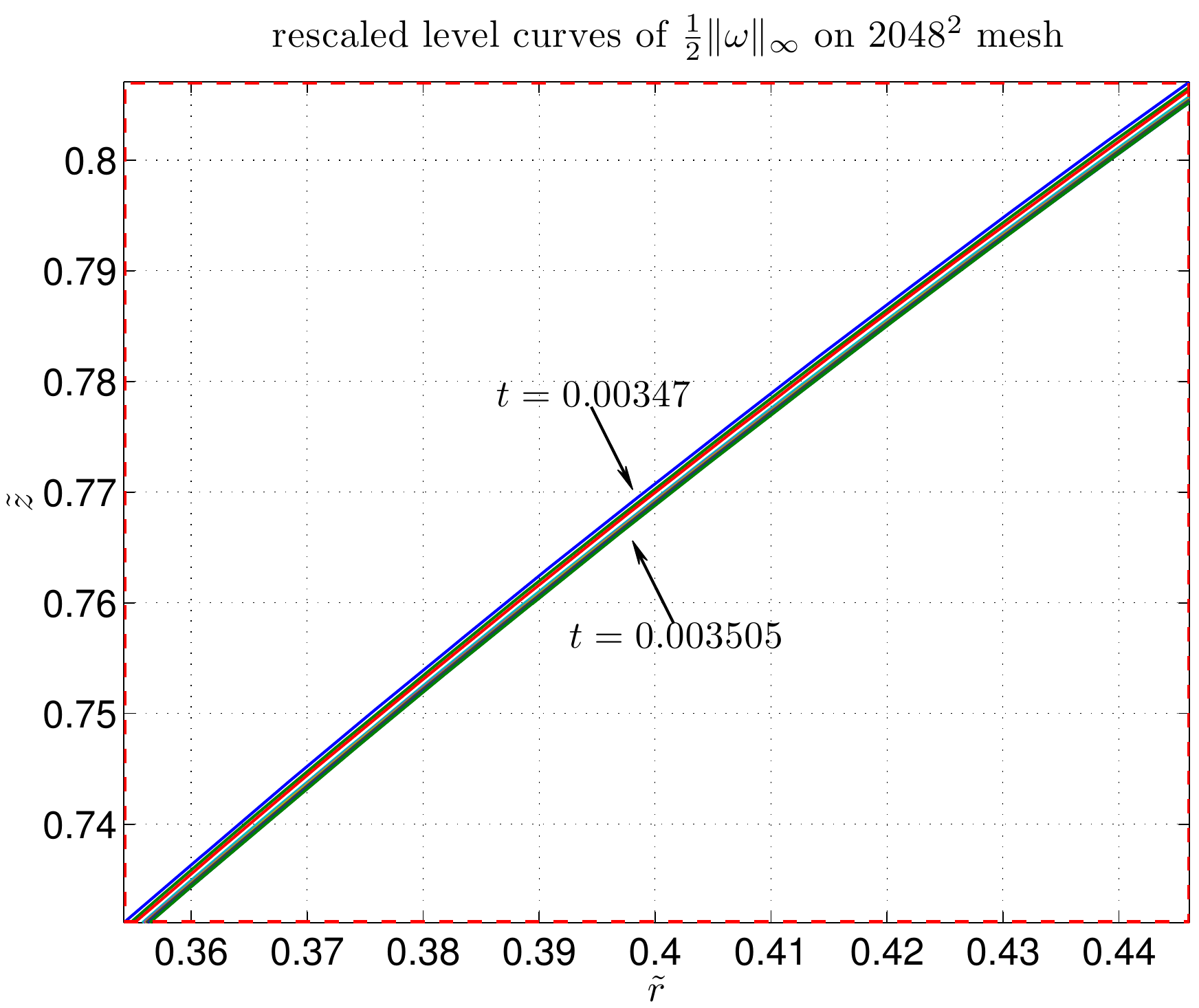}
    \label{fig_ssim_contv_sz}
  }
  \caption{The level curves $C_{\infty}(t)$ at various time instants in (a) linear-linear and (b) log-log scale (against the variables $(1-r)$ and $z$). The rescaled level curves and their zoom-in view are shown in (c) and (d).}
  \label{fig_ssim_contv}
\end{figure}

The identification of a locally self-similar solution requires three basic ingredients: first, the center of self-similarity, $\tilde{x}_{0}$, around which the self-similar structure is developed; second, a neighborhood of $\tilde{x}_{0}$ in which the
self-similar behavior is observed; third, a self-similar profile based on which the self-similar solution is determined. In our computations, the center of self-similarity must be $\tilde{q}_{0} = (1,0)^{T}$, the location of the maximum vorticity, since
this is the point at which the solution is about to blow up. To identify a ``self-similar neighborhood'' of $\tilde{q}_{0}$, we consider again the neighborhood of the maximum vorticity (see \eqref{eqn_d_inf}):
\begin{displaymath}
  D_{\infty}(t) = \Bigl\{ (r,z) \in D(1,\tfrac{1}{4} L)\colon \abs{\omega(r,z,t)} \geq \tfrac{1}{2} \norm{\omega(\cdot,t)}_{\infty} \Bigr\},
\end{displaymath}
and plot in Figure \ref{fig_ssim_contv}\subref{fig_ssim_contv_lin} the boundary of $D_{\infty}(t)$:
\begin{equation}
  C_{\infty}(t) = \Bigl\{ (r,z) \in D(1,\tfrac{1}{4} L)\colon \abs{\omega(r,z,t)} = \tfrac{1}{2} \norm{\omega(\cdot,t)}_{\infty} \Bigr\},
  \label{eqn_c_inf}
\end{equation}
at the nine time instants
\begin{equation}
  \Bigl\{ 0.00347,\ 0.00348,\ 0.00349,\ 0.0035,\ 0.003501,\ 0.003502,\ 0.003503,\ 0.003504,\ 0.003505 \Bigr\}.
  \label{eqn_ssim_t}
\end{equation}
We note that the curves $C_{\infty}(t)$ shrink very rapidly toward $\tilde{q}_{0}$ and the shape of $C_{\infty}(t)$ remains roughly the same at the first few time instants when the curves are still visible in the figure. To reveal more clearly the
asymptotic behavior of $C_{\infty}(t)$ at the later times, we plot these curves in Figure \ref{fig_ssim_contv}\subref{fig_ssim_contv_log} in log-log scale against the variables $(1-r)$ and $z$. The results show that the shape of $C_{\infty}(t)$ indeed
remains roughly the same at all nine time instants. Motivated by this observation, we then rescale each curve $C_{\infty}(t)$ according to the rule
\begin{displaymath}
  \tilde{r} = 1 - \frac{1-r}{d_{r}(C_{\infty}(t))},\qquad \tilde{z} = \frac{z}{d_{z}(C_{\infty}(t))},
\end{displaymath}
where
\begin{displaymath}
  d_{r}(C_{\infty}(t)) = \max_{(r,z) \in C_{\infty}(t)} \abs{1-r},\qquad d_{z}(C_{\infty}(t)) = \max_{(r,z) \in C_{\infty}(t)} \abs{z}.
\end{displaymath}
The results show that the rescaled curves $\tilde{C}_{\infty}(t)$ collapse \emph{almost perfectly} to a single curve (Figure \ref{fig_ssim_contv}\subref{fig_ssim_contv_s}), which confirms the existence of a self-similar neighborhood of $\tilde{q}_{0}$.
The small variations among the different curves $\tilde{C}_{\infty}(t)$ as shown in Figure \ref{fig_ssim_contv}\subref{fig_ssim_contv_sz} are manifestations of the local (inexact) nature of the self-similarity.

\subsubsection{Existence of Self-Similar Profiles}\label{ssec2_ssim_prof}
\begin{figure}[h]
  \centering
  \subfigure[near the $r$-axis]{
    \includegraphics[scale=0.415]{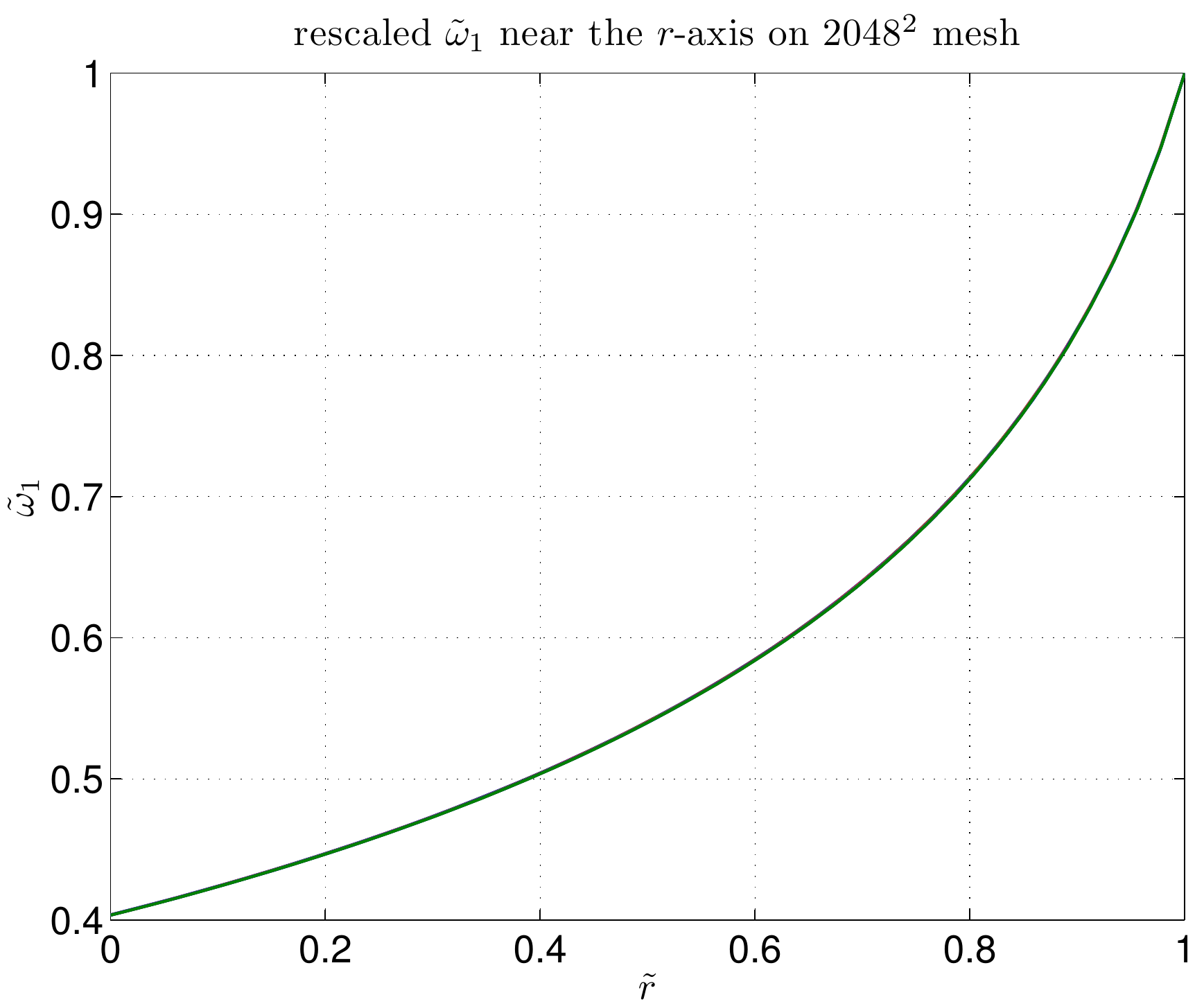}
    \label{fig_ssim_w_r}
  }\quad
  \subfigure[along the wall $r = 1$]{
    \includegraphics[scale=0.415]{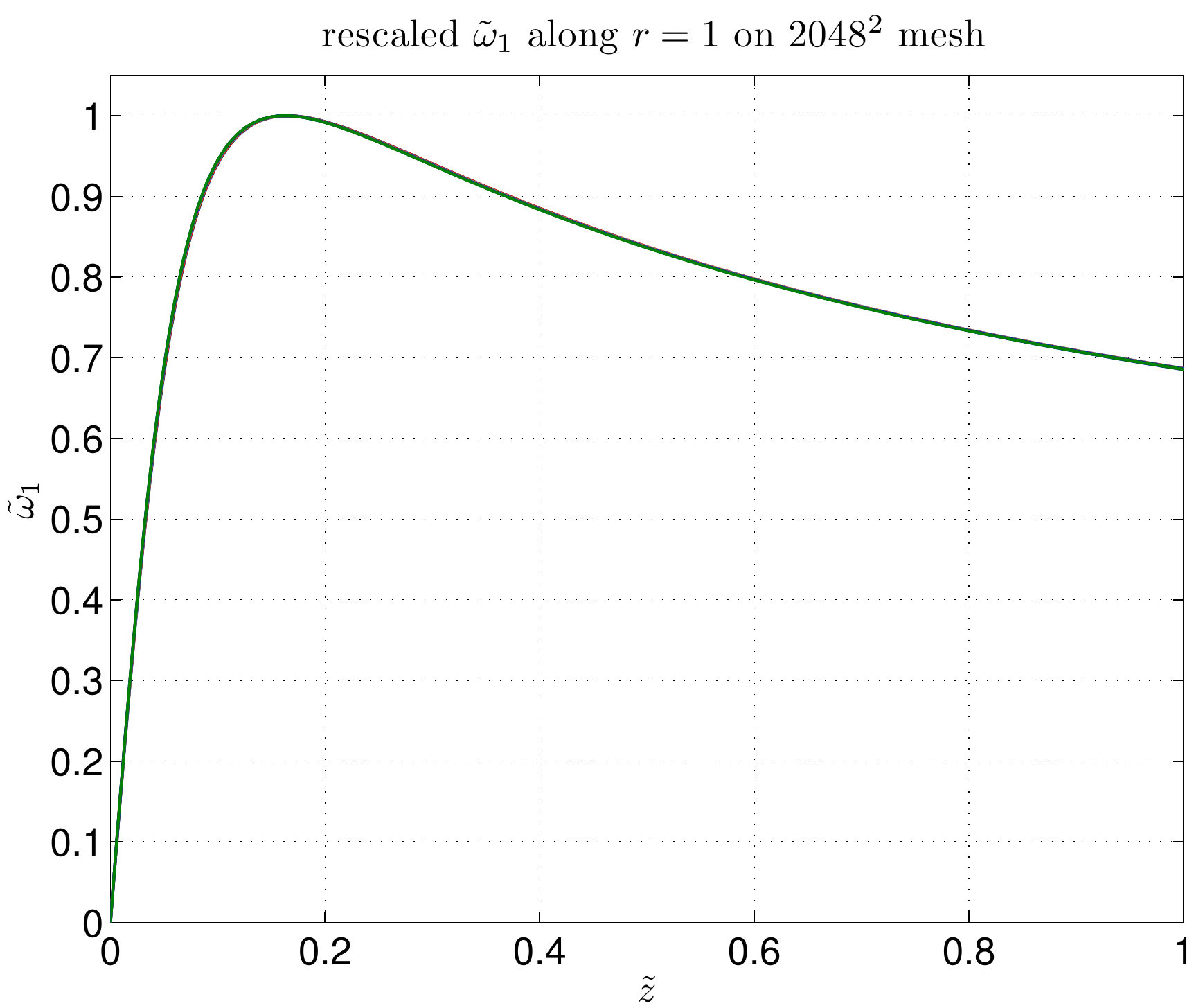}
    \label{fig_ssim_w_z}
  }
  \caption{The 1D self-similar profiles of $\omega_{1}$ (a) near the $r$-axis and (b) along the wall $r = 1$, obtained by rescaling the solutions at the nine time instants given by \eqref{eqn_ssim_t}.}
  \label{fig_ssim_w}
\end{figure}

\begin{figure}[h]
  \centering
  \subfigure[$t = 0.0034$]{
    \includegraphics[scale=0.415]{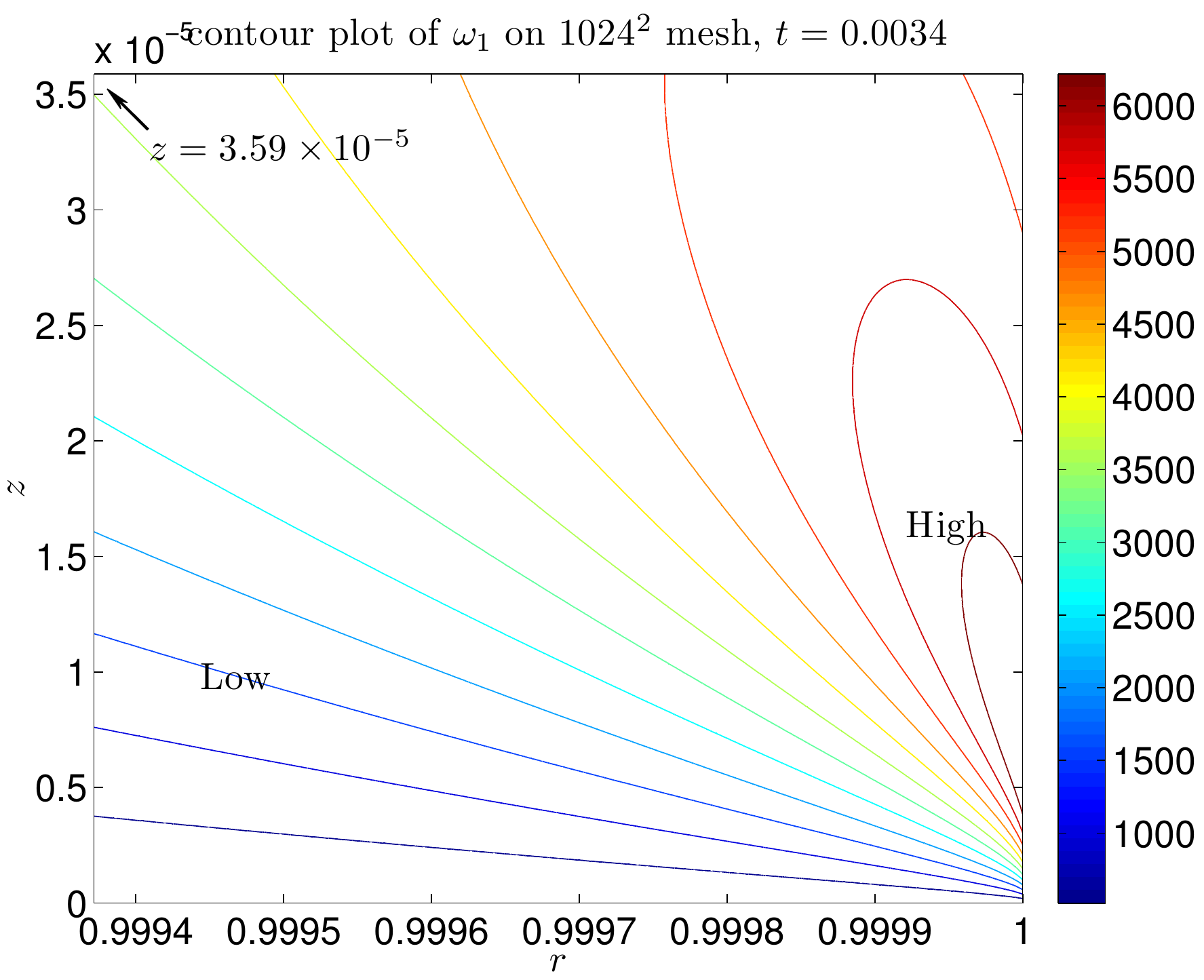}
  }\quad
  \subfigure[$t = 0.003505$]{
    \includegraphics[scale=0.415]{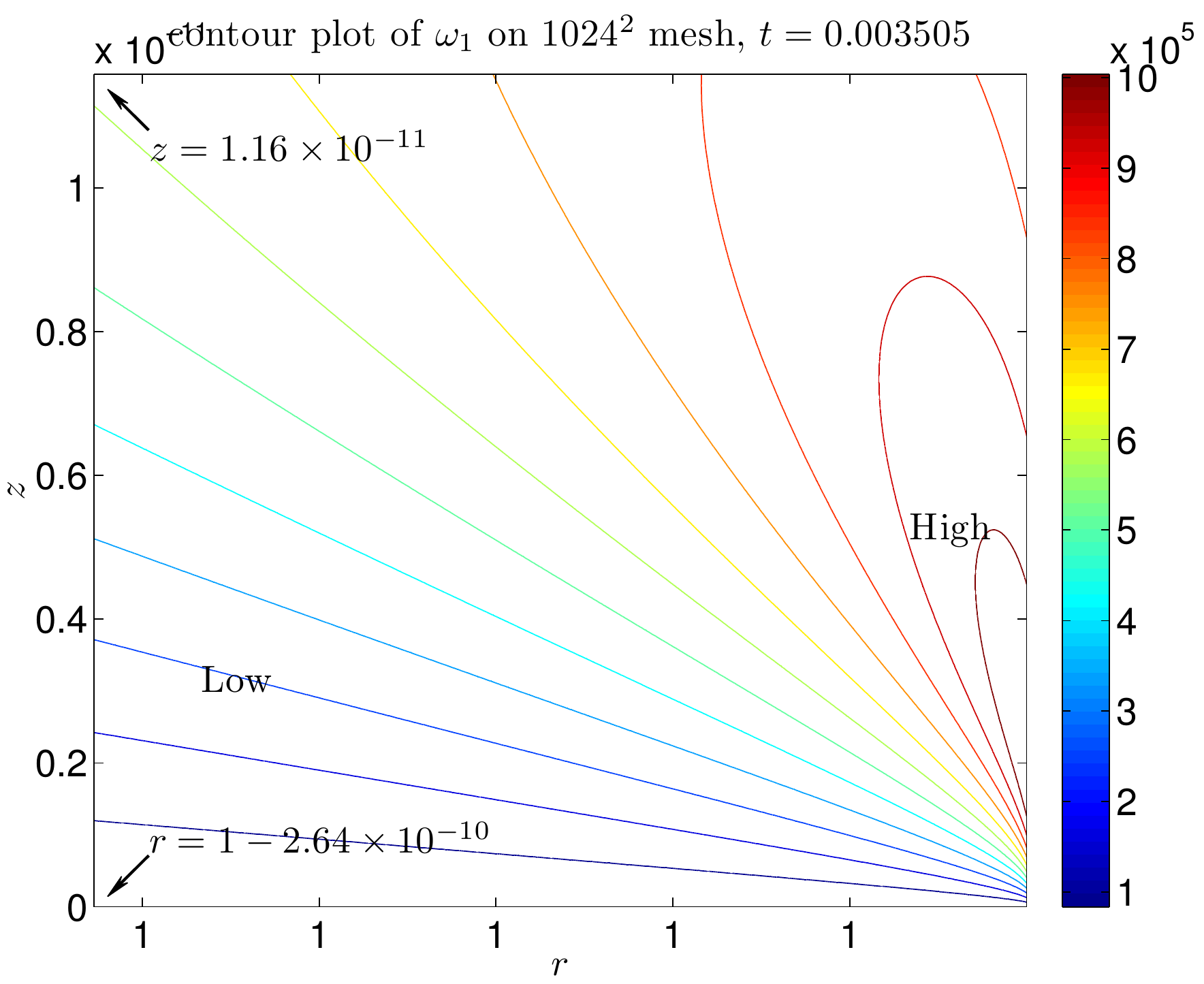}
  }
  \caption{The 2D contour plot of $\omega_{1}$ near the location of the maximum vorticity at (a) $t = 0.0034$ and (b) $t = 0.003505$, both computed on the $1024 \times 1024$ mesh.}
  \label{fig_cont_w1}
\end{figure}

By employing a procedure completely similar to that described in the previous section, we examine the solution $(u_{1},\omega_{1},\psi_{1})$ in the self-similar neighborhood $D_{\infty}(t)$ and confirm the existence of self-similar profiles. For the
purpose of illustration, we plot in Figure \ref{fig_ssim_w} the 1D self-similar profiles of $\omega_{1}$ along selected 1D $r$- and $z$-mesh lines, and in Figure \ref{fig_cont_w1} the 2D contour plots of $\omega_{1}$ near the location of the maximum
vorticity at $t = 0.0034$ and $0.003505$. Similar plots are also obtained for $u_{1},\ \psi_{1}$ and are omitted here for the sake of brevity.

\subsubsection{Self-Similar Analysis}\label{ssec2_ssim_fit}
Given the existence of self-similar profiles in the self-similar neighborhood $D_{\infty}(t)$, we conclude that the solution $(u_{1},\omega_{1},\psi_{1})$ develops a locally self-similar structure near the point of blowup $\tilde{q}_{0}$. This motivates
the ansatz
\begin{subequations}\label{eqn_ssim_fit_v}
\begin{align}
  u_{1}(\tilde{x},t) & \sim [t_{s}-t]^{\gamma_{u}} U \biggl( \frac{\tilde{x}-\tilde{q}_{0}}{[t_{s}-t]^{\gamma_{l}}} \biggr), \label{eqn_ssim_fit_v_u} \\
  \omega_{1}(\tilde{x},t) & \sim [t_{s}-t]^{\gamma_{\omega}} \Omega \biggl( \frac{\tilde{x}-\tilde{q}_{0}}{[t_{s}-t]^{\gamma_{l}}} \biggr), \label{eqn_ssim_fit_v_w} \\
  \psi_{1}(\tilde{x},t) & \sim [t_{s}-t]^{\gamma_{\psi}} \Psi \biggl( \frac{\tilde{x}-\tilde{q}_{0}}{[t_{s}-t]^{\gamma_{l}}} \biggr),\qquad \tilde{x} \to \tilde{q}_{0},\ t \to t_{s}^{-}, \label{eqn_ssim_fit_v_phi}
\end{align}
\end{subequations}
where $\tilde{x} = (r,z)^{T}$ is a point on the $rz$-plane, $\tilde{q}_{0} = (1,0)^{T}$ is the location of the maximum vorticity, $(U,\Omega,\Psi)$ are self-similar profiles, and $\gamma_{u},\ \gamma_{\omega}$, etc. are scaling exponents. Upon
substitution of \eqref{eqn_ssim_fit_v} into \eqref{eqn_eat}, we obtain the dominant balance
\begin{align*}
  \gamma_{u} - 1 & = \gamma_{u} + \gamma_{\psi} - 2\gamma_{l}, \\
  \gamma_{\omega} - 1 & = \gamma_{\omega} + \gamma_{\psi} - 2\gamma_{l} = 2\gamma_{u} - \gamma_{l}, \\
  \gamma_{\psi} - 2\gamma_{l} & = \gamma_{\omega},
\end{align*}
which, after simplification, yields the one-parameter family of scaling laws
\begin{equation}
  \gamma_{u} = -1 + \tfrac{1}{2} \gamma_{l},\qquad \gamma_{\omega} = -1,\qquad \gamma_{\psi} = -1 + 2\gamma_{l}.
  \label{eqn_ssim_constr}
\end{equation}
Table \ref{tab_ssim_fit_at} summarizes the scaling exponents estimated from the numerical data. It is clear from this table that the estimated exponents satisfy the relations \eqref{eqn_ssim_constr}.
\begin{table}[h]
  \centering
  \caption{Scaling exponents of the self-similar solution \eqref{eqn_ssim_fit_v}.}
  \label{tab_ssim_fit_at}
  \begin{tabular}{*{8}{>{$}c<{$}}}
    \toprule
    \text{Mesh size} & \hat{\gamma}_{l} & \hat{\gamma}_{u} & \hat{\gamma}_{\omega} & \hat{\gamma}_{\psi} & -1+\frac{1}{2} \hat{\gamma}_{l} & -1 + 2 \hat{\gamma}_{l} & \hat{\gamma}_{u} - \hat{\gamma}_{l} \\
    \midrule
    1024 \times 1024 & 2.7359 & 0.4614 & -0.9478 & 4.7399 & 0.3679 & 4.4717 & -2.2745 \\
    1280 \times 1280 & 2.9059 & 0.4629 & -0.9952 & 4.8683 & 0.4530 & 4.8118 & -2.4430 \\
    1536 \times 1536 & 2.9108 & 0.4600 & -0.9964 & 4.8280 & 0.4554 & 4.8215 & -2.4508 \\
    1792 \times 1792 & 2.9116 & 0.4602 & -0.9966 & 4.8294 & 0.4558 & 4.8232 & -2.4514 \\
    2048 \times 2048 & 2.9133 & 0.4604 & -0.9972 & 4.8322 & 0.4567 & 4.8266 & -2.4529 \\
    \bottomrule
  \end{tabular}
\end{table}%
In addition, the scaling exponent $\gamma_{l}$ satisfies $\gamma_{l} \geq \frac{2}{5}$, the minimum exponent for a blowup to occur in view of the conservation of energy \citep{constantin1994}. Since in our computations the velocity $u$ is observed to be
uniformly bounded, the scaling exponent $\gamma_{l}$ needs also to satisfy $\gamma_{l} \geq 1$ for a blowup to occur \citep{constantin1994}. It is clear from Table \ref{tab_ssim_fit_at} that this constraint is satisfied by our numerical data.

Finally, the fitting results shown in Table \ref{tab_ssim_fit_at} imply that
\begin{displaymath}
  \omega^{r} = -ru_{1,z} = O(t_{s}-t)^{-2.45},\quad \omega^{\theta} = r\omega_{1} = O(t_{s}-t)^{-1},\quad \omega^{z} = 2u_{1} + ru_{1,r} = O(t_{s}-t)^{-2.45},
\end{displaymath}
which confirms the scaling law $\norm{\omega(\cdot,t)}_{\infty} = O(t_{s}-t)^{-2.45}$ and hence the existence of a finite-time singularity.

\subsection{Understanding the Blowup}\label{ssec_intrp}
\begin{figure}[h]
  \centering
  \subfigure[local flow field]{
    \includegraphics[scale=0.415]{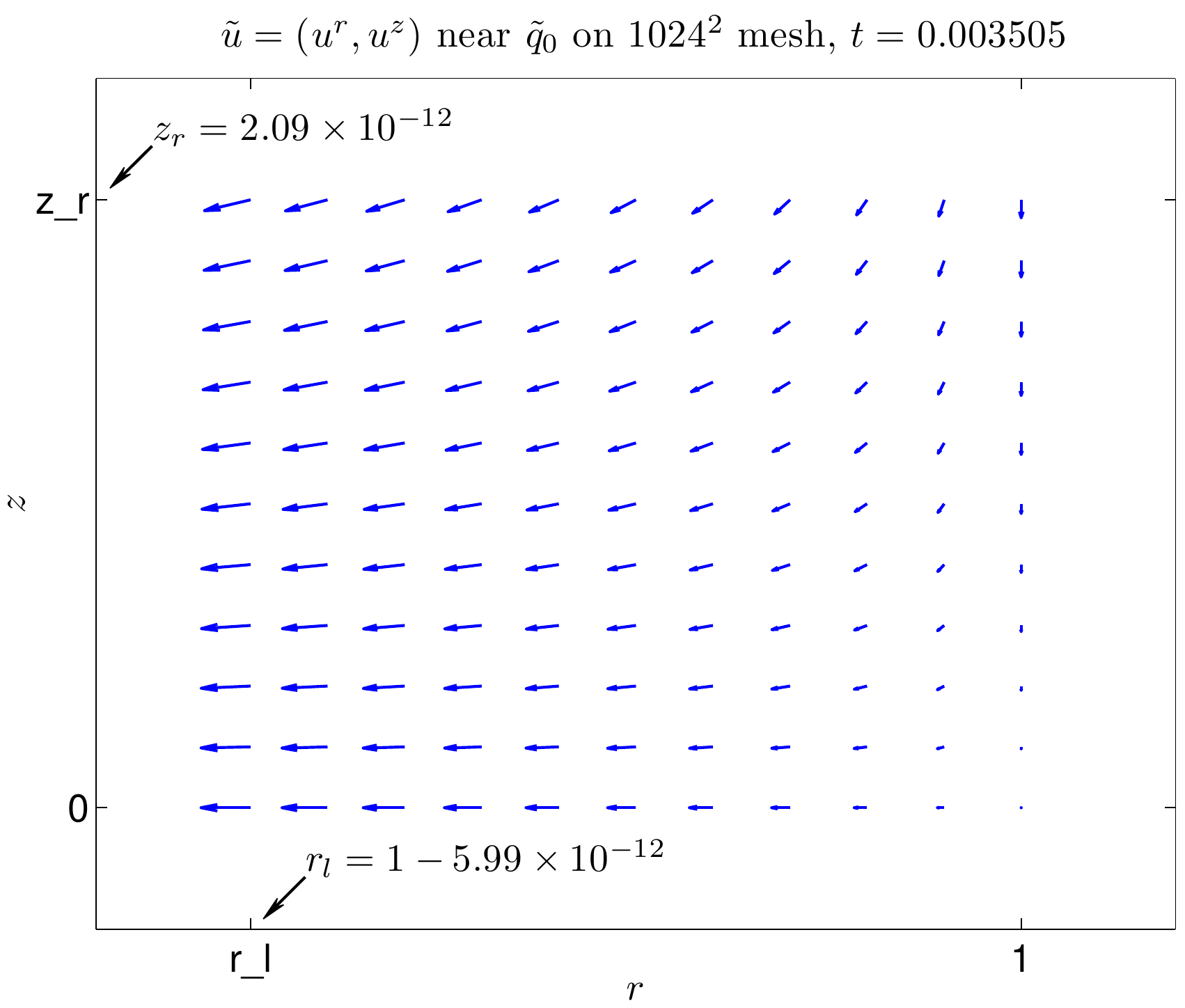}
    \label{fig_flow_loc}
  }\quad
  \subfigure[global vorticity dynamics]{
    \psfrag{z0}{\tiny $z=0$}
    \psfrag{zpL}{\tiny $z=\frac{1}{4} L$}
    \psfrag{zmL}{\tiny $z=-\frac{1}{4} L$}
    \includegraphics[scale=0.65]{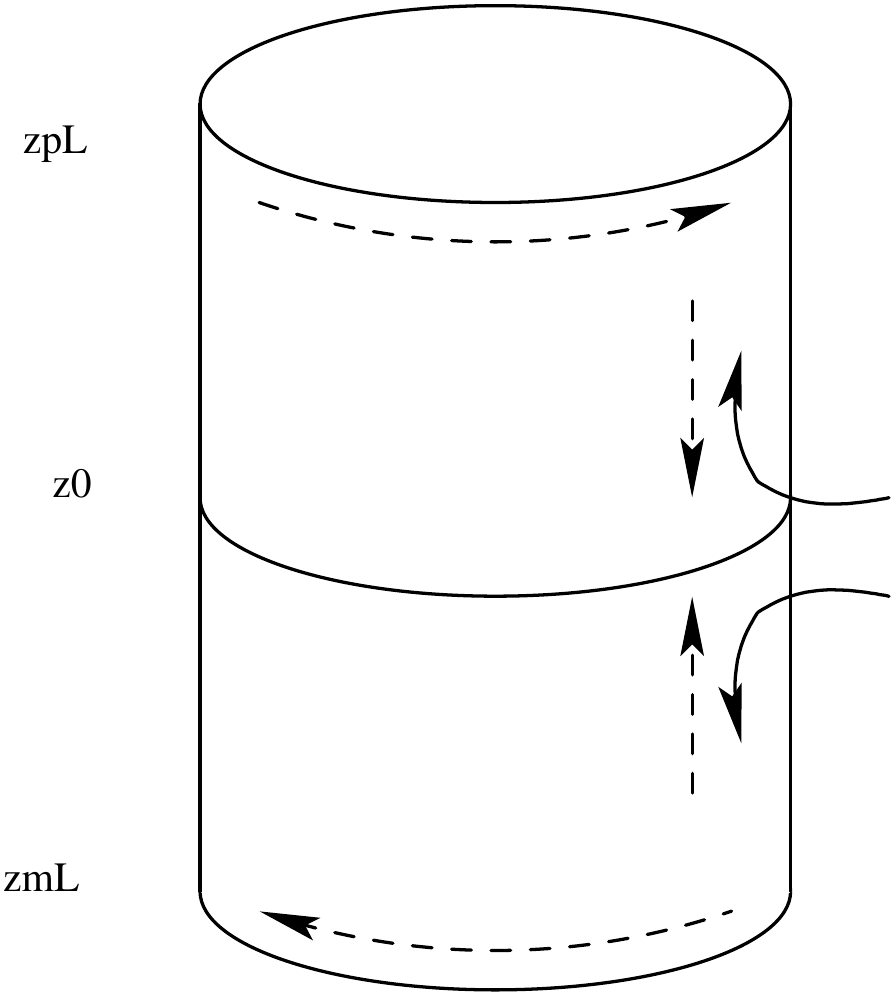}
    \label{fig_flow_glob}
  }
  \caption{Understanding the blowup: (a) Local velocity field near the point of the maximum vorticity. (b) Global vorticity kinematics of the 3D Euler singularity. The vortex lines (solid) end at the wall and are brought to sections of zero circulation
  by the axial flow (straight dashed lines). The curved dash lines indicate vortical circulation. See also Figure 5 in \citet{childress1987} and Figure 7 in \citet{childress2004}.}
  \label{fig_flow}
\end{figure}

For the specific initial data \eqref{eqn_eat_ic} considered in this paper, it is observed that the initial angular velocity $ru_{1}^{0}$ is monotonically increasing in both $r$ and $z$ within the quarter cylinder $D(1,\frac{1}{4} L)$. It turns out that
this property is preserved by the equations \eqref{eqn_eat} (for reasons yet to be determined), thus $u_{1,z}$ and consequently $\omega_{1}$ (see \eqref{eqn_eat_w}) remain positive for all times. The positivity of $\omega_{1}$ and the homogeneous
boundary condition of $\psi_{1}$ together imply the positivity of $\psi_{1}$ (see \eqref{eqn_eat_psi}), which in turn implies that
\begin{displaymath}
  u^{z} = 2\psi_{1} + r \psi_{1,r} = \psi_{1,r} \leq 0\quad \text{on}\quad r = 1,
\end{displaymath}
(indeed, by the strong maximum principle the strict inequality holds). This shows that the flow has a compression mechanism near the corner $\tilde{q}_{0} = (1,0)^{T}$ (Figure \ref{fig_flow}\subref{fig_flow_loc}; recall $u^{z}$ is odd at $z = 0$), which
seems to be responsible for the generation of the finite-time singularity observed at $\tilde{q}_{0}$. Indeed, as far as the formation of a singularity is concerned, the precise form of the initial data seems to be immaterial. As long as $ru_{1}^{0}$ has
the desired symmetry properties and is monotonically increasing in both $r$ and $z$ in the quarter cylinder $D(1,\frac{1}{4} L)$, the solution of the initial-boundary value problem \eqref{eqn_eat}--\eqref{eqn_eat_ibc} should develop a singularity in
finite-time, in much the same way as the solution described in this paper does.

From a physical point of view, the blowup can be deduced from vorticity kinematics applied to the initial flow. The gradient of circulation down the tube, $2\pi r u_{z}^{\theta}$, creates a $\theta$-component of vorticity (see \eqref{eqn_eat_w}). This
component in turn creates the flow $u^{r},\ u^{z}$ (see \eqref{eqn_eat_psi}--\eqref{eqn_eat_urz}) which is toward the symmetry plane $z = 0$ on the solid wall $r = 1$. Since vortex lines threading through the wall are carried by this flow, their points
of intersection with the wall move toward the symmetry plane $z = 0$ and then collapse onto $z = 0$ in finite time (see Figure \ref{fig_flow}\subref{fig_flow_glob}).
This is similar to what was observed by \citet{childress1987} in the study of a model problem, which was derived as the leading-order approximation to a stretched version of the Taylor-Green initial value problem for the 3D Euler equations. The model
closely resembles the axisymmetric Euler equations but the fluid inertia ($D_{t} u^{r}$) in the radial transport equation is missing. Since the variable $u^{\theta}$ studied in \citet{childress1987} occurs as coefficients in the asymptotic expansions,
the blowup of its $z$-derivatives merely indicates the breakdown of the expansions and the return of the flow to three-dimensionality. It does not imply the loss of regularity of the underlying solutions.

\begin{figure}[h]
  \centering
  \includegraphics[scale=0.415]{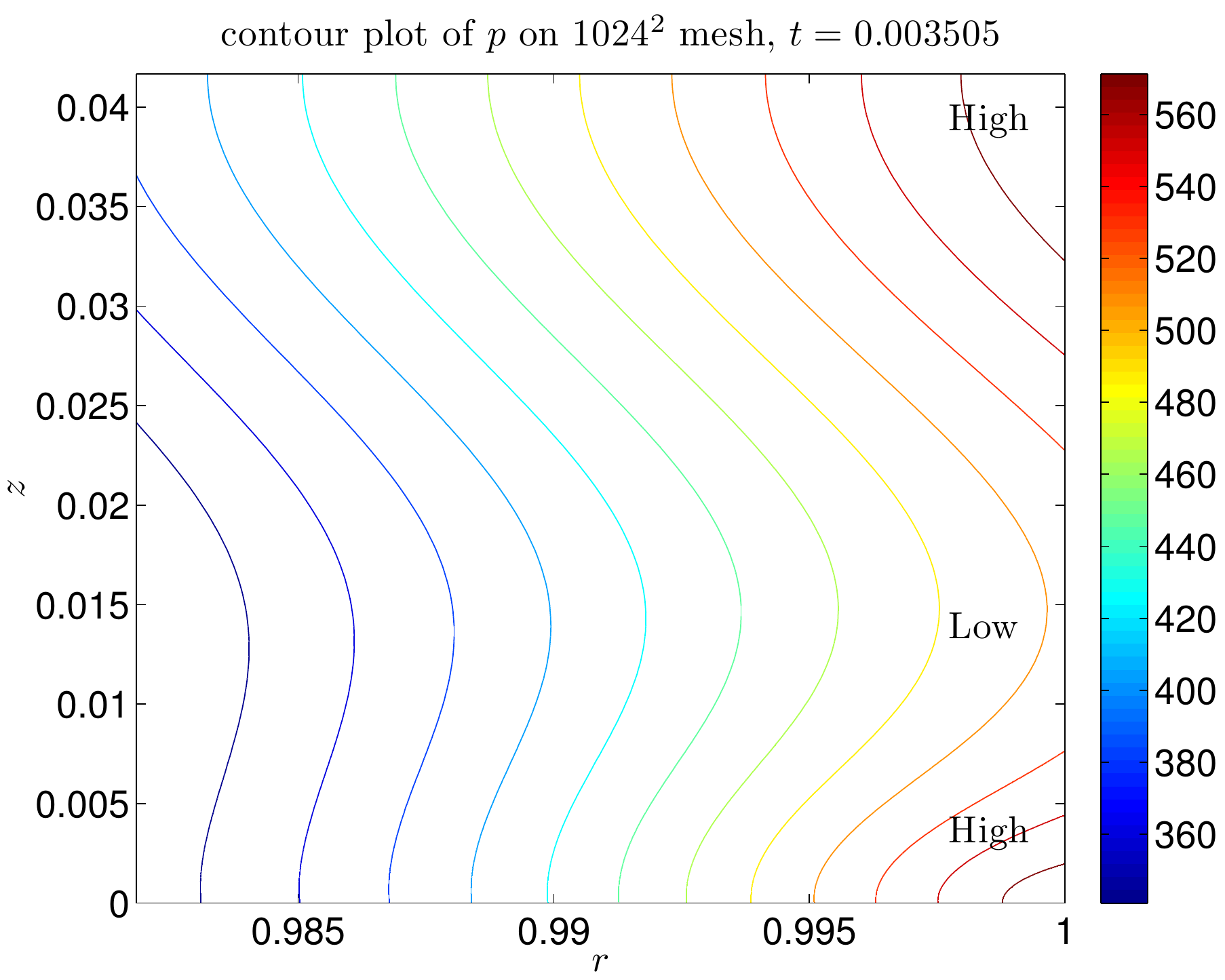}
  \caption{The contour plot of the pressure $p$ near the location of the maximum vorticity at $t = 0.003505$. Note the unfavorable axial pressure gradient near $z = 0$.}
  \label{fig_cont_p_4}
\end{figure}

Despite the apparent similarity between our computations and the model studied by \citet{childress1987}, there is a fundamental difference between the two scenarios. More precisely, in \citet{childress1987}, it was observed that the absence of radial
momentum transfer creates a \emph{favorable} pressure gradient, which sets up an axial flow near the solid wall toward the symmetry plane $z = 0$. In our case, however, it is observed that the pressure gradient near the solid wall $r = 1$ and the
symmetry plane $z = 0$ is \emph{unfavorable} in the sense that it tends to push fluids \emph{away} from $z = 0$ (Figure \ref{fig_cont_p_4}). Thus unlike the scenario described in \citet{childress1987}, it must be the radial fluid inertia, not the
pressure gradient, that is responsible for the finite-time blowup observed at the corner $\tilde{q}_{0} = (1,0)^{T}$.

\section{Conclusion and Future Work}\label{sec_conclu}
In this paper, we have numerically studied the 3D Euler equations in axisymmetric geometries and have discovered a class of potentially singular solutions from carefully chosen initial data. By using a specially designed yet highly effective adaptive
mesh, we have resolved the nearly singular solution with high accuracy and have advanced the solution to a point asymptotically close to the predicted singularity time. Detailed analysis based on the blowup/non-blowup criteria of Beale-Kato-Majda,
Constantin-Fefferman-Majda, and Deng-Hou-Yu provides convincing evidence for the existence of a singularity. Local analysis also suggests the existence of a self-similar structure in the blowing-up solution.

Our computations not only lead to potential resolution of one of the greatest open problems in mathematical fluid dynamics concerning the finite-time blowup of the 3D Euler equations, but also lead to potential resolution of a related open problem
concerning the global regularity of the 2D Boussinesq equations. The Boussinesq equations describe the motion of variable-density, stratified flows under the influence of gravitational forces, and are known to be qualitatively similar to the 3D
axisymmetric Euler equations away from the symmetry axis. Since the singularity discovered in our Euler computations lies on the solid boundary of the cylinder, the solution of the 2D Boussinesq equations resulting from similar initial data is expected
to develop a singularity in finite time. This has been confirmed in a separate computation and is the subject of a forthcoming paper.

Motivated by the observation that the Euler/Boussinesq singularity is likely a consequence of a compression flow along the solid wall, we have derived a 1D model
\begin{subequations}\label{eqn_eat_1d}
\begin{align}
  u_{t} + v u_{z} & = 0,\qquad z \in (0,L), \label{eqn_eat_1d_u} \\
  \omega_{t} + v \omega_{z} & = u_{z}, \label{eqn_eat_1d_w}
\end{align}
with the nonlocal, zero-mean velocity $v$ defined by
\begin{equation}
  v_{z}(z) = H\omega(z) = \frac{1}{L} \PV\int_{0}^{L} \omega(y) \cot \bigl[ \mu (z-y) \bigr]\,dy,\qquad \mu = \pi/L.
  \label{eqn_eat_1d_vz}
\end{equation}
\end{subequations}
This 1D model can be viewed as the ``restriction'' of the 3D axisymmetric Euler equations \eqref{eqn_eat} to the wall $r = 1$, with the identification
\begin{displaymath}
  u(z) \sim u_{1}^{2}(1,z),\qquad \omega(z) \sim \omega_{1}(1,z),\qquad v(z) \sim \psi_{1,r}(1,z).
\end{displaymath}
With the local approximation of the Poisson equation \eqref{eqn_eat_psi}:
\begin{displaymath}
  -\bigl[ \partial_{r}^{2} + \partial_{z}^{2} \bigr] \psi_{1} = \omega_{1}(1,z),
\end{displaymath}
and the matching condition $\limsup\limits_{r \to -\infty} \psi_{1} \leq C$, it can be shown that
\begin{displaymath}
  \psi_{1,rz}(1,z) = H\omega_{1}(1,z).
\end{displaymath}
This is precisely equation \eqref{eqn_eat_1d_vz} which provides the key relation needed to close \eqref{eqn_eat_1d_u}--\eqref{eqn_eat_1d_w}.

\begin{figure}[h]
  \centering
  \subfigure[$\log(\log \norm{(u^{1/2})_{z}}_{\infty})$ vs. $\log(\log \norm{\omega}_{\infty})$]{
    \includegraphics[scale=0.415]{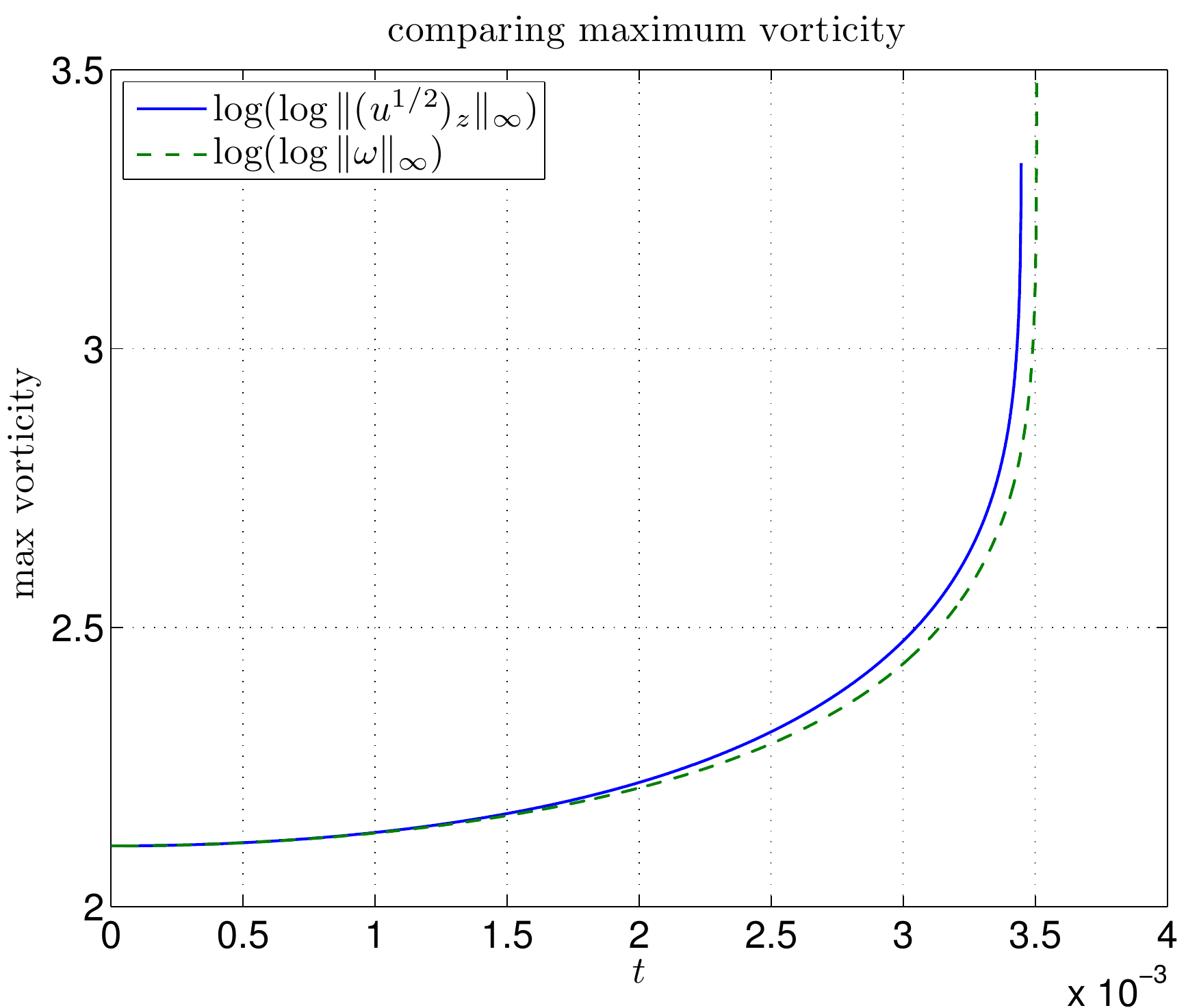}
  }\quad
  \subfigure[$u^{1/2}(z)$ vs. $u_{1}(1,z)$]{
    \includegraphics[scale=0.415]{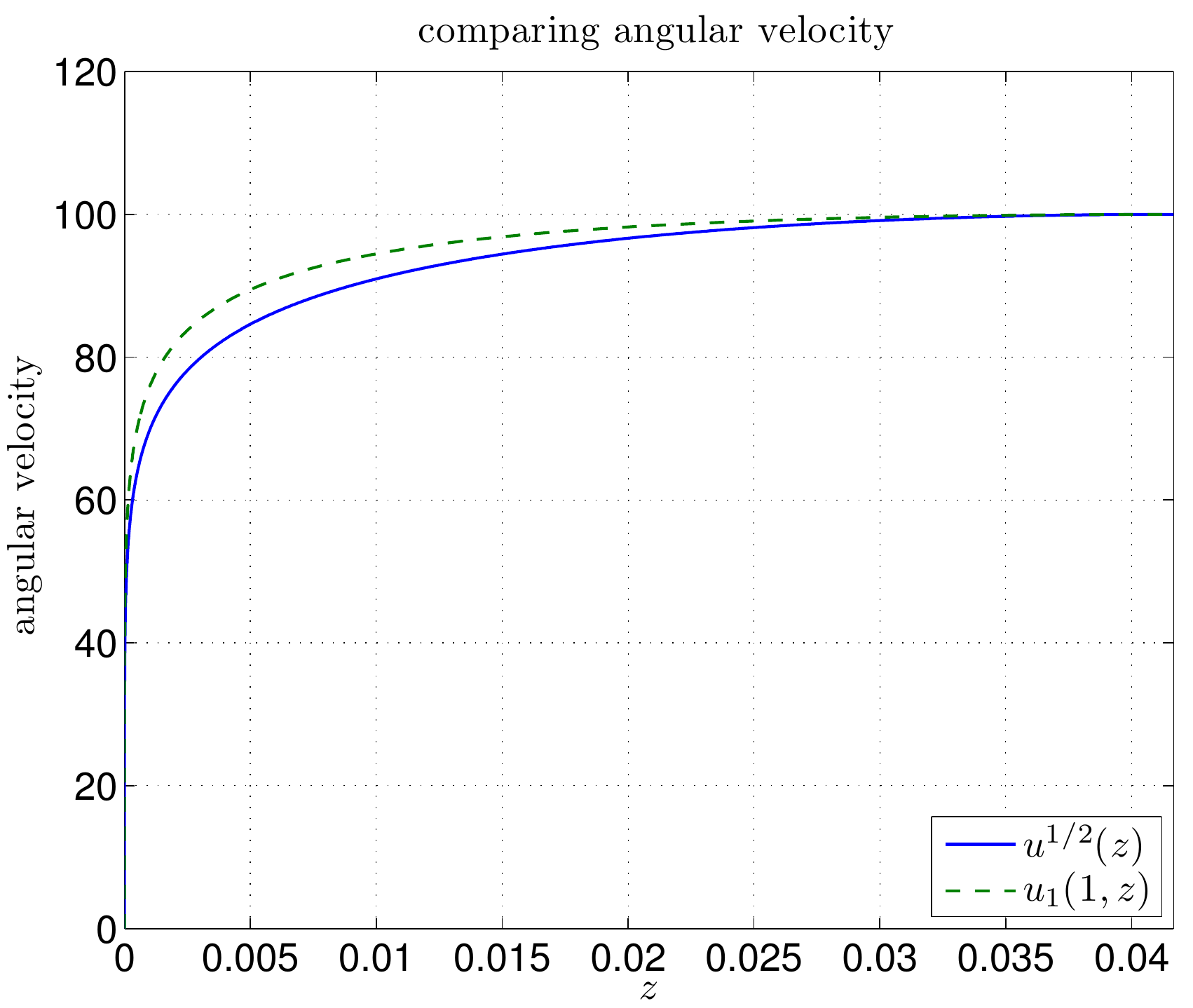}
  }
  \subfigure[$\omega(z)$ vs. $\omega_{1}(1,z)$]{
    \includegraphics[scale=0.415]{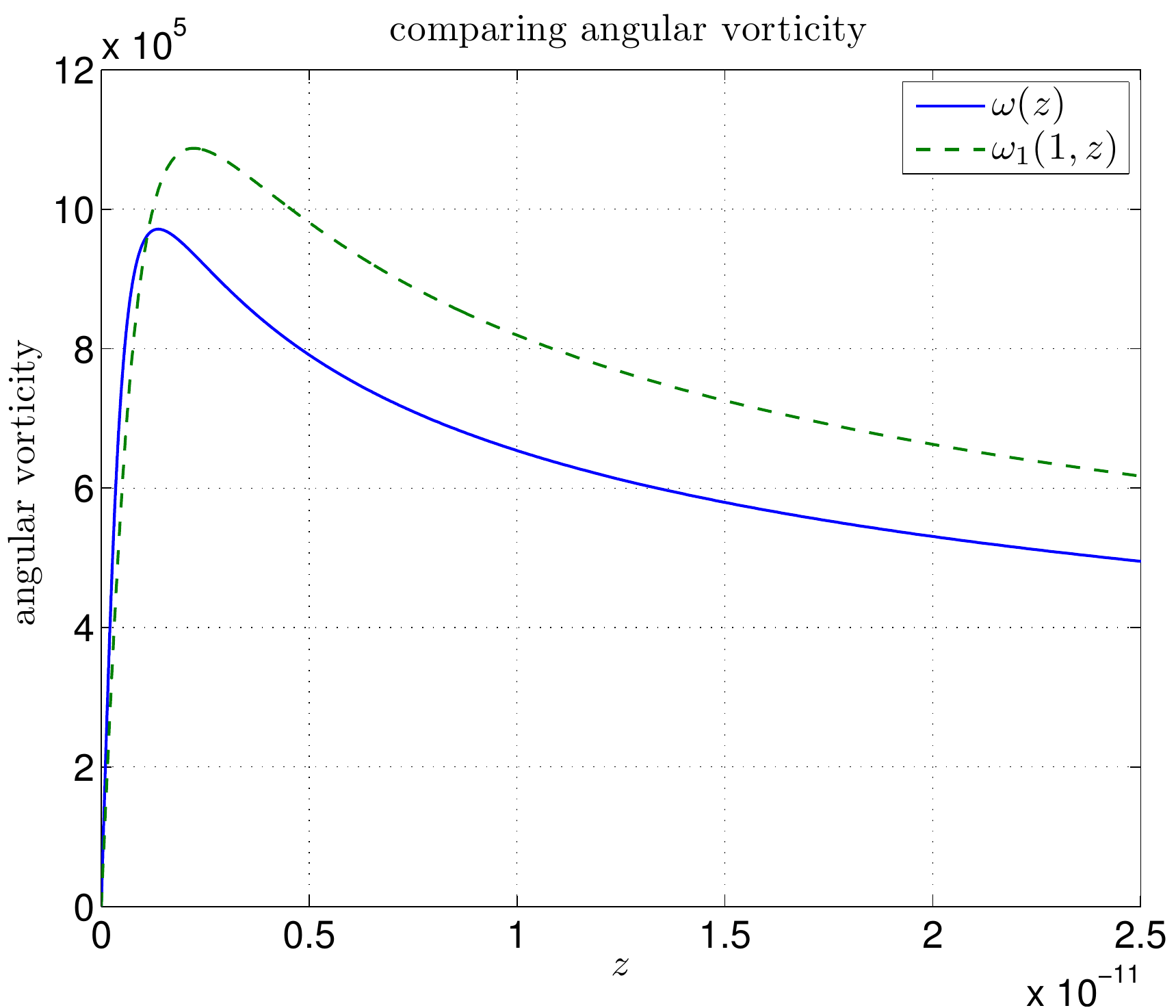}
  }\quad
  \subfigure[$v(z)$ vs. $\psi_{1,r}(1,z)$]{
    \includegraphics[scale=0.415]{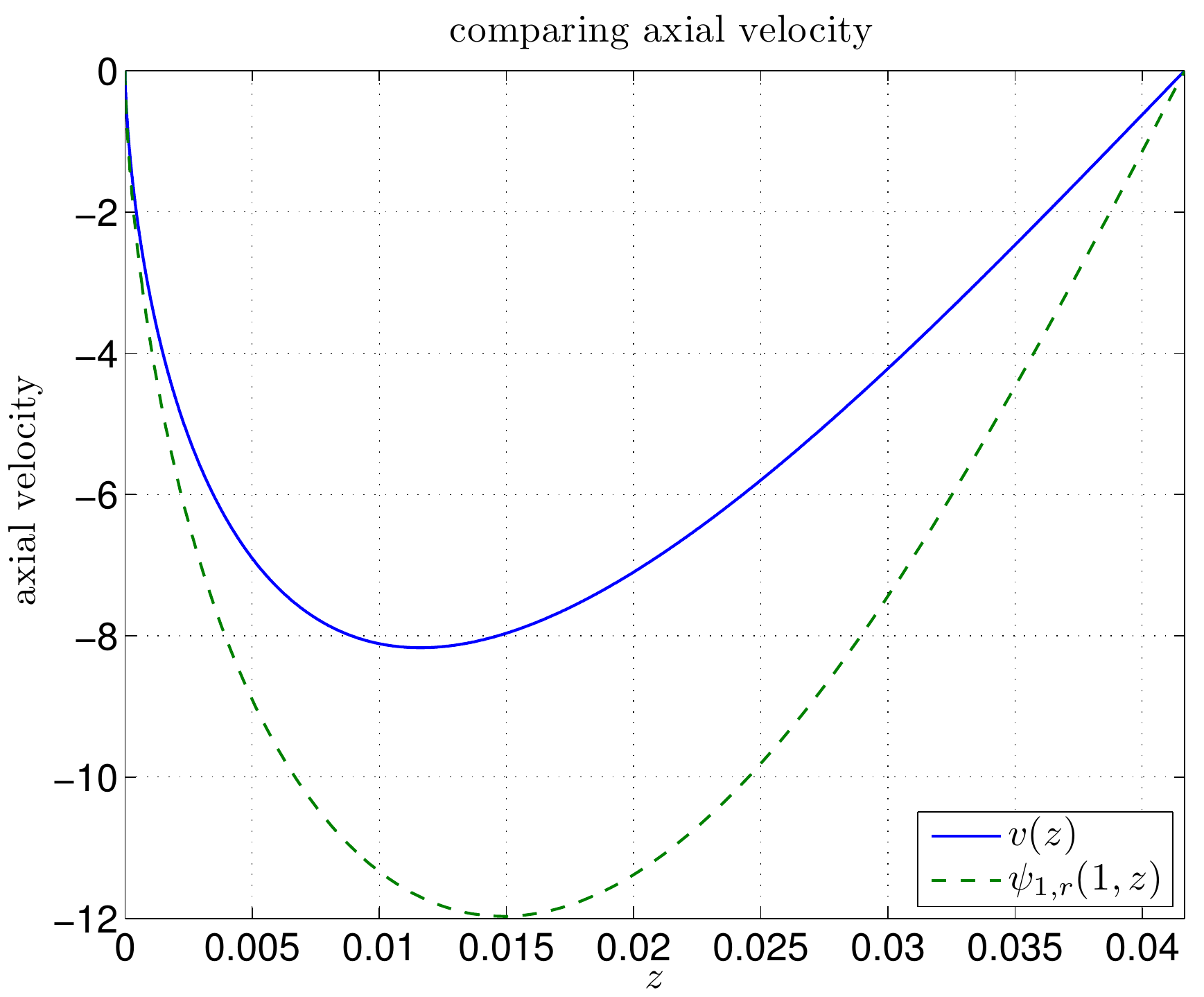}
  }
  \caption{Comparison of numerical solutions of the 1D model \eqref{eqn_eat_1d} with those of the 3D axisymmetric Euler equations \eqref{eqn_eat}: (a) maximum vorticity, (b) angular velocity, (c) angular vorticity, and (d) axial velocity. In all the
  plots the solution of the 1D model is computed at $t = 0.003447$ and that of the 3D Euler is computed at $t = 0.003505$.}
  \label{fig_cmp_1d}
\end{figure}

We have numerically solved problem \eqref{eqn_eat_1d} with the initial data
\begin{displaymath}
  u^{0}(z) = 10^{4} \sin^{2} \Bigl( \frac{2\pi}{L} z \Bigr),\qquad \omega^{0}(z) = 0,
\end{displaymath}
and discovered that the solution develops a singularity in $(u^{1/2})_{z}$ in finite-time, in much the same way as the solution to the 3D Euler equations \eqref{eqn_eat} does (Figure \ref{fig_cmp_1d}). The details of these computations, as well as the
analysis on the well-posedness and finite-time blowup of the 1D model \eqref{eqn_eat_1d}, will be reported in a forthcoming paper \citep{hl2013}.

\section*{Acknowledgement}
The authors would like to gratefully acknowledge the computing resources provided by the SHC cluster at Caltech Center for Advanced Computing Research (CACR) and the Brutus cluster at ETH Z\"{u}rich (ETHZ). The authors gratefully acknowledge the
excellent support provided by the staff members at SHC, especially Sharon Brunett, and the support provided by Prof. Petros Koumoutsakos at ETHZ, who kindly allowed us to use his computing resources. This research was supported in part by an NSF FRG
Grant DMS-1159138 and a DOE Grant DE-FG02-06ER25727.

\appendix
\section{Construction of the Adaptive Mesh}\label{app_mm}
The mesh mapping functions $r(\rho),\ z(\eta)$ are defined via an analytic function $\mu$,
\begin{displaymath}
  r(\rho) = \mu(\rho;\alpha^{r},\sigma^{r}),\qquad z(\eta) = \mu(\eta;\alpha^{z},\sigma^{z}),
\end{displaymath}
where $\alpha^{r},\ \sigma^{r}$ etc. are parameters and
\begin{equation}
  \mu_{s}(s;\alpha,\sigma) = \alpha_{0} + \alpha_{1} \re^{-\pi s^{2}/\sigma_{1}^{2}} + \alpha_{2} \re^{-\pi (s-1)^{2}/\sigma_{2}^{2}},\qquad s \in [0,1].
  \label{eqn_mm}
\end{equation}
The particular form of the function $\mu$ is chosen to meet the following goals. First, it should map the interval $[0,1]$ onto another interval, say $[0,L]$, in a one-to-one manner. Second, given any subset $[a,b]$ of $[0,L]$ and any $\delta \in (0,1)$,
it should place at least $\delta$-fraction of the mesh points in $[a,b]$ and maintain a uniform mesh on $[a,b]$. In our computations, the interval $[0,L]$ will be the entire computational domain along either the $r$- or the $z$-dimension, and $[a,b]$ a
small neighborhood of the maximum vorticity along that dimension. The mesh mapping functions constructed this way will always place enough points near the maximum vorticity where the solution is most singular and where resolution is most needed.

The one-to-one correspondence of the map generated by $\mu$ is equivalent to the positivity of $\mu_{s}$, which can be ensured provided that $\alpha_{0} > 0$ and $\alpha_{1},\, \alpha_{2} \geq 0$. To place the required amount of mesh points in the
interval $[a,b]$ and ensure a uniform mesh on $[a,b]$, we observe that
\begin{displaymath}
  \mu_{s}(s;\alpha,\sigma) = \alpha_{0} + \alpha_{1} \re^{-\pi s^{2}/\sigma_{1}^{2}} + \alpha_{2} \re^{-\pi (s-1)^{2}/\sigma_{2}^{2}} \approx \alpha_{0},\qquad s \in [2\sigma_{1},1-2\sigma_{2}],
\end{displaymath}
in view of the rapid decay of the Gaussians away from their centers. Therefore, if we choose $(\sigma_{1},\sigma_{2})$ such that $1-2\sigma_{1}-2\sigma_{2} = \delta$ and map the interval $[2\sigma_{1},1-2\sigma_{2}]$ onto $[a,b]$, the resulting mesh will
have the desired properties.

We remark that there are other ways to construct adaptive meshes with similar point concentration properties, for example the popular sine transform
\begin{align*}
  \mu(s) = s + \frac{\alpha}{\pi} \sin(\pi s),&\qquad s \in [0,1],\ \alpha \in (-1,1), \\
  \intertext{the Chebyshev (cosine) transform}
  \mu(s) = \cos(s),&\qquad s \in [0,\pi], \\
  \intertext{and the ``negative Gaussian'' transform}
  \mu(s) = s - \alpha \re^{-(s-s_{0})^{2}/\sigma^{2}},&\qquad s \in [0,1],\ \alpha > 0.
\end{align*}
The drawback of these ``traditional'' mapping functions is that the resulting mesh has unlimited resolution at only a \emph{single} point, and this point must be one of the end points in the case of sine/cosine transforms. This is inadequate when
unlimited resolution is demanded in an \emph{interval}, such as in our case, and the mapping function proposed in \eqref{eqn_mm} solves this problem.

The mapping function $\mu$ defined by \eqref{eqn_mm} is constructed using the following procedure. First, the parameters $(\sigma_{1},\sigma_{2})$, which specify the amount of points to be distributed to the intervals $[0,a]\ (2\sigma_{1}),\ [a,b]\
(1-2\sigma_{1}-2\sigma_{2})$, and $[b,L]\ (2\sigma_{2})$, are supplied by the users and are fixed throughout the computations. To ensure a meaningful distribution, these parameters must satisfy
\begin{subequations}\label{eqn_mm_constr}
\begin{equation}
  0 < \sigma_{1}, \sigma_{2} < \tfrac{1}{4}.
  \label{eqn_mm_constr_sigma}
\end{equation}
Next, the parameters $(\alpha_{0},\alpha_{1},\alpha_{2})$ are determined from the equations
\begin{equation}
  \mu(0) = 0,\qquad \mu(2\sigma_{1}) = a,\qquad \mu(1-2\sigma_{2}) = b,\qquad \mu(1) = L,
  \label{eqn_mm_constr_gamma}
\end{equation}
\end{subequations}
which ensure that $[0,1]$ is mapped onto $[0,L]$ and $[2\sigma_{1},1-2\sigma_{2}]$ is mapped onto $[a,b]$. If $(\sigma_{1},\sigma_{2})$ are reasonably small, as we shall assume in what follows, \eqref{eqn_mm_constr_gamma} may be replaced by the
approximate equations
\begin{align*}
  2\sigma_{1} \alpha_{0} + \tfrac{1}{2} \sigma_{1} \alpha_{1} & = a, \\
  (1-2\sigma_{2}) \alpha_{0} + \tfrac{1}{2} \sigma_{1} \alpha_{1} & = b, \\
  \alpha_{0} + \tfrac{1}{2} \sigma_{1} \alpha_{1} + \tfrac{1}{2} \sigma_{2} \alpha_{2} & = L,
\end{align*}
which can be readily solved to give
\begin{equation}
  \alpha_{0} = \frac{b-a}{1-2\sigma_{1}-2\sigma_{2}},\qquad \alpha_{1} = \frac{2}{\sigma_{1}}\, (a-2\sigma_{1} \alpha_{0}),\qquad \alpha_{2} = \frac{2}{\sigma_{2}}\, (L-b-2\sigma_{2} \alpha_{0}).
  \label{eqn_mm_alpha}
\end{equation}
Note that $\alpha_{0} > 0$ since $b > a$ (by construction) and $\sigma_{1}+\sigma_{2} < \frac{1}{2}$ (by constraint \eqref{eqn_mm_constr_sigma}). If $\alpha_{1}$ and $\alpha_{2}$ as given by \eqref{eqn_mm_alpha} are both nonnegative, then a unique,
strictly increasing mesh mapping function $\mu$ satisfying \eqref{eqn_mm_constr} results. If not, then the values of $\alpha_{i}$'s need to be adjusted to maintain the strict monotonicity of $\mu$. Consider first the case where $\alpha_{1}$ as given by
\eqref{eqn_mm_alpha} is negative. In this case the left end point $a$ of the ``singularity interval'' $[a,b]$ is too close to $\mu = 0$ (so close that $a < 2\sigma_{1} \alpha_{0}$), and the interval $[0,a]$ must be merged with $[a,b]$ to form a larger
singularity interval $[0,b]$. The mesh mapping function is modified accordingly by setting $\alpha_{1} = 0$ in \eqref{eqn_mm}:
\begin{displaymath}
  \mu_{s}(s;\alpha,\sigma) = \alpha_{0} + \alpha_{2} \re^{-\pi (s-1)^{2}/\sigma_{2}^{2}},\qquad s \in [0,1],
\end{displaymath}
and the values of $\alpha_{0},\ \alpha_{2}$ are recomputed from the constraints:
\begin{align*}
  \mu(1-2\sigma_{2}) = b,\qquad \mu(1) = L.
\end{align*}
Replacing these equations by
\begin{align*}
  (1-2\sigma_{2}) \alpha_{0} & = b, \\
  \alpha_{0} + \tfrac{1}{2} \sigma_{2} \alpha_{2} & = L,
\end{align*}
and solving for $\alpha_{0},\ \alpha_{2}$ yields
\begin{displaymath}
  \alpha_{0} = \frac{b}{1-2\sigma_{2}} > 0,\qquad \alpha_{2} = \frac{2}{\sigma_{2}}\, (L-\alpha_{0}).
\end{displaymath}
If $\alpha_{2}$ computed this way is still negative, then the right end point $b$ of the (extended) singularity interval is too close to $\mu = L$ (so close that $(1-2\sigma_{2}) L < b$), and the interval $[b,L]$ must be merged with $[0,b]$ to form a
larger singularity interval $[0,L]$. In this case the mesh mapping function simply takes the form
\begin{displaymath}
  \mu_{s}(s;\alpha,\sigma) = \alpha_{0} = L,\qquad s \in [0,1],
\end{displaymath}
and the adaptive mesh reduces to a uniform mesh. The case where $\alpha_{1} \geq 0,\ \alpha_{2} < 0$ in \eqref{eqn_mm_alpha} can be handled in a similar way.

\section{Construction of the B-Spline Subspace}\label{app_bsp}
Consider the finite-dimensional subspace of weighted uniform B-splines of even order $k$:
\begin{displaymath}
  V_{h} := V_{w,h}^{k} = \text{span} \Bigl\{ w(\rho) b_{j,h_{r}}^{k}(\rho) b_{i,h_{z}}^{k}(\eta) \Bigr\} \cap V,
\end{displaymath}
where $w(\rho) = 1-\rho^{2},\ h_{r} = 1/N,\ h_{z} = 1/M$, and
\begin{align*}
  V & = \text{span} \Bigl\{ \phi \in H^{1}[0,1]^{2}\colon \phi(-\rho,\eta) = \phi(\rho,\eta), \\
  &\hspace{1.56in} \phi(1,\eta) = 0,\ \phi(\rho,\ell-\eta) = -\phi(\rho,\ell+\eta),\ \forall \ell \in \mathbb{Z} \Bigr\}.
\end{align*}
The functions $b_{\ell,h}^{k}(s) = b^{k}((s/h) - (\ell-k/2))$ are shifted and rescaled uniform B-splines of order $k$ where $b^{k}$, the ``reference'' uniform B-splines, satisfy the recursion (H\"{o}llig 2003)
\begin{align*}
  b^{1}(s) & = \chi_{[0,1)}(s) =
  \biggl\{\! \begin{array}{ll}
    1, & \text{if $0 \leq s < 1$} \\
    0, & \text{otherwise}
  \end{array}, \\
  b^{k}(s) & = \int_{s-1}^{s} b^{k-1}(\tau)\,d\tau,\qquad k \geq 2.
\end{align*}
A basis of the subspace $V_{h}$ can be conveniently chosen as
\begin{displaymath}
  B_{w,i,j}(\rho,\eta) := w(\rho) B_{j}(\rho) B_{i}(\eta),\qquad 1 \leq i \leq M-1,\ 0 \leq j \leq N+k/2-1,
\end{displaymath}
where
\begin{displaymath}
  B_{j}(\rho) = \frac{b_{j,h_{r}}^{k}(\rho) + b_{j,h_{r}}^{k}(-\rho)}{1+\delta_{j0}},\qquad B_{i}(\eta) = \sum_{\ell=-\infty}^{\infty} \bigl[ b_{i,h_{z}}^{k}(2\ell+\eta) - b_{i,h_{z}}^{k}(2\ell-\eta) \bigr].
\end{displaymath}
If we write
\begin{displaymath}
  \psi_{h}(\rho,\eta) = \sum_{i,j} c_{ij} B_{w,i,j}(\rho,\eta),
\end{displaymath}
then the finite-dimensional variational problem
\begin{displaymath}
  a(\psi_{h},\phi_{h}) = f(\phi_{h}),\qquad \forall \phi_{h} \in V_{h},
\end{displaymath}
can be transformed to an equivalent linear system $Ax = b$, which in component form reads
\begin{displaymath}
  \sum_{i,j} a(B_{w,i,j},B_{w,m,n}) c_{ij} = f(B_{w,m,n}).
\end{displaymath}
In our computations, the entries of $A,\ b$ are approximated using composite 6-point Gauss-Legendre quadrature rules. This essentially reproduces the exact values of the stiffness matrix $A$, and hence ensures the \emph{uniform $V_{h}$-ellipticity} of
the approximate bilinear forms and the convergence of the discrete approximations \citep{ciarlet2002}. The large sparse linear system resulting from the above discretization is solved using the PaStiX
package\footnote{\texttt{https://gforge.inria.fr/projects/pastix}}, a parallel sparse direct solver based on the super-nodal (left-looking) method \citep{hrr2002}.

\section{Description of the Test Problem}\label{app_test}
The finite-element, finite-difference hybrid adaptive method described in Section \ref{sec_method} is applied to a forced axisymmetric Euler system:
\begin{subequations}\label{eqn_eat_f}
\begin{align}
  u_{1,t} + u^{r} u_{1,r} + u^{z} u_{1,z} & = 2 u_{1} \psi_{1,z} + F^{u}, \label{eqn_eat_f_u} \\
  \omega_{1,t} + u^{r} \omega_{1,r} + u^{z} \omega_{1,z} & = (u_{1}^{2})_{z} + F^{\omega}, \label{eqn_eat_f_w} \\
  -\bigl[ \partial_{r}^{2} + (3/r) \partial_{r} + \partial_{z}^{2} \bigr] \psi_{1} & = \omega_{1}, \label{eqn_eat_f_psi}
\end{align}
\end{subequations}
where the forcing terms $F^{u},\ F^{\omega}$ are generated from a smooth test solution:
\begin{subequations}\label{eqn_test_sol}
\begin{align}
  \tilde{u}_{1}(r,z,t) & = \xi(r,T-t) \sin \bigl[ \tfrac{1}{2} \pi \zeta(z,T-t) \bigr], \label{eqn_test_sol_u} \\
  \tilde{\psi}_{1}(r,z,t) & = 30\, (1-r^{2}) \xi(r,T-t) \sin \bigl[ \pi \zeta(z,T-t) \bigr], \label{eqn_test_sol_w} \\
  \tilde{\omega}_{1}(r,z,t) & = -\bigl[ \partial_{r}^{2} + (3/r) \partial_{r} + \partial_{z}^{2} \bigr] \tilde{\psi}_{1}(r,z,t). \label{eqn_test_sol_psi}
\end{align}
\end{subequations}
The solution as given by \eqref{eqn_test_sol} develops a singularity at a finite time $T$ with locally self-similar profiles determined by the functions $\xi,\ \zeta$, which in our case are chosen to be
\begin{displaymath}
  \xi(r,t) = t^{2} \exp \biggl( -\frac{1-r^{2}}{10 t^{2}} \biggr),\qquad \zeta(z,t) = \tanh \biggl( \frac{2z}{5L t^{2}} \biggr).
\end{displaymath}
The velocity component $\tilde{u}_{1}$ of the test solution contains a sharp front near $\tilde{q}_{0} = (1,0)^{T}$, which would become a shock with finite strength at $t = T$ if the scaling factor $t^{2}$ in $\xi(r,t)$ is absent (this scaling factor is
introduced to mitigate the stiff forcing terms $F^{u},\ F^{\omega}$). Meanwhile, the vorticity component $\tilde{\omega}_{1}$ contains a sharp peak propagating toward $\tilde{q}_{0}$, which would blow up at $t = T$ without the $t^{2}$ factor. This
particular test solution closely resembles the behavior of the potentially singular Euler solution computed from \eqref{eqn_eat}--\eqref{eqn_eat_ibc}, and it provides an excellent benchmark on the performance of the numerical method described in Section
\ref{sec_method}.

The forced system \eqref{eqn_eat_f} is complemented with the initial data:
\begin{displaymath}
  u_{1}^{0}(r,z) = \tilde{u}_{1}(r,z,0),\qquad \omega_{1}^{0}(r,z) = \tilde{\omega}_{1}(r,z,0),\qquad \psi_{1}^{0}(r,z) = \tilde{\psi}_{1}(r,z,0),
\end{displaymath}
and boundary conditions \eqref{eqn_eat_bc_z}--\eqref{eqn_eat_bc_r}. It is solved with $T = 0.03,\ L = \frac{1}{6}$ on the quarter cylinder $D(1,\frac{1}{24})$ to $t = 0.029$, at which time the errors are computed and reported in Table \ref{tab_test_err}.

\section{Derivation of the Lower Bound for $K_{\rho}(q_{0})$}\label{app_k_lbd}
Consider the quantity
\begin{displaymath}
  K_{\rho}(q_{0},t) = \int_{\abs{y} \leq \rho} \abs{D(\hat{y},\xi(q_{0}+y,t),\xi(q_{0},t))} \cdot \abs{\omega(q_{0}+y,t)}\,\frac{dy}{\abs{y}^{3}}.
\end{displaymath}
To obtain a lower bound for the above integral, we consider the set $N_{0,t} = V_{0,t} - q_{0}$ where
\begin{gather*}
  V_{0,t} = \Bigl\{ (x_{1},x_{2},x_{3}) \in \mathbb{R}^{3}\colon ({\textstyle \sqrt{x_{1}^{2}+x_{2}^{2}}},x_{3}) \in D_{\infty}(t),\ \tan^{-1} (x_{2}/x_{1}) \in (-d_{1},0) \Bigr\}, \\
  d_{1} = d_{1}(D_{\infty}(t)) = \min_{\tilde{y} \in C_{\infty}(t)} \abs{\tilde{y} - \tilde{q}_{0}},\qquad C_{\infty}(t) = \Bigl\{ (r,z) \in D(1,\tfrac{1}{4} L)\colon \abs{\omega(r,z,t)} = \tfrac{1}{2} \norm{\omega(\cdot,t)}_{\infty} \Bigr\}.
\end{gather*}
In words, $V_{0,t}$ is the ``cylindrical shell'' obtained by rotating the set $D_{\infty}(t)$ about the symmetry axis $r = 0$, starting from the angle $\theta = 0$ and ending at the angle $\theta = -d_{1}$. Since the diameter of $D_{\infty}(t)$ shrinks
rapidly to 0 as $t$ approaches $t_{s}$, we deduce that $N_{0,t} \subseteq B_{\rho}(0)$ for $t$ sufficiently close to $t_{s}$, and hence (recall $\abs{\omega} \geq \frac{1}{2} \norm{\omega}_{\infty}$ on $D_{\infty}(t)$)
\begin{displaymath}
  K_{\rho}(q_{0},t) \geq \frac{1}{2} \norm{\omega(\cdot,t)}_{\infty} \int_{N_{0,t}} \abs{D(\hat{y},\xi(q_{0}+y,t),\xi(q_{0},t))}\,\frac{dy}{\abs{y}^{3}}.
\end{displaymath}
To continue, we observe that
\begin{displaymath}
  \omega^{\theta} = r\omega_{1} = 0,\quad \omega^{z} = 2u_{1} + ru_{1,r} = 0\qquad \text{on}\qquad z = 0,
\end{displaymath}
due to the odd symmetry of $u_{1},\ \omega_{1}$ at $z = 0$ (see Section \ref{sec_eqn}). This means that the direction vectors $\xi(q_{0}+y)$, when restricted to the plane $z = 0$, all point in the radial direction $-(\cos\theta,\sin\theta,0)^{T}$ and
hence are closely aligned with $\xi(q_{0}) = (-1,0,0)^{T}$ provided that $\abs{\theta} \ll 1$. Consequently, $\abs{D}$ is small near the intersection of the planes $z = 0$ and $\theta = 0$. In addition, for a point $x = (\cos\theta,\sin\theta,z)^{T}$
lying on the solid wall $r = 1$, the vector $y = x-q_{0}$ satisfies
\begin{displaymath}
  \hat{y} \cdot \xi(q_{0}) = \frac{(\cos\theta-1,\sin\theta,z)}{\abs{(\cos\theta-1,\sin\theta,z)}} \cdot (-1,0,0) = \frac{1-\cos\theta}{\sqrt{2-2\cos\theta+z^{2}}} \approx \frac{\theta^{2}}{2\sqrt{\theta^{2}+z^{2}}},
\end{displaymath}
provided that $\abs{\theta} \ll \min\{\abs{z},1\}$. This shows that $\hat{y} \cdot \xi(q_{0})$ and hence $\abs{D}$ is small near the intersection of the wall $r = 1$ and the plane $\theta = 0$. Motivated by these observations, we choose to estimate
$\abs{D}$ on the set $N_{1,t} = V_{1,t}-q_{0} \subseteq N_{0,t}$ where
\begin{gather*}
  V_{1,t} = \Bigl\{ (x_{1},x_{2},x_{3}) \in V_{0,t}\colon ({\textstyle \sqrt{x_{1}^{2}+x_{2}^{2}}},x_{3}) \in S_{t}+\tilde{q}_{0} \Bigr\}, \\
  S_{t} = \Bigl\{ (\tilde{y}_{1},\tilde{y}_{2})\colon \phi = \tan^{-1} (\tilde{y}_{2}/\tilde{y}_{1}) \in (\tfrac{11}{20} \pi,\tfrac{3}{4} \pi),\ \abs{\tilde{y}} \in (\tfrac{1}{2} \rho(\phi),\rho(\phi)) \Bigr\},
\end{gather*}
where $\rho(\phi),\ \phi \in [\frac{1}{2} \pi,\pi]$, denotes a parametrization of the curve $C_{\infty}(t)$ in polar coordinates (see Figure \ref{fig_ssim_contv}\subref{fig_ssim_contv_lin} for an illustration of $C_{\infty}(t)$). Note that
$S_{t}+\tilde{q}_{0}$ lies in the interior of the set $D_{\infty}(t)$ and stays away from the rays $z = 0,\ r = 1$ where $\abs{D}$ is small.

Now we estimate
\begin{displaymath}
  K_{\rho}(q_{0},t) \geq \frac{1}{2} \norm{\omega(\cdot,t)}_{\infty} \ell_{D,q_{0}}(t) \int_{N_{1,t}} \frac{dy}{\abs{y}^{2}},
\end{displaymath}
where
\begin{displaymath}
  \ell_{D,q_{0}}(t) = \inf_{y \in N_{1,t}} \frac{1}{\abs{y}} \abs{D(\hat{y},\xi(q_{0}+y,t),\xi(q_{0},t))}.
\end{displaymath}
For each $y \in N_{1,t}$ and $x \in V_{1,t}$ such that $y = x-q_{0}$, we have
\begin{displaymath}
  \abs{y} = \abs{x-q_{0}} \leq \abs{x-\tilde{x}} + \abs{\tilde{x}-\tilde{q}_{0}} \leq d_{1} + \rho(\phi) \leq 2\rho(\phi),\qquad \phi = \pi + \tan^{-1} \frac{x_{3}}{r-1},
\end{displaymath}
where as usual $\tilde{x} = (r,x_{3})^{T}$ denotes the projection of $x$ onto the $rz$-plane. It then follows that
\begin{displaymath}
  \int_{N_{1,t}} \frac{dy}{\abs{y}^{2}} \geq \frac{1}{4} \int_{N_{1,t}} \frac{dy}{\rho^{2}(\phi)} \geq \frac{1}{8}\, d_{1} \int_{11\pi/20}^{3\pi/4} \rho^{-2}(\phi) \int_{\rho(\phi)/2}^{\rho(\phi)} s\,ds\,d\phi = \frac{3}{320}\, d_{1} \pi,
\end{displaymath}
where in the second inequality above we have used the fact that, for any $y \in N_{1,t}$, the distance between the point $x = y+q_{0}$ and the symmetry axis is greater than $\frac{1}{2}$ for $t$ sufficiently close to $t_{s}$. This leads to the estimate
\eqref{eqn_k_lbd_m}:
\begin{displaymath}
  K_{\rho}(q_{0},t) \geq \frac{3\pi}{640}\, d_{1}(D_{\infty}(t)) \norm{\omega(\cdot,t)}_{\infty} \ell_{D,q_{0}}(t).
\end{displaymath}

\end{document}